\PassOptionsToPackage{table}{xcolor}

\documentclass[acmsmall,screen,nonacm]{acmart}

\acmJournal{TAAS}
\acmVolume{0}
\acmNumber{0}
\acmArticle{0}
\acmYear{2026}
\acmMonth{0}
\copyrightyear{2026}
\setcopyright{none}                   
\usepackage{amsmath,amssymb}
\usepackage{graphicx}
\usepackage{xcolor}
\usepackage{url}
\usepackage{booktabs}
\usepackage{tabularx}
\usepackage{multirow}
\usepackage{adjustbox}
\usepackage{enumitem}
\usepackage{subcaption}
\usepackage[capitalize]{cleveref}
\usepackage{tikz}
\usetikzlibrary{arrows.meta,positioning,fit,backgrounds,calc}

\newcommand{\smp}{MELD}

\title{MELD: A Protocol for Merging Knowledge Across Distributed Agentic Memories}

\author{Lauri Lov\'{e}n}
\orcid{0000-0001-9475-4839}
\authornote{Corresponding author.}
\affiliation{%
  \institution{Future Computing Group, University of Oulu}
  \city{Oulu}
  \country{Finland}
}
\email{lauri.loven@oulu.fi}

\author{Jaakko Sauvola}
\affiliation{%
  \institution{University of Oulu}
  \city{Oulu}
  \country{Finland}
}

\author{Jukka Riekki}
\affiliation{%
  \institution{University of Oulu}
  \city{Oulu}
  \country{Finland}
}

\author{Sasu Tarkoma}
\affiliation{%
  \institution{University of Helsinki and University of Oulu}
  \city{Helsinki}
  \country{Finland}
}

\renewcommand{\shortauthors}{Lov\'{e}n et al.}

\ccsdesc[500]{Computer systems organization~Self-organizing autonomic computing}
\ccsdesc[300]{Theory of computation~Distributed computing models}
\ccsdesc[300]{Computing methodologies~Natural language processing}
\ccsdesc[100]{Networks~Middleware for databases}

\keywords{Knowledge merging, distributed agent memory, autonomic systems,
self-management, conflict-free replicated data types, eventual consistency,
semantic publish/subscribe, multi-agent systems, agentic AI, computing continuum}

\begin{document}

\begin{abstract}
Autonomous agents share a transport and can call each other's tools, but they cannot share what they know: no protocol lets two agents' memories reconcile a fact phrased two ways, link related facts held apart, or reconcile contradictory knowledge without silently discarding either claim. We present \smp{}, a self-managing coherence mechanism for a federation of agent memories whose run-time model is the knowledge graph itself. Each brain admits every incoming claim through a five-outcome procedure (\emph{insert}, \emph{merge}, \emph{relate}, \emph{conflict}, or \emph{reject}), decided from three signals (scoped claim-key identity, embedding similarity, and a natural-language-inference verdict) under context and freshness gates, and acting through exactly one auditable, authenticated \emph{Patch}, the only object that mutates state. A binding onto standard publish/subscribe transport with a per-claim status CRDT keeps sovereign brains coherent in claim \emph{status} without a coordinator: self-healing after partitions and under lossy routing, and self-protecting against silent rewrite by a peer, under a benign-fault model. \smp{} does not adjudicate truth; a detected contradiction is preserved for later adjudication, never silently resolved. On HotpotQA distractor, distributed merge is recall-non-inferior to a centralized store under a pre-specified equivalence test and recall-superior to naive union at about $11\%$ less live storage; the merge classifier separates at AUC $0.968$ with a $0.013$ false-merge rate on adjudicated candidate pairs; the status CRDT reconverges in $30/30$ real partition-heal trials where last-writer-wins manages $11/30$; and semantic routing delivers about $3\times$ fewer messages at matched recall. We evaluate on a real computing continuum spanning an operator-grade 5G edge, national HPC, and a local tier, with empirically calibrated thresholds.

\end{abstract}

\maketitle

\section{Introduction}
\label{sec:intro}

\subsection{The missing knowledge-merge layer}
\begin{figure*}[t]
  \centering
  \includegraphics[width=0.86\textwidth]{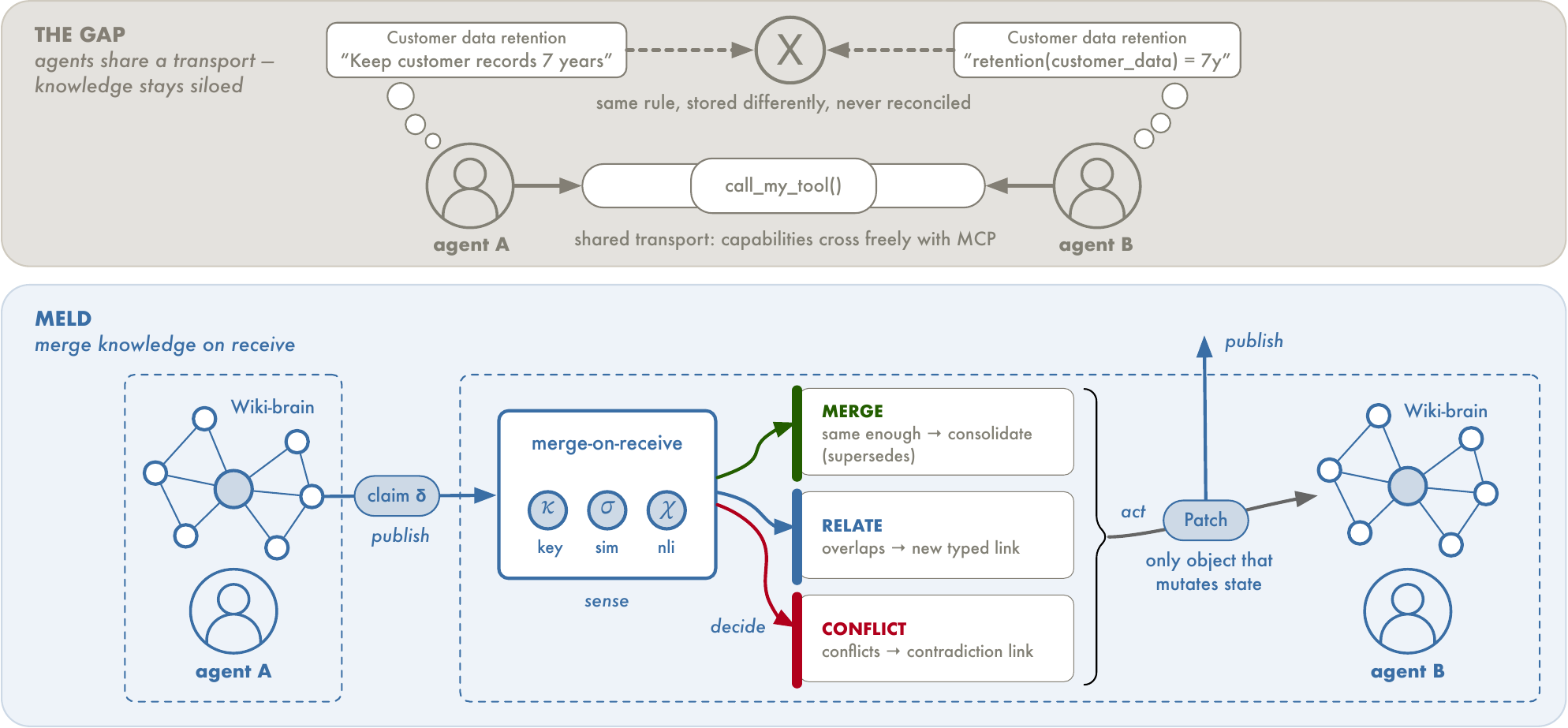}
  \caption{\smp{} at a glance. Top: a shared transport lets capabilities cross
  (e.g.\ via MCP), but the same fact stored two ways stays unreconciled. Bottom:
  each brain merges on receive: it senses three signals ($\kappa$ scoped claim-key,
  $\sigma$ similarity, $\chi$ inference), decides among same-enough,
  overlaps, conflicts, or insert (novel), and acts by one auditable
  Patch, the only object that mutates state (tabulated in the electronic supplement, Appendix~N).}
  \label{fig:hero}
\end{figure*}

Multi-agent infrastructure now provides connectivity, content-based and learned
routing, and, with tool-invocation protocols such as the Model Context Protocol
(MCP)~\cite{anthropic2024mcp}, a
standardized way for agents to expose and call one another's tools and resources.
It does not provide a way for them to share what they know (Figure~\ref{fig:hero}).

Each non-trivial agent accumulates a private, evolving memory: a knowledge
store of concepts and claims, the evidence grounding them, the contexts in which
they hold, and the typed relations among them (associations,
entity coreference, supersession, contradiction, precedent). Recent agent-memory
systems have converged on this shape (a versioned, typed-link knowledge
graph rather than a flat document store) in Zep/Graphiti's bi-temporal knowledge
graph~\cite{rasmussen2025zep}, A-MEM's linked atomic notes~\cite{xu2025amem},
GraphRAG's entity graph~\cite{edge2024graphrag},
HippoRAG~\cite{gutierrez2024hipporag}, and Mem0~\cite{chhikara2025mem0}. We call
one agent's such store its \emph{wiki brain}, after the distributed wiki-brain
vision that motivates this work. Two agents working in the same domain will,
over time, hold overlapping, complementary, and sometimes incompatible
knowledge, a divergence recent multi-agent memory systems observe but address
with access control rather than a merge protocol~\cite{rezazadeh2025collaborativememory}. There is today no protocol that lets two wiki brains reconcile a fact
they both hold but phrase or scope differently, record a relation between facts
that neither holds in isolation, or reconcile contradictory
knowledge without silently discarding either claim. Knowledge stays siloed in each agent, is re-derived redundantly, or
is flattened by ad-hoc last-writer-wins overwrites that silently discard
information and lose provenance.

The gap is not the interchange of knowledge, which is now being standardized: the
Open Knowledge Format~\cite{google2026okf} gives curated knowledge a vendor-neutral
representation an agent can consume, but it leaves reconciliation open. A format fixes how a
claim is written down but does not say what a brain should do when a peer's claim arrives
that says the same thing in other words, says something adjacent, or says the opposite.
Knowledge is merged, related, surfaced as a contradiction, or silently lost at
that moment of arrival rather than at parse time.

A tool-invocation protocol cannot fill this gap, because it is the wrong kind of object.
MCP moves capability (``here is a tool you may invoke''): its verbs are call and read,
and its state is the tool surface. It has no notion of two parties holding the same fact
differently, and hence none of merging them. The missing protocol must move and reconcile
knowledge (``here is a claim you may merge into your brain''), and its central
operation is reconciliation of two replicas that may hold the same surface term with the same,
a related, or an incompatible meaning, which is the gap this paper addresses.

The pieces such a protocol needs have only recently become deployable components: cheap
inference and high-quality embeddings make semantic similarity a practical online primitive,
routable over a federated publish/subscribe
transport~\cite{loven2026neuralrouter,loven2026neuralpubsub}; agentic memories have converged
on versioned, typed-link knowledge graphs~\cite{rasmussen2025zep,xu2025amem,edge2024graphrag};
and natural-language inference is now reliable enough to flag contradictions between claims at
scale. The missing piece is the protocol that composes them into merge.

\subsection{\smp{} in outline}
We present \smp{} (a name for the semantic meld, or merge, of knowledge): a
state-synchronization protocol for many sovereign agent wiki brains. Where MCP exposes tools and
resources and agent-to-agent protocols expose tasks~\cite{google2025a2a},
\smp{} exposes meanings, claims, and merges. \smp{} is built
around one operation that no tool protocol has: when a brain receives knowledge
from a peer, it does not apply that knowledge directly; it runs a merge decision
procedure that classifies the incoming knowledge against what it already holds
and emits an auditable record of the outcome (Figure~\ref{fig:hero}): \emph{same-enough}
(the same fact, merged), \emph{overlaps} (related but distinct, linked rather than collapsed),
or \emph{conflicts} (incompatible, the contradiction kept as a first-class object rather than
silently resolved). A single operation, PUBLISH of a graph delta, carries the knowledge;
the merge-on-receive decision supplies the semantics and is the load-bearing new idea.

In autonomic-computing terms, this merge-on-receive loop makes \smp{} a
\emph{self-managing coherence mechanism} for a federation of agent memories, and the
adapted run-time model is the knowledge graph itself. Each brain runs a per-delta
admission loop: it senses three signals (claim-key identity over content and scope, embedding similarity,
and an inference verdict), decides via the five-outcome admission procedure gated by Context, authority, and
freshness, and acts by emitting and applying a Patch that mutates its own memory
(Section~\ref{sec:merge}); the electronic supplement (Appendix~N) tabulates this loop element by element.
The loop closes on knowledge admission (the
inference verdict drives the sole object that mutates state), but it is not a
self-optimizing controller, as the decision thresholds are calibrated offline and
held fixed, with no run-time feedback that re-tunes them. Two self-* properties rest on it: \emph{self-healing}, the status CRDT reconverging a partitioned, reordered federation to one per-claim status without a coordinator (Section~\ref{sec:eval-c}); and \emph{self-protection}, authority-gated admission keeping a canonical brain from being silently rewritten by a peer (Section~\ref{sec:cons-trust}). Section~\ref{sec:consistency} states how each is bounded: the digest that frees self-healing from relying on the semantic-routing path for complete delivery, and the benign-fault (non-Byzantine) threat model self-protection assumes. We adopt neither a MAPE-K vocabulary nor a central manager: \smp{} realizes adaptation as a wire protocol and a per-brain decision procedure, in the lineage of autonomic computing~\cite{kephart2003vision} and self-adaptive software~\cite{salehie2009self}, and of recent self-adaptive LLM-based multi-agent systems~\cite{nascimento2023selfadaptive}, but as a protocol for knowledge merge rather than a managed-element framework.

\smp{} stands on a deliberately minimal substrate that public agent-memory systems largely provide: each wiki brain need only be a typed-link agent memory, a set of content-addressed nodes with typed links among them, exposing per-node claims as individually addressable assertions. Such stores are now standard (Section~\ref{sec:bg-layers} surveys Zep/Graphiti, GraphRAG, HippoRAG, Mem0, and A-MEM); \smp{} additionally relies on per-claim status and validity intervals to make supersession non-lossy, first-class today only in Zep/Graphiti's bi-temporal model~\cite{rasmussen2025zep}, and supplies them at the protocol layer for stores that track only current state. \smp{} defines its data model self-contained against this shape (Section~\ref{sec:bg-datamodel}) and depends on no particular memory implementation. \smp{}'s own contribution is the wire protocol, the merge decision procedure, and the Patch object; we evaluate distributed agents merging into a shared brain against a centralized memory store, with MCP as the contrast object and the public agent-memory stores treated as the substrate \smp{} synchronizes. We do not re-derive their results.

\subsection{Contributions}
\label{sec:contributions}
This paper makes four contributions, all at the protocol/systems layer and all
self-standing over a public, versioned typed-link agent memory as substrate:

\begin{enumerate}
  \item \textbf{A wire protocol for knowledge merge.} A single wire operation,
  PUBLISH (contribute a graph delta of concepts, claims, and typed links), under a
  content-based delivery contract; status changes (supersession, overrule, revocation)
  ride as append-only status links, and the only new wire object is the Patch.
  Prior framings name ``subscribe/publish/update is the protocol'' as a slogan;
  \smp{} specifies the message types, the graph-delta payload, and the
  merge-result semantics that make it an actual protocol (Section~\ref{sec:protocol}).

  \item \textbf{An operational merge decision procedure.} An executable procedure
  that, for each incoming claim, decides \emph{insert} (novel), same-enough,
  overlaps, conflicts, or \emph{reject} (an admission-gate drop) from three
  standard signals (claim-key identity over content and scope, embedding similarity, and
  natural-language inference) gated by context compatibility, authority, and freshness.
  The procedure first adjudicates whether an incoming claim has any candidate at all;
  candidate selection is an exhaustive scan within the receiving brain, whose cost we
  measure directly (Section~\ref{sec:eval}). The novelty is the combining procedure
  rather than the borrowed signals. Once a candidate pair exists, the pair-adjudication
  core is a hierarchy (exact claim-key match in a compatible frame first, scoped to content and discrete
  scope so a match cannot cross a jurisdiction boundary, then embedding-plus-gates) rather
  than a bare similarity threshold, and it never silently resolves a detected
  contradiction (Section~\ref{sec:merge}).

  \item \textbf{The Patch as a first-class, auditable wire object.} The result of
  a merge is reified as an authenticated, versioned \emph{Patch}: it records the decision,
  the target, the emitted deltas, and which gates fired, so that a merge is
  itself replayable and auditable across the federation. The Patch, not the raw
  incoming delta, is the only thing that mutates local state, and it is itself
  publishable (Sections~\ref{sec:protocol} and~\ref{sec:merge}).

  \item \textbf{A binding of merge semantics onto a standard publish/subscribe
  transport, with status convergence.} A mapping of graph deltas (node, link, and
  claim-version sets) onto a standard content-based publish/subscribe fabric (Apache
  Kafka in our deployment), with semantic routing as an optional content-matching
  layer; an append-only wire discipline that preserves the premises of a status
  conflict-free replicated data type end-to-end; and an authority-gated merge-admission
  policy over a trust hierarchy of canonical brains. The result keeps many sovereign brains coherent
  in real time: per-claim status converges via a
  self-contained status conflict-free replicated data type built on standard CRDT
  theory, while the semantic graph structure each claim sits in is adjudicated locally,
  per delivery (Sections~\ref{sec:consistency} and~\ref{sec:impl}).
\end{enumerate}

We evaluate \smp{} by having $N$ agents sync knowledge into a shared global brain, an aggregate built by decentralized merge, and answer queries against the merged result, measured against a centralized single-store baseline, on the real three-tier computing continuum the protocol targets (a 5G test-network edge, national HPC, and a local host), with \smp{} additionally deployed live across all three (Section~\ref{sec:eval}). The main result is that decentralized merge is recall-non-inferior to a centralized store and recall-superior to naive union at lower live storage; we report a separate experiment for the merge classifier, the status CRDT's partition-heal convergence, semantic routing, and fan-out latency. Numbers and baselines are deferred to Section~\ref{sec:eval}.

Five limits frame these contributions: merge is evaluated relative to a centralized
baseline; consistency converges on claim status without addressing truth; the merge graph itself
is order-sensitive, unlike per-claim status; first-class relational
structure rests on an open empirical premise; and the merge thresholds are calibrated
rather than derived. Section~\ref{sec:disc-limits} states and develops each.

\section{Background}
\label{sec:background}

This section fixes two things the rest of the paper relies on: the minimal
substrate \smp{} assumes and where that substrate already exists in public
systems (Section~\ref{sec:bg-layers}); and a self-contained statement of the data
model \smp{} synchronizes (Section~\ref{sec:bg-datamodel}). The contrast with
tool- and task-level agent protocols is drawn in Section~\ref{sec:related}.

\subsection{The substrate \smp{} assumes}
\label{sec:bg-layers}
\smp{} is a protocol above an agent-memory store, and it is deliberately
undemanding about that store. The only substrate it requires is a versioned,
typed-link agent memory: a store of content-addressed nodes, typed and versioned
links among them, and per-node claims with a status and a validity interval. This shape is approximated, to differing degrees, by the public stores introduced in Section~\ref{sec:intro}: Zep/Graphiti realizes it most fully, with typed links and validity intervals first class on every edge~\cite{rasmussen2025zep}; A-MEM's associative links are untyped~\cite{xu2025amem}; and GraphRAG~\cite{edge2024graphrag}, HippoRAG~\cite{gutierrez2024hipporag}, and Mem0's graph variant~\cite{chhikara2025mem0} type their links but leave per-claim versioning and status derivable rather than first-class. The emerging Open Knowledge Format~\cite{google2026okf} sits at the same layer as an interchange encoding: it carries provenance, trust, and lifecycle fields, but its links are untyped and its lifecycle is per concept rather than per claim. \smp{} supplies typed links, status, and validity intervals at the protocol layer for stores that track only current state, treating any such store as a wiki brain it synchronizes rather than mandates or reinvents. The new objects (the wire protocol, the merge decision procedure, and the trust-hierarchy admission policy) live at \smp{}'s own protocol/systems layer.

\subsection{The wiki-brain data model \smp{} synchronizes}
\label{sec:bg-datamodel}
An agent's knowledge is a versioned, typed-link knowledge graph, which we call its
wiki brain. We state the model self-contained, recapping only what the
protocol touches. The model is the shape the stores above approximate; it restates no single
system.

A brain holds nodes (content-addressed, carrying an owner identity, an origin tag, an authority level, a confidence, an access-control scope, and a retention state) and first-class links (typed, versioned, authenticated, weighted relations such as association, entity coreference, supersession, partial overrule, grounding, and contradiction). A node carries one or more claims: addressable assertions, each with a claim key derived from its canonical content together with its discrete scope, a confidence, a validity interval, and a status drawn from a monotone lifecycle (active, deprecated, overruled, revoked). Assertions are appended, not overwritten: a new version supersedes an old one through a link rather than by mutation, so provenance and history are preserved. Each node carries a staleness state (a content hash, a fetch time, a freshness budget, and a latest flag) that governs whether it should be re-fetched or allowed to decay.

Two properties of this model matter for the protocol. First, the natural unit of
exchange is a graph delta (a version-set of nodes, claims, and links),
not a flat document. Second, because links are first class and typed, the result of
reconciling two brains can itself be represented in the graph (as an
association, a coreference, or a contradiction link) rather than collapsed into a
single overwritten value. \smp{} exploits both.

\section{The \smp{} protocol}
\label{sec:protocol}

\smp{} is a state-synchronization protocol for sovereign agent wiki brains. Its
transport surface is a single operation (PUBLISH of a graph delta under a
content-based delivery contract), and its semantics describe what a brain does when a
delta is delivered to it. This section specifies the wire surface
(Section~\ref{sec:proto-verbs}) and the typed objects it carries
(Section~\ref{sec:proto-primitives}), including the Patch, the protocol's
distinctive new wire object. The decision procedure that a delivered delta triggers
is the subject of Section~\ref{sec:merge}.

\subsection{A single wire operation}
\label{sec:proto-verbs}
\smp{} reduces prior ``subscribe, publish, update'' framings to one wire operation, \texttt{PUBLISH(graph-delta)}, whose payload is a
graph delta (a version-set of nodes, claims, and typed links), the unit of
exchange throughout \smp{}. The slogan's ``subscribe'' and ``update'' are not
separate verbs: a delta is delivered by content-based interest match (a transport
concern, scoped by cell and authority, under the at-least-once / idempotent /
eventually-complete contract of Section~\ref{sec:consistency}), and a status change
(supersession, overrule, revocation) is itself a PUBLISH of an append-only status
link, never an in-place edit. The append-only form preserves the status-CRDT
premises end to end.

A receiving brain does not apply an incoming delta as delivered; the delta is decomposed by type. Each incoming claim runs the merge decision procedure (Section~\ref{sec:merge}); status links (supersede, overrule, revoke) are admitted append-only into the status CRDT after MAC verification, through a Patch of the \textsc{status} kind, and filtered at materialization-read (Section~\ref{sec:consistency}); node metadata and other typed links follow the store's own append rule. In every case, local state is mutated by a Patch and by nothing else.

\subsection{Typed objects as concrete message and state types}
\label{sec:proto-primitives}
\smp{} names typed objects over the wiki-brain data model; only the Patch is new. The
node, claim, and typed-link types map directly onto the data model
(Section~\ref{sec:bg-datamodel}); Evidence and Context are roles over those
types, not new wire objects: a Claim's grounding attribute (its provenance; an
evidence-overlap signal and an evidence-weighted tie-break are future work, Section~\ref{sec:disc-future})
and the derived validity/scope frame the merge gate reads (Section~\ref{sec:merge}). The typed-link form
is first-class in Zep/Graphiti~\cite{rasmussen2025zep} and present in
the graph-based public stores~\cite{edge2024graphrag,gutierrez2024hipporag,chhikara2025mem0};
A-MEM supplies the node-and-link core through untyped associative links~\cite{xu2025amem}.
We adopt this shape and do not re-specify it.

\begin{description}[leftmargin=1.2em,labelsep=0.5em,font=\bfseries]
  \item[Concept] a node: a named meaning or entity the brain holds, content addressed and carrying owner, authority, confidence, access scope. The
  candidate merge node at the entity or topic level.
  \item[Claim] an addressable assertion inside a node, with a claim key derived from canonical content and discrete scope, confidence, validity interval, and status. The claim is the merge key:
  same-enough is decided at claim granularity, not at whole-node granularity.
  \item[Mapping] a typed, authenticated, versioned, weighted link asserting a relation
  between two meanings or claims across brains (association, entity coreference,
  supersession, partial overrule). A Mapping is the relate-don't-merge output:
  a typed link recording a cross-brain association or coreference.
  \item[Patch] the reified result of running the decision procedure on an incoming
  delta.
\end{description}

\paragraph{The Patch is the new wire object.}
A Patch's decision field carries a claim-admission outcome or the \textsc{status} kind for an already-authored status-link append (Section~3.1). A Patch is an authenticated, versioned record
\[
  \begin{aligned}
    \texttt{Patch} = \langle\, &\mathrm{decision},\ \mathrm{target},\
      \mathrm{emitted\text{-}deltas},\\
    &\mathrm{gates\text{-}that\text{-}fired},\ \mathrm{version} \,\rangle ,
  \end{aligned}
\]
where $\mathrm{decision} \in \{\textsc{insert}, \textsc{merge}, \textsc{relate}, \textsc{conflict},
\textsc{reject}\}$, $\mathrm{target}$ is the affected endpoint, and
$\mathrm{gates\text{-}that\text{-}fired}$ records which signals drove the decision
(claim-key match, embedding score, inference outcome, freshness gate). The Patch is the only object that mutates local state, and it is
itself publishable, so a global brain can audit and replay merges across the
federation. Replay re-runs the decision procedure against the consumer's own held
state, so a replayed decision can differ from the one recorded in the Patch if the
consumer's state has since changed. The Patch preserves the original decision and the
gates that produced it for audit, and does not guarantee that replay reproduces it. The five
decisions emit, respectively: an \textsc{active} claim-status link for a novel
\textsc{insert}; a claim-status link for a \textsc{merge} (\textsc{active} for an
exact-key match; \textsc{deprecated} under a \textsc{supersedes} link for a
consolidated paraphrase); a Mapping link for a
\textsc{relate}; a first-class contradiction link for a \textsc{conflict} (with no
silently chosen winner); and a recorded no-op for a \textsc{reject} (the admission gate
alone: a stale or overruled delta). Reifying the merge result as a
first-class, auditable, replayable wire object, rather than letting a brain mutate
silently on receipt, is a contribution of \smp{}.

\begin{figure*}[t]
  \centering
  \resizebox{0.90\textwidth}{!}{%
  \begin{tikzpicture}[
    font=\footnotesize,
    head/.style={draw=black!55, line width=0.5pt, rounded corners=2pt,
                 align=center, inner sep=3pt, font=\scriptsize, fill=white},
    life/.style={black!35, line width=0.5pt, dash pattern=on 2pt off 2pt},
    msg/.style={-{Latex[length=1.8mm]}, black!70, line width=0.7pt},
    xmsg/.style={-{Latex[length=1.8mm]}, black!88, line width=1.0pt},
    ml/.style={font=\scriptsize, text=black!80, align=center, inner sep=1.5pt},
    sl/.style={font=\scriptsize, text=black!55, align=center, inner sep=1pt},
    note/.style={font=\scriptsize, text=black!55, align=center, inner sep=1pt},
    dom/.style={font=\scriptsize\bfseries, text=black!55},
    verd/.style={draw=black!55, line width=0.6pt, rounded corners=2pt,
                 align=center, inner sep=2.5pt, font=\scriptsize\bfseries,
                 fill=white, minimum height=0.5cm},
    averd/.style={verd, draw=black!85, line width=1.0pt, fill=black!82, text=white}
  ]
    \def\yB{-7.9}
    \coordinate (US) at (0,0);     
    \coordinate (A)  at (3.4,0);   
    \coordinate (T)  at (5.8,0);   
    \coordinate (C)  at (8.2,0);   
    \coordinate (B)  at (10.6,0);  

    \begin{scope}[on background layer]
      \fill[black!8, rounded corners=3pt] (-1.4,0.55) rectangle (1.3,\yB-0.15);
      \fill[black!4, rounded corners=3pt] (2.1,0.55) rectangle (12.6,\yB-0.15);
    \end{scope}
    \draw[black!50, dash pattern=on 3pt off 2pt, line width=0.7pt]
      (1.7,0.5) -- (1.7,\yB-0.1);
    \node[dom] at (0,0.85) {US domain};
    \node[dom] at (7.35,0.85) {EU domain};
    \node[note, rotate=90, text=black!45, font=\scriptsize] at (1.7,-7.0)
      {admin boundary};

    \node[head] (hUS) at (US) {US domain\\aggregate $B_{US}$};
    \node[head] (hA)  at (A)  {Edge agent\\(EU)};
    \node[head] (hT)  at (T)  {Team agent\\(EU)};
    \node[head] (hC)  at (C)  {EU authority $C$\\(canonical)};
    \node[head] (hB)  at (B)  {\textbf{EU aggregate $B$} (receiver)\\holds $c_1$ active};
    \foreach \n in {hUS,hA,hT,hC,hB}{\draw[life] (\n.south) -- ($(\n.south)+(0,\yB)$);}

    \def\ym{-1.2}
    \draw[msg] ($(hA.south)+(0,\ym)$) -- ($(hB.south)+(0,\ym)$);
    \node[ml, above] at ($(hA.south)!0.5!(hB.south)+(0,\ym)$)
      {\texttt{PUBLISH} $c_1'$: ``\dots retain customer data for seven years''};
    \node[sl, below] at ($(hA.south)!0.5!(hB.south)+(0,\ym)$)
      {$\sigma{=}0.93$,\ $\chi{\neq}$contradicts,\ ctx EU${\approx}$EU};
    \node[verd, anchor=west] at ($(hB.south)+(0.35,\ym)$)
      {\textsc{merge}\\[-1pt]{\scriptsize\mdseries Patch: \textsc{supersedes} $c_1$ ($c_1'$ deprecated)}};

    \def\yr{-3.0}
    \draw[xmsg] ($(hUS.south)+(0,\yr)$) -- ($(hB.south)+(0,\yr)$);
    \node[ml, above] at ($(hUS.south)!0.5!(hB.south)+(0,\yr)$)
      {\texttt{PUBLISH} same rule, scope $\langle$US$\rangle$ (across the boundary)};
    \node[sl, below] at ($(hUS.south)!0.5!(hB.south)+(0,\yr)$)
      {$\sigma{=}0.93$,\ $\chi{\neq}$contradicts,\ ctx EU${\not\approx}$US};
    \node[verd, anchor=west] at ($(hB.south)+(0.35,\yr)$)
      {\textsc{relate}\\[-1pt]{\scriptsize\mdseries Patch: Mapping link (not fused)}};

    \def\yc{-4.8}
    \draw[msg] ($(hT.south)+(0,\yc)$) -- ($(hB.south)+(0,\yc)$);
    \node[ml, above] at ($(hT.south)!0.5!(hB.south)+(0,\yc)$)
      {\texttt{PUBLISH} $c_2$: ``\dots delete\dots after two years''\,$\langle$EU$\rangle$};
    \node[sl, below] at ($(hT.south)!0.5!(hB.south)+(0,\yc)$)
      {$\sigma{=}0.86$,\ $\chi{=}$\textbf{contradicts},\ ctx EU${\approx}$EU};
    \node[verd, anchor=west] at ($(hB.south)+(0.35,\yc)$)
      {\textsc{conflict}\\[-1pt]{\scriptsize\mdseries Patch: $c_1{\leftrightarrow}c_2$ edge, no winner}};

    \def\yo{-6.6}
    \draw[msg] ($(hC.south)+(0,\yo)$) -- ($(hB.south)+(0,\yo)$);
    \node[ml, above] at ($(hC.south)!0.5!(hB.south)+(0,\yo)$)
      {\texttt{PUBLISH} overrule$(c_2)$ link};
    \node[sl, below] at ($(hC.south)!0.5!(hB.south)+(0,\yo)$)
      {domain-scoped: applied};
    \node[averd, anchor=west] at ($(hB.south)+(0.7,\yo)$)
      {$c_2 \Rightarrow$ \textsc{overruled}\\[-1pt]{\scriptsize\mdseries join, any order}};
  \end{tikzpicture}}
  \caption{Running example as a sequence trace (threaded through
  Sections~\ref{sec:protocol}--\ref{sec:consistency}); it instantiates the
  hybrid, cross-administrative-boundary deployment of
  Section~\ref{sec:disc-deploy}. Two admin domains (US, EU) each run a domain
  aggregate and peer across the dashed boundary. Four reasons make the brains
  differ, each driving one mechanism: locality (each tier ingests its own
  source), data minimization (a privacy-relevant property, not a guarantee:
  only a claim crosses the wire, never the
  customer data, message~1), sovereignty (the Context gate keeps the US
  rule related, not fused when the US aggregate peers across the boundary,
  even at identical similarity; the scope is part of the claim key, so the two are
  not even key-identical, Section~\ref{sec:merge-signals}; message~2, the one
  arrow that crosses), and
  latency/partition (the status CRDT converges $c_2$ to \textsc{overruled}
  regardless of delivery order, message~4). Receiver $B$ runs the decision
  procedure on each delta and emits exactly one Patch. The overrule (accent) is
  applied because it carries a MAC under the federation group key (asserted
  origin: EU authority $C$) and stays
  within the domain; a US-origin overrule of an EU claim would append but never
  materialize (the authorization predicate, Section~\ref{sec:consistency}). All signal values reproduce from the reference
  implementation.}
  \label{fig:running}
\end{figure*}
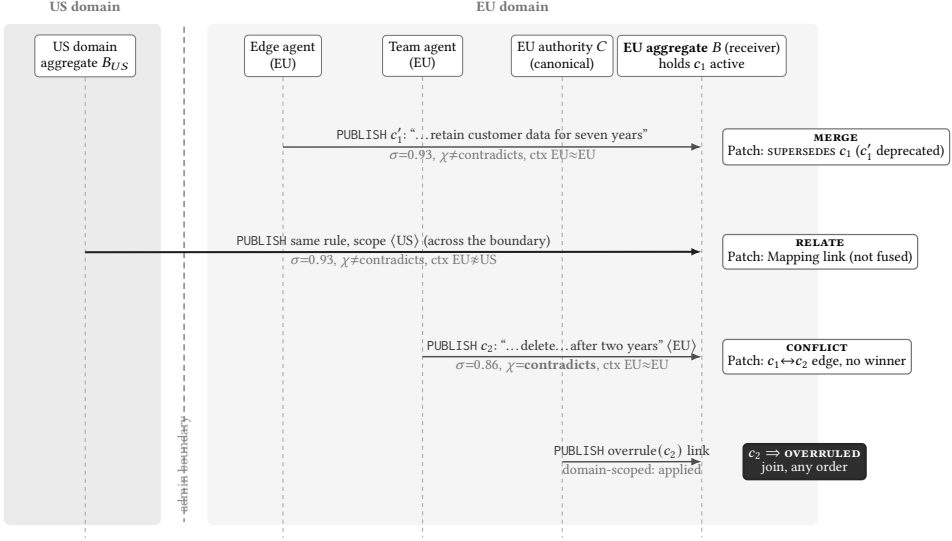

\textbf{Running example: governance agents across the continuum.}
We thread one example through this section and the next two
(Figure~\ref{fig:running}). A multinational runs data-governance agents across two
administrative domains, EU and US, each operating its own domain
aggregate (a per-domain global brain) that peers with the other across the
boundary, the hybrid deployment of Section~\ref{sec:disc-deploy}. The EU
domain holds an edge agent $A$, a team agent, the EU domain aggregate $B$
(our receiver), and the EU canonical authority $C$ of regulator-grade authority;
the US domain runs its own aggregate $B_{US}$. Their brains hold different
knowledge, and the reasons are the continuum's own drivers, each of which the
protocol must answer: locality (each tier and domain ingests its own policy
source), data sovereignty (a rule is scoped to a jurisdiction and may not
be silently fused across a domain boundary), latency and partition (edge
tiers run intermittently disconnected and cannot consult $C$ synchronously on each
receive), and data minimization (a privacy-relevant property, not a privacy
guarantee). Data minimization is part of why \smp{} exists in this form: the
domains may not pool the underlying customer records, so a claim and a Patch cross
the wire, and the boundary, rather than the underlying data. Brain
$B$ already holds the claim $c_1 = $ ``Customer data must be retained for seven
years,'' with Context $\langle\text{jurisdiction}{:}\,\text{EU}\rangle$ and status
active, its Evidence the governing handbook clause. When the EU edge agent
$A$ now \texttt{PUBLISH}es a paraphrase of the same requirement, $B$ does not apply
it directly: it runs the decision procedure of Section~\ref{sec:merge} and emits one
Patch. Section~\ref{sec:merge} resolves that delta and two more (the US domain
aggregate peering its rule across the boundary, and a contradicting publish
from an EU team agent); Section~\ref{sec:consistency} converges the status after
$C$ overrules within the EU domain.

\section{Merge semantics}
\label{sec:merge}

The procedure returns one of five outcomes for an incoming claim: insert
(novel, no matching candidate), same-enough, overlaps,
conflicts, or reject (an admission-gate drop). Once a candidate
exists, a four-verdict pair-adjudication classifier decides among \textsc{merge},
\textsc{relate}, \textsc{conflict}, and \textsc{no-relation} (no actionable relation), which maps to
\textsc{insert} at the protocol level (Section~\ref{sec:merge-procedure}).
\smp{}'s contribution is the decision procedure that
computes which outcome holds for an incoming claim, and the Patch it emits. The
procedure runs per incoming claim $b$ against the claims the receiving brain
already holds, delivered by a PUBLISH, within a candidate-matched Concept.

\subsection{Three signals}
\label{sec:merge-signals}
The decision procedure reads three signals over the pair $(a,b)$. The signals
themselves are standard, drawn from prior art, and \smp{}'s contribution is not the
signals but the safety-preserving decision procedure that combines them
(Section~\ref{sec:merge-procedure}):
\begin{itemize}
  \item $\kappa = [\,\mathrm{claim\text{-}key}(a) = \mathrm{claim\text{-}key}(b)\,]$, with
  $\mathrm{claim\text{-}key}(x) = H\!\big(\mathrm{canon}(x)\ \|\ \mathrm{scope}(x)\big)$:
  the key hashes a claim's canonical content together with its discrete scope
  (its cell), so $\kappa = 1$ iff two assertions are byte-identical after
  canonicalization and carry the same scope (standard content addressing,
  extended to the frame). Only the discrete scope enters the key; the
  validity interval and freshness stay in the Context gate, because hashing a
  continuous field would break key identity on any timestamp difference. Binding
  scope into identity makes data sovereignty structural: the same sentence
  published under a different jurisdiction is a different claim, so the exact-key
  fast path cannot fuse across an administrative boundary and the guarantee of
  Section~\ref{sec:merge-sameenough} does not depend on a downstream gate. Exact and
  cheap, with no model call. Because independently authored brains rarely produce
  byte-identical keys for the same fact (canonicalization is local and key
  assignment is decentralized), $\kappa$ is an opportunistic fast path, not a
  load-bearing identity test: when keys diverge (the common case), same-enough falls
  through to the semantic gate ($\sigma$ with the $\chi$ guard,
  Section~\ref{sec:merge-sameenough}), so a mis-canonicalized key degrades to a
  re-decided \textsc{relate}/\textsc{merge}, never a permanent split. The ablation
  confirms the semantic path carries paraphrase matching ($\kappa$ alone scores
  \textsc{merge}-recall $0$, Section~\ref{sec:eval-m2}), so scoping the key costs the
  evaluated procedure nothing.
  \item $\sigma = \cos\!\big(\mathrm{emb}(a), \mathrm{emb}(b)\big)$, the dense
  semantic similarity of the claim texts under a sentence
  encoder~\cite{reimers2019sbert}.
  \item $\chi = \mathrm{NLI}(a,b)$, a natural-language-inference
  verdict~\cite{bowman2015snli} used as a contradiction detector: reads only whether $b$ contradicts $a$, not the finer
  entails/neutral distinction.
\end{itemize}
A fourth signal, grounding-evidence overlap (Jaccard overlap of two claims'
provenance), applies where claims carry grounding; we find a fixed rule for it
inherits the inference signal's contradiction-miss rate, so safe use needs a learned,
inference-quality-aware combination. The core procedure here keeps to the three
deployment-universal signals; the grounding signal and the learned combination are
future work (Section~\ref{sec:disc-future}).
The embedding similarity and the inference verdict are online primitives; \smp{}
composes them into a five-outcome admission procedure (Section~\ref{sec:merge-procedure})
whose pair-adjudication core is a four-way classifier, gated by Context, authority,
and freshness, that emits a Patch.

\subsection{The decision procedure}
\label{sec:merge-procedure}
Let $\theta_{\mathrm{merge}}$ be a calibrated merge threshold and
$\sigma_{\mathrm{lo}} \le \theta_{\mathrm{merge}}$ a relatedness floor. Write
$\mathrm{Context}(a) \approx \mathrm{Context}(b)$ when the two claims share a
compatible frame (same cell, overlapping validity interval, compatible freshness),
$R_{\min}$ for the minimum retention below which a node is treated as stale, and
$\mathrm{auth}(b\!\to\!a)$ for the authority-admission predicate of
Section~\ref{sec:cons-trust}, which fails exactly when the incoming source is
strictly less authoritative than the local target. The procedure runs on an
incoming claim $b$ against the claims the receiving brain already holds, and returns
exactly one of five outcomes. It is a first-match cascade: the rules are
evaluated in the order written and the first rule whose guard holds returns, so the
outcomes are mutually exclusive by construction, and rule~R6 carries no guard, so the
procedure is total.
{\small
\begin{align*}
\textbf{R1}\ \textsc{reject} \ \ &\text{if}\ \ R(b) < R_{\min}\ \vee\ \mathrm{status}(b) \ne \textit{active};\\[2pt]
\textbf{R2}\ \textsc{insert}\ (\text{novel}) \ \ &\text{if no local $a$ has}\ \ \kappa(a,b) = 1\ \vee\ \sigma(a,b) \ge \sigma_{\mathrm{lo}};\\[3pt]
  &\text{otherwise let $a^\star$ be the best such candidate, and read}\\
  &\text{$\kappa,\sigma,\chi$ on the pair $(a^\star,b)$:}\\[3pt]
\textbf{R3}\ \textsc{conflict} \ \ &\text{if}\ \ \chi = \textit{contradicts};\\[2pt]
\textbf{R4}\ \textsc{merge}\ (\text{same-enough}) \ \ &\text{if}\ \ \kappa = 1\ \wedge\ \mathrm{Context}(a^\star) \approx \mathrm{Context}(b);\\[2pt]
\textbf{R5}\ \textsc{merge}\ (\text{same-enough}) \ \ &\text{if}\ \ \sigma \ge \theta_{\mathrm{merge}}\ \wedge\ \mathrm{Context}(a^\star) \approx \mathrm{Context}(b)\\
  &\ \ \ \ \ {}\wedge\ \mathrm{auth}(b\!\to\!a^\star);\\[2pt]
\textbf{R6}\ \textsc{relate}\ (\text{overlaps}) \ \ &\text{otherwise}.
\end{align*}}
The candidate $a^\star$ is deterministic given the store state, and selection is
gate-aware. Candidates are the receiving brain's active claims only (a deprecated or
overruled claim is never a merge target); an exact-key match, which under the scoped
key implies the same content in the same scope, is selected first; otherwise
$a^\star$ is the active claim of maximum similarity $\sigma$, ties resolved to the
earliest-admitted claim. When the provisional candidate (the exact-key hit if one exists, otherwise the
maximum-similarity claim) does not contradict $b$ but fails a merge-admission gate,
and another active claim satisfies all the gates at $\sigma \ge \theta_{\mathrm{merge}}$,
the procedure adjudicates that claim instead, with its own inference check, so a
contradiction there still surfaces; an ineligible candidate, exact-key or semantic,
cannot shadow an admissible merge. A contradicting provisional candidate is always
the adjudication target. Selection is an
exhaustive scan of the local store; its measured cost is in
Section~\ref{sec:eval-m2} and the electronic supplement (Appendix~M).
Admission adjudicates a single relationship, reading at most two candidate pairs
when the gate-aware fallback fires: \smp{} performs single-target admission, and it
does not attempt exhaustive relational reconciliation. A
contradiction is therefore discovered when the contradicting claim is the admission
target; a second-best candidate that also contradicts the incoming claim is not
tested at admission, and completeness of the semantic link set is out of scope.

MAC verification precedes the procedure: a delta whose MAC does not verify
against the shared group key is dropped before any signal is read
(Section~\ref{sec:cons-threat}). That drop is not a \textsc{reject}. It happens before
R1, and an unverifiable delta never reaches adjudication.

\textsc{reject} (R1) is the admission gate, and nothing else. A stale or overruled
delta is rejected before any semantic signal is consulted, so such deltas never
propagate. \textsc{reject} never means ``no match found'': that case is R2. When the
receiving brain holds no claim at or above the relatedness floor there is nothing to
adjudicate against, so the incoming claim is admitted as a new node with status
active and no link. A federation whose receivers dropped unmatched claims could
not acquire knowledge at all, which is why novelty is an outcome of the procedure rather
than a failure of it.

R3 precedes the merge rules, and it is gated on the relatedness floor
$\sigma_{\mathrm{lo}}$ rather than on the high merge threshold $\theta_{\mathrm{merge}}$:
a contradiction need not be near-identical in surface form to be a contradiction, so a
lexically divergent contradiction (similarity in
$[\sigma_{\mathrm{lo}},\theta_{\mathrm{merge}})$) is surfaced rather than dropped, while
\textsc{merge} keeps the high bar. Because R3 is evaluated first, a contradicting pair in
that band is a \textsc{conflict} and not a \textsc{relate}; the two rules cannot both
fire. The evaluation (Section~\ref{sec:eval-m2}) measures the effect: this placement
raises \textsc{conflict} recall without changing the false-merge rate, because it only
re-routes contradictions toward the contradiction link, never toward a merge.

R4 and R5 are the two same-enough paths, and they are gated differently. The exact-key
path R4 is exempt from the authority predicate because the claim key hashes canonical
content together with the claim's discrete scope (Section~\ref{sec:merge-signals}):
$\kappa = 1$ means the peer already holds the byte-identical assertion in the same
scope, so there is no cross-boundary rewrite to guard against. The key does not carry
the validity interval or freshness, so R4 still requires a compatible frame; an
exact-key pair whose Contexts are incompatible, for example identical scoped
assertions with disjoint validity intervals, falls through to R6 and relates. The similarity path R5 carries no such
guarantee, so it carries both gates: Context compatibility, and the authority admission
of Section~\ref{sec:cons-trust}, which withholds an automatic merge from a source
strictly less authoritative than the local target and lets R6 admit it as a candidate
instead.

\begin{figure*}[!t]
  \centering
  \resizebox{0.72\textwidth}{!}{%
  \begin{tikzpicture}[
    font=\footnotesize,
    node distance=5mm,
    sig/.style={align=left, inner sep=1pt, font=\scriptsize, text=black!75},
    gate/.style={draw=black!55, line width=0.5pt, rounded corners=2pt,
                 align=center, inner sep=3pt, outer sep=0pt, minimum height=0.62cm},
    rgate/.style={gate, draw=black!85, line width=1.0pt},
    verd/.style={align=center, inner sep=2pt, font=\scriptsize\bfseries},
    patch/.style={draw=black!85, fill=black!82, text=white, line width=1.0pt,
                  rounded corners=2pt, align=center, inner sep=3pt,
                  minimum height=0.6cm, font=\footnotesize},
    e/.style={-{Latex[length=1.8mm]}, black!65, line width=0.6pt},
    el/.style={font=\scriptsize, text=black!55, inner sep=1pt}
  ]
    \node[sig] (sigs) at (0,0)
      {signals on $(a,b)$:\\
       $\kappa$ key identity\\
       $\sigma$ similarity\\
       $\chi$ contradiction\\
       Context, authority gates\\
       {\scriptsize MAC verified upstream}};

    \node[rgate] (rej) at (4.0,1.6)
      {\textbf{admission gate} (first):\\$R<R_{\min}$ \,or\, status$(b) \ne$ active};
    \node[verd, text=black!70] (vrej) at (9.0,1.6) {\textsc{reject}};
    \draw[e] (sigs.east) -- (rej.west);
    \draw[e] (rej.east) -- (vrej.west) node[midway, el, above]{stale / overruled};

    \node[gate] (cand) at (4.0,0.4)
      {any local candidate?\\$\kappa=1 \vee \sigma\ge\sigma_{\mathrm{lo}}$};
    \node[verd] (vins) at (9.0,0.4) {\textsc{insert}\\{\scriptsize novel}};
    \draw[e] (rej.south) -- (cand.north) node[midway, el, right]{pass};
    \draw[e] (cand.east) -- (vins.west) node[midway, el, above]{no (novel)};

    \node[gate] (con) at (4.0,-0.9)
      {$\chi=$ contradicts ?};
    \node[verd] (vconf) at (9.0,-0.9) {\textsc{conflict}\\{\scriptsize conflicts}};
    \draw[e] (cand.south) -- (con.north) node[midway, el, right]{yes};
    \draw[e] (con.east) -- (vconf.west) node[midway, el, above]{yes};

    \node[gate] (key) at (4.0,-2.2)
      {$\kappa=1 \wedge \mathrm{Context}\approx$ ?\\(key: content $\|$ scope; frame compatible)};
    \node[gate] (sim) at (4.0,-3.5)
      {else $\sigma\ge\theta_{\mathrm{merge}}$ ?\\gated by Context $\wedge$ authority};
    \draw[e] (con.south) -- (key.north);
    \draw[e] (key.south) -- (sim.north);

    \node[verd] (vmerge) at (9.0,-2.85) {\textsc{merge}\\{\scriptsize same-enough}};
    \node[verd] (vrel)   at (9.0,-4.6)  {\textsc{relate}\\{\scriptsize overlaps}};
    \draw[e] (key.east) -- (vmerge.west) node[midway, el, above]{yes};
    \draw[e] (sim.east) -- (vmerge.west) node[midway, el, below]{yes};
    \draw[e] (sim.south) -- (vrel.north) node[midway, el, right]{otherwise};

    \node[patch] (patch) at (7.2,-5.8) {one \textbf{Patch}\\{\scriptsize the only state mutation; publishable}};
    \draw[black!80, line width=0.9pt] (vrej.east)   -- (10.8,1.6);
    \draw[black!80, line width=0.9pt] (vins.east)   -- (10.8,0.4);
    \draw[black!80, line width=0.9pt] (vconf.east)  -- (10.8,-0.9);
    \draw[black!80, line width=0.9pt] (vmerge.east) -- (10.8,-2.85);
    \draw[black!80, line width=0.9pt] (vrel.east)   -- (10.8,-4.6);
    \draw[black!80, line width=0.9pt] (10.8,1.6)    -- (10.8,-5.8);
    \draw[e, black!80, line width=0.9pt] (10.8,-5.8) -- (patch.east);
  \end{tikzpicture}}
  \caption{The merge decision procedure, a first-match cascade over five outcomes. The signals $\kappa,\sigma,\chi$ are read on the adjudicated candidate pair (a second pair is read only when the gate-aware fallback of Section~4.2 fires); a delta whose MAC fails verification, or is stale or overruled, is dropped before any semantic test. A claim with no candidate at or above the relatedness floor is novel and \textsc{insert}ed; \textsc{reject} is the admission gate alone, never ``no match found''. Contradiction is tested next, so a contradicting pair is \textsc{conflict}, not \textsc{relate}. Same-enough is then a hierarchy: claim-key identity in a compatible frame ($\kappa=1 \wedge \mathrm{Context} \approx$) decides first, else similarity $\sigma\ge\theta_{\mathrm{merge}}$ gated by Context and authority. Anything remaining \textsc{relate}s; every path emits exactly one Patch, the only object that mutates local state.}
  \label{fig:merge}
\end{figure*}
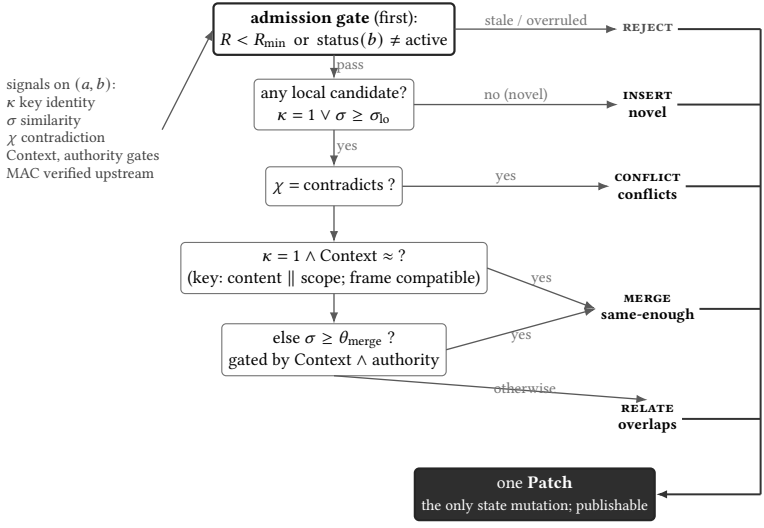

The non-contradiction gate is only as strong as the contradiction signal $\chi$: a
missed contradiction (an inference false negative) is not caught here, so the
guarantee is that \smp{} never silently resolves a detected contradiction
(Section~\ref{sec:cons-threat} treats the adversarial evasion of this gate), and the
residual false-merge rate is measured by the evaluation (Section~\ref{sec:eval}).
Figure~\ref{fig:merge} renders this control flow: the signals read once, the admission
gate first, the novelty test, the same-enough hierarchy, and the single Patch that every
path emits.

\subsection{The same-enough predicate}
\label{sec:merge-sameenough}
Same-enough is a hierarchy, not a bare similarity threshold. The primary test
is claim-key identity in a compatible frame ($\kappa = 1$, Context~$\approx$): exact, cheap, and decisive absent a detected contradiction. Only when keys
differ does the procedure fall back to embedding similarity over
$\theta_{\mathrm{merge}}$, and even then the merge is gated by two
conditions: non-contradiction ($\chi \ne \textit{contradicts}$) and Context
compatibility. Continuing the running example (Figure~\ref{fig:running}), the edge
agent's paraphrase ``We are required to retain customer data for seven years'' shares
no claim key with $c_1$ ($\kappa = 0$), but embeds at $\sigma = 0.93$ and does not
contradict $c_1$ in a matching EU Context, so it clears the gate and
\textsc{merge}s. The merge is a \emph{status} consolidation, not a node rewrite: the
Patch retains the paraphrase, marks it \textsc{deprecated} under an append-only
\textsc{supersedes} link to $c_1$, and leaves $c_1$ canonical and \textsc{active}, so a
status-aware recall surfaces the single representative $c_1$ while the duplicate is kept
for audit. The merge adds one append-only link and makes no destructive edit, so it is
monotone on the status CRDT (Section~\ref{sec:consistency}) and does not mutate content. The Context gate
then stops a high-similarity false merge across an
administrative boundary. When the US domain aggregate $B_{US}$ peers its retention
rule into the EU domain, it peers the very same sentence, scoped
$\langle\text{jurisdiction}{:}\,\text{US}\rangle$ and therefore carrying a
different claim key (Section~\ref{sec:merge-signals}). The pair still embeds at
$\sigma = 0.93$ and does not contradict, but the frames differ (EU${\not\approx}$US), so it
\textsc{relate}s rather than merges. The US rule enters the EU view as a related,
distinct claim. It is not collapsed into $c_1$, and data sovereignty is preserved across
the boundary by construction. The same gate stops the textbook frame error (``Paris''
the city versus the mythological figure); here it keeps an EU and a US retention rule
distinct. The dangerous error a merge protocol must avoid is the \emph{false merge}
(two distinct facts collapsed into one), and the gate hierarchy drives its rate down
without sacrificing recall on genuine paraphrases.

\subsection{Overlaps and conflicts as first-class graph structure}
\label{sec:merge-outcomes}
\paragraph{Overlaps emit a Mapping.}
A \textsc{relate} decision emits a typed, weighted Mapping link (weighted by
$\sigma$). The two claims stay distinct, and a later graph walk can traverse the
relation. This is the principled handling of the same meaning in a different
Context: the protocol records the relationship instead of collapsing the frames.

\paragraph{Conflicts emit a contradiction link.}
R3 fires on the inference verdict alone, before any Context test: a \textsc{conflict}
is a semantic contradiction irrespective of scope. The contradiction link
records both claims' Contexts, so a cross-frame contradiction is preserved as
information (two frames may legitimately disagree) rather than treated as an
operational fault, and whether a contradiction binds within a compatible frame
is part of the deferred adjudication (Section~\ref{sec:consistency}).
A \textsc{conflict} decision represents the conflict as a first-class
contradiction link and does not pick a winner. In the running example, an EU team
agent publishes, within the EU domain, $c_2 = $ ``Customer data must be
deleted after two years'' ($\sigma = 0.86$ against $c_1$, $\chi = \textit{contradicts}$,
same EU Context):
\smp{} records a first-class $c_1 \leftrightarrow c_2$ contradiction link and chooses
no winner. The pair is a genuine policy tension (a retention duty against a
right-to-erasure rule) and not a surface typo. Note that $\sigma = 0.86$ sits
below $\theta_{\mathrm{merge}} = 0.90$, so the two would not have merged in any
case, and \smp{} surfaces the contradiction only because the conflict gate keys on the
relatedness floor $\sigma_{\mathrm{lo}}$ rather than the merge bar
(Section~\ref{sec:merge-procedure}). Resolution is deferred to the
trust hierarchy or a deterministic tie-break (authority, then recency, then
evidence weight; Section~\ref{sec:consistency}). \smp{} states explicitly that
the normatively correct tie-break is an open question: it represents
and converges the status of a conflict, but does not claim to settle the
conflict-of-law problem.

\subsection{The Patch as the merge output}
\label{sec:merge-patch}
Every run of the procedure emits exactly one Patch (Section~\ref{sec:proto-primitives}),
recording the decision, the target endpoint, the emitted deltas, and which gates
fired. The Patch is the only thing that mutates local state and is itself
publishable, so the federation can audit and replay any merge.

The thresholds $\theta_{\mathrm{merge}}$ and $\sigma_{\mathrm{lo}}$ are
calibrated on held-out claim-pair data, not derived; Section~\ref{sec:eval}
reports their sensitivity rather than claiming derived optima.

\section{Consistency and trust}
\label{sec:consistency}

\smp{} keeps many sovereign brains coherent in the sense it controls: per-claim
status reaches a strong-eventually-consistent agreement without a
coordinator, while the merge graph itself, decided per pair by the procedure of
Section~\ref{sec:merge}, is adjudicated locally at each brain and is not itself
claimed coherent in that same sense. \smp{} achieves the status guarantee by
defining a self-contained status conflict-free replicated data type (CRDT) over
per-claim status, grounded in standard CRDT theory
(Section~\ref{sec:cons-crdt}); preserving that CRDT's premises end to end
through an append-only wire discipline that is \smp{}'s own contribution
(Section~\ref{sec:cons-wire}); and admitting merges through an authority-gated trust
hierarchy (Section~\ref{sec:cons-trust}).

\subsection{Status convergence versus truth}
\label{sec:cons-crdt}
\smp{}'s controlled variable is the per-claim status. Epistemic truth (which of
two contradictory claims is correct) is deliberately outside the loop, a
world-model concern that sits above \smp{} (Section~\ref{sec:disc-future}). This is
a design boundary, not a shortfall: the mechanism converges the status it controls and
represents, rather than silently resolves, the truth it does not.

Each claim carries a status on the totally ordered status chain
\[
  \textit{active} \sqsubseteq \textit{deprecated} \sqsubseteq \textit{overruled}
  \sqsubseteq \textit{revoked}.
\]
A brain's local status state is a grow-only set (G-Set) of incoming, append-only
overrule and supersede links, and merge is set union of these link sets. Set union is
commutative, associative, and idempotent, so the link set is a state-based
CRDT and reaches strong eventual consistency~\cite{shapiro2011crdt,kleppmann2017jsoncrdt}:
any two brains that have received the same set of links hold the identical link set,
independent of delivery order. The per-claim status is then a deterministic
function of that converged set, the join (the maximum along the status chain above) of
the status effects its links carry; because the function is deterministic, brains that
agree on the link set agree on every status. The links are only ever added, never
removed, so this is the grow-only instance of the standard state-based-CRDT
construction~\cite{shapiro2011crdt}: convergence rests on the union, and the status
chain supplies the deterministic materialization on top. This order-independence is a
property of the per-claim status; the merge graph itself (built by the per-pair
decision procedure, Section~\ref{sec:merge}, against already-held state) is not claimed
order-independent, and the protocol's guarantees rest only on status convergence. The
electronic supplement (Appendix~M) measures this directly: under randomized delivery
orders the status layer converges while link structure and retrieval-visible state can
differ across orders.

Strong eventual consistency guarantees
that all brains agree on which claims are active, deprecated, or overruled.
It does not guarantee that any overrule was correct, and it does not
resolve a conflict between two mutually contradictory active claims. That
conflict-of-law problem (which of two contradictory active claims should govern) is
open: \smp{} represents the conflict (as a first-class
contradiction link, Section~\ref{sec:merge-outcomes}) and converges its
status, but defers the normative resolution rule.

\paragraph{Running example concluded.}
The EU canonical authority $C$ resolves the standing $c_1 \leftrightarrow c_2$
contradiction by publishing an authority \textsc{overrules} status link on $c_2$
(Figure~\ref{fig:running}). Because
the edge tiers reconnect at different times, the two status links on $c_2$ (an earlier
deprecate from brain $B$ and $C$'s overrule) reach brains in either order,
yet the status CRDT joins them identically on every brain, $\mathrm{join}(\textit{deprecated},
\textit{overruled}) = \textit{overruled}$, with no coordinator. That $C$'s overrule is
applied follows the domain-scoped authority hierarchy of
Section~\ref{sec:cons-trust}: $C$ holds regulator-grade EU authority, which a peer EU
agent or any US-domain brain does not, so the two domains stay coherent without either
rewriting the other. The federation settles on $c_1$ active (the paraphrase deprecated
beneath it under a \textsc{supersedes} link), the US rule related but distinct,
and $c_2$ overruled. The result is the same on every tier, and no customer record
ever crosses a domain boundary.

\subsection{An append-only wire discipline that preserves the CRDT premises}
\label{sec:cons-wire}
A CRDT's convergence rests on premises that hold for local state but can be violated
by a careless transport: a destructive overwrite on the wire, or the propagation of
a stale or overruled delta, breaks them. \smp{}'s contribution here is the wire-level
discipline that preserves the premises end to end:
\begin{itemize}
  \item \textbf{Append-only status links.} A status change on the wire is always an
  appended status link, never a destructive overwrite (Section~\ref{sec:proto-verbs}).
  The automatic merge emits only \textsc{active} and \textsc{deprecated} effects; the
  \textsc{overruled} and \textsc{revoked} effects are authority-authored links
  (Section~\ref{sec:cons-trust}). This append-only form is the premise that makes the
  grow-only link set monotone across the federation and not only within one brain.
  \item \textbf{A staleness and overrule gate at materialization-read.} A brain admits every received overrule and supersede link into its grow-only set unconditionally, and applies the staleness/overrule gate, dropping a link whose retention is below threshold or whose own carried status is no longer active, only when it computes a claim's effective status. The same placement carries the authorization predicate: $\mathrm{authorized}(\ell)$ holds when the link's immutable origin metadata (asserted cell and authority level; benign-fault, Section~\ref{sec:cons-threat}) places its author inside the claim's scope with sufficient authority; an unauthorized link, like a stale one, is appended and auditable but never materializes. The gate reads only fields the link carries, so it is a pure function of the converged set and order-independent; an admission-time gate keyed on receipt-time local status would instead make the admitted set order-dependent and diverge, which the evaluation demonstrates directly (Section~\ref{sec:eval-c}). Applying the gate at read prevents stale or overruled links from inflating a status. This status-link gate is distinct from the merge-decision \textsc{reject} of Section~\ref{sec:merge-procedure}: the two apply to different objects (status links versus incoming claims), and only the latter, firing on a stale or overruled delta, ever drops an input.
\end{itemize}

\paragraph{Delivery contract.}
The convergence above asks little of the transport, by design: \smp{} assumes
at-least-once, idempotent, eventually-complete delivery and requires
no ordering or exactly-once guarantee. Re-adding a link id is idempotent (the
grow-only set absorbs duplicates) and materialization is order-free, so a log-based fabric's
cheap default mode (Apache Kafka's at-least-once, per-partition ordering) suffices. The
premise that does bite is eventual completeness: a link a brain never receives is missing
state, so semantic routing, which delivers only to interested peers at a recall below
one (Section~\ref{sec:eval-r}), is backstopped by a periodic anti-entropy digest~\cite{demers1987epidemic}. Each brain
summarizes its grow-only status-link set as a per-claim, order-independent digest (the sorted
link ids and a set-hash); two brains exchange digests, take the per-claim symmetric difference,
and each pulls and applies (idempotent union) the links it lacks, so both reach the same link
set (and the same per-claim status) regardless of routing recall. This makes per-claim
status convergence independent of the completeness of the semantic-routing path: \smp{} achieves the
eventually-complete delivery its CRDT needs rather than assuming it of the transport. We
evaluate the digest under injected routing loss (Section~\ref{sec:eval-c}); self-healing thus
holds for partition-and-reconnect and under lossy routing, provided each link survives
on at least one reachable peer (a link lost by every brain is the disconnection case,
recovered by log re-consumption).

The convergence guarantee follows from standard CRDT theory; the wire discipline that
keeps its premises true across a real transport is new in \smp{}.

\subsection{Trust hierarchy and canonical brains}
\label{sec:cons-trust}
Brains differ in authority. Each node and link carries an authority level, and a
precedent sub-graph records which claims supersede or overrule which. A
canonical brain is one whose nodes carry high authority (for example an
institution-public brain holding a founding or precedent record). \smp{} uses this
authority field (part of the wiki-brain data model of
Section~\ref{sec:bg-datamodel}) through two new protocol-level policies:
\begin{itemize}
  \item \textbf{Conflict tie-break (deferred resolution).} When a contradiction must
  be resolved, the intended order consults authority first (a canonical brain's active
  claim dominates a peer's), then recency, then evidence weight. \smp{} specifies
  this order but defers automatic resolution (Section~\ref{sec:merge-outcomes}); it does
  not claim the order is normatively settled (Section~\ref{sec:cons-crdt}).
  \item \textbf{Authority-gated merge admission.} A similarity merge from a
  low-authority brain into a canonical brain is admitted as a candidate (a
  \textsc{relate}) rather than an automatic \textsc{merge}. An exact-key match in a compatible frame still
  merges, and that exemption is structural rather than a convenience: the claim key binds
  canonical content to the claim's discrete scope (Section~\ref{sec:merge-signals}), so an
  exact-key match is the same assertion in the same scope and consolidating it rewrites
  nothing the peer does not already hold byte-identically. The exemption then leans
  on candidate selection, so the rule is stated there too: a local claim whose
  effective status is no longer active is not an admission target. A deprecated or
  overruled local claim cannot be revived by an exact-key match from a peer, and the
  status lifecycle stays monotone (Section~\ref{sec:cons-crdt}). Together these keep a
  canonical brain from being silently rewritten by an arbitrary peer.
\end{itemize}
The authority field and the precedent sub-graph are part of the wiki-brain data model
(Section~\ref{sec:bg-datamodel}); the authority-gated admission policy and the
tie-break order are \smp{}'s contribution.

\subsection{Threat model and security scope}
\label{sec:cons-threat}
We state the trust assumptions explicitly. \smp{} MAC-authenticates every published delta over a shared group key, and a receiving brain verifies the MAC before admission. This authenticates a delta as originating within the keyed group and defeats transport tampering: the (keyless) transport cannot silently forge a delta, and the MAC attests authorship of the authenticated content, not the receiver-computed merge decision (a re-published Patch is re-adjudicated by each consuming brain, since the decision procedure re-runs on the underlying delta, so a benign-but-misconfigured brain cannot propagate a wrong merge decision on its MAC alone). A keyed message-authentication code (HMAC) does not, however, attribute a delta to a specific member of the group: any key holder can produce the same tag, so the audit trail is attributable to the keyed group as a whole, not to an individual sender. Per-sender attribution requires either per-brain keys or an asymmetric suite (e.g.\ Ed25519, likewise sub-millisecond); both are deployment upgrades we do not evaluate. The evaluated HMAC primitive authenticates a delta within the keyed group at sub-millisecond cost (Section~\ref{sec:eval-scope}). Key distribution is out-of-band; authority levels are provisioned out-of-band by an administrator and bound to keys in the same directory. The trusted computing base is each brain and its key; the transport (brokers, routing) is trusted for availability, not content integrity, which the MACs protect. Under this model the consistency and trust mechanisms defend against benign faults (crashes, partitions, reordered and delayed delivery, and stale or overruled deltas) and detect, via the group-attributable Patch trail (though not prevent), authority misconfiguration: the gate protects a canonical brain from low-authority peers, not the federation from a wrongly-provisioned authority. This is the regime the evaluation exercises (Section~\ref{sec:eval-c}). The mechanisms do not defend against a Byzantine participant: a valid-key holder can publish well-formed false claims; the authority field is asserted, so without a trust root binding keys to authority levels a malicious brain can claim to be canonical; and an encoder-aware adversary can craft a contradiction below $\sigma_{\mathrm{lo}}$ to evade the conflict gate (Section~\ref{sec:merge-procedure}). Hardening these (a PKI binding keys to authority, per-sender attribution, and an adversarially robust contradiction gate) is a self-contained security extension, with Byzantine eventual consistency~\cite{kleppmann2020byzantine} the natural starting point, that we scope out; here the guarantees are stated for benign faults.

\subsection{Error model and correction}
\label{sec:cons-errors}
The decision procedure makes two error types, a false merge (two distinct
claims consolidated) and a contradiction miss (a real conflict merged silently),
bounded empirically at false-merge $0.013$ and contradiction-miss $0/8$ on a
naturalistic sample (Section~\ref{sec:eval-m2}). Neither is destructive, so both are correctable: a \textsc{merge} only appends a \textsc{supersedes} status link and retains the consolidated claim, and every decision is a replayable Patch, so a surfaced error is corrected in place by appending a corrective link (a later contradiction, an authority \textsc{overrule}, or a re-publication) that the status CRDT converges like any other (Section~\ref{sec:cons-crdt}), and nothing needs to be reconstructed because nothing was overwritten. \smp{} cannot detect a silent
error alone: detection is bounded by the embedding and inference signals, and systematic
re-checking by an offline curation pass over the Patch trail is future work
(Section~\ref{sec:disc-future}).

\section{Implementation}
\label{sec:impl}

\smp{} is mostly a merge engine plus binding glue over mature components rather than a
system built from scratch. The remainder of this section gives the deployment model
(Section~\ref{sec:impl-deploy}), the binding of \smp{} onto a learned
publish/subscribe transport (Section~\ref{sec:impl-transport}), the brain store
(Section~\ref{sec:impl-store}), a ledger of what is reused versus new
(Section~\ref{sec:impl-reuse}), and the reproducibility setup, including the numeric
operating point (Section~\ref{sec:impl-repro}).

\subsection{Deployment model}
\label{sec:impl-deploy}
The electronic supplement (Appendix~N) illustrates the three-layer stack and the
deployment topology. The deployment frame is a premise rather than a contribution: harnesses (execution)
are kept distinct from agents (capability and task logic), which are kept distinct
from agent brains (the evolving knowledge a brain accumulates from both harness and
agent experience). Above many agent brains sits a platform-level \emph{global brain}, created, merged, and tracked on demand under a trust hierarchy (Section~\ref{sec:cons-trust}). On one harness, the agent speaks MCP to tools and \smp{} to brains (Section~\ref{sec:related}).

A brain is one versioned, typed-link agent-memory store (Section~\ref{sec:bg-datamodel}): an instance holding nodes and first-class links, as Zep/Graphiti provides first-class and other public stores approximate~\cite{rasmussen2025zep,xu2025amem}. Each brain PUBLISHes graph deltas, content-routed to interested brains; the global brain is the many-to-one aggregation target, with the merge procedure as the sink. ``Global'' names its scope rather than centralized control: it is an aggregation role, not a coordinator. It runs the same merge-on-receive procedure as any brain, holds no special authority, and is reproducible at any subscriber, so a deployment may run several. \smp{} stays peer-to-peer: the status CRDT converges the sovereign brains with no coordinator (Section~\ref{sec:cons-crdt}). This separates the global brain, an optional aggregate of decentralized merge, from the centralized single-store baseline it is measured against (Section~\ref{sec:eval-h}), the sole required store.

\subsection{Sync over a standard publish/subscribe transport}
\label{sec:impl-transport}
\smp{} rides a standard content-based publish/subscribe fabric and is agnostic to the
specific broker; our deployment uses Apache Kafka, the federated messaging substrate
also used by the companion federated-pub/sub work~\cite{loven2026neuralpubsub}. The
binding is the fourth contribution: a PUBLISH(graph-delta) becomes a transport event
keyed by the delta's interest, and an arriving event becomes a merge-on-receive call.
Routing a delta to the interested brains uses the fabric's native topic/subject matching
by default, with an optional content-based semantic matching layer (embed, threshold, cluster; pre-calibrated, frozen, embedding-based in our deployment) that improves selectivity where interests are not cleanly separable by topic;
the LLM-matching variant of the same layer, and its cost--accuracy envelope, are
characterised separately~\cite{loven2026neuralrouter}. The federated brokers tolerate broker death and partition, and \smp{} maps the transport's bounded-staleness propagation directly onto the graph-delta layer.

The binding is deliberately modest in scope. The transport routes opaque events;
mapping merge-protocol semantics onto it is new, and the integration is evidenced by a
binding plus a routing ablation (Section~\ref{sec:eval}) against broadcast, random, and
round-robin baselines, rather than by a validated optimality claim. Section~\ref{sec:eval-r}
states what that ablation does and does not establish.

\subsection{The brain store: a versioned typed-link memory}
\label{sec:impl-store}
A brain store is a thin \smp{} facade over a versioned, typed-link agent-memory
store such as those public systems provide (a versioned, typed-link memory as in
Zep/Graphiti~\cite{rasmussen2025zep}, with A-MEM~\cite{xu2025amem} supplying the
untyped node-and-link core), which supplies content-addressed, append-only
persistence, scope- and authority-aware routing, retention computation, and
access checks. \smp{} builds on such a store wholesale and does not reinvent the
data model or its persistence. The only \smp{}-new code over the store is the
graph-delta apply and extract path (applying a Patch's emitted deltas; extracting a
delta for PUBLISH) and the graph-walk read path the evaluation uses for retrieval.

For wire types, \smp{} defines a links-first-class schema (node, claim, and typed-link types, with freshness derived from a claim's status) as the on-wire representation of the adopted data-model types, with the Patch the only new type; the schema is carried in \smp{}'s own tree.

\subsection{What is reused versus new}
\label{sec:impl-reuse}
The net new code is small and bounded: (1) the protocol layer (the PUBLISH wire
operation, the typed message types, the Patch); (2) the merge engine (three-signal classifier, Context gate,
status CRDT recompute, reject gate); (3) the binding from a PUBLISH to a transport event
and an arriving event to a merge-on-receive; and (4) the evaluation drivers. Reused
wholesale are the transport (brokers, semantic routing, federation, the fault-injection
harness), the memory store (persistence, routing, retention, capabilities), and the
embedding/encoder plumbing. The status enum and precedent lifecycle are reused and extended
with the grow-only-set merge and deterministic recompute (Section~\ref{sec:cons-crdt}),
whose convergence follows from the standard state-based-CRDT conditions~\cite{shapiro2011crdt}.

\subsection{Reproducibility}
\label{sec:impl-repro}
The decision-procedure thresholds are calibrated rather than derived, and we state them numerically: at the operating point used throughout, $\theta_{\mathrm{merge}}{=}0.90$ and $\sigma_{\mathrm{lo}}{=}0.12$, with retention floor $R_{\min}{=}0.10$; the similarity $\sigma$ is all-MiniLM-L6-v2 cosine and the inference verdict $\chi$ is a cross-encoder NLI, specifically \texttt{cross-encoder/nli-deberta-v3-small} (a DeBERTa-v3-small cross-encoder fine-tuned on MNLI). Every reported number comes from a seeded run, deterministic given its seed, and the model-family runs record their model tags, library versions, and commit. The protocol implementation, the merge decision procedure and status CRDT, the gold set, the experiment and cross-tier deployment drivers, and the testbed harness are released as an open repository (\url{https://github.com/Future-Computing-Group/meld-experiments}, archived at Zenodo, DOI \href{https://doi.org/10.5281/zenodo.21878274}{10.5281/zenodo.21878274}). The deterministic results (including the merge classifier, the gate and conflict ablations, the order-independence demonstrations, and the cross-tier protocol drivers) reproduce end-to-end from the repository; the model-family numbers of the headline experiment and the 5G-testbed numbers require external GPU and edge infrastructure, and the repository carries the drivers, model tags, and run configurations needed to reproduce them there.

\section{Evaluation}
\label{sec:eval}

This section asks four questions about \smp{} as an adaptive coherence mechanism, in the order that matters for a self-adaptive system, each answered by a controlled experiment on real computing-continuum infrastructure:
\begin{enumerate}[leftmargin=1.4em]
  \item \textbf{Does \smp{} adapt safely under conflicting knowledge?} The merge decision procedure's classification quality, in particular its false-merge rate, the dangerous error (Section~\ref{sec:eval-m2}).
  \item \textbf{Does it recover from partitions and message loss?} The status CRDT's order-independent reconvergence after a real partition heal and under lossy routing (Section~\ref{sec:eval-c}).
  \item \textbf{Does authority gating prevent unsafe state change?} A canonical brain is not silently rewritten by a peer (Sections~\ref{sec:cons-trust} and~\ref{sec:eval-extra}).
  \item \textbf{What adaptation errors remain?} The residual false-merge and contradiction-miss rates, threshold non-transfer across encoders, and the Byzantine case left out of scope (Sections~\ref{sec:eval-m2} and~\ref{sec:disc-limits}).
\end{enumerate}
Two further axes establish that this coherence costs little: utility (whether decentralizing memory preserves what agents can recall, Section~\ref{sec:eval-h}) and efficiency (sync latency, routing selectivity, node-scaling; Sections~\ref{sec:eval-s},~\ref{sec:eval-r}, and~\ref{sec:eval-scale}). The reported accuracy is not, and need not be, cutting-edge. The contribution is a new distributed protocol and mechanism: the question is whether it keeps sovereign brains coherent in per-claim status, with the semantic graph structure locally adjudicated, preserving recall where a centralized store needs neither property. The substrate must supply a competence floor: the harness must give the agent non-trivial competence on the task (non-zero, non-saturated accuracy), so the measured delta is a real comparison. Two near-zero scores would not distinguish the conditions, and HotpotQA distractor avoids that failure mode. An at-a-glance inventory of all experiments (dataset, tier, headline result) is in the electronic supplement (Appendix~I).

Four easily-conflated properties stay distinct: merge correctness and operational convergence (Sections~\ref{sec:eval-m2} and~\ref{sec:eval-c}) are the contribution; retrieval quality (recall@$k$, Section~\ref{sec:eval-h}) is the utility check; answer correctness sits at merge-vs-centralized parity. The measurable delta is in coherence, retrieval, and storage. Reasoning quality is out of scope.

\textbf{Shared topology.}
In every experiment, $N$ agent processes each own a brain store and consume a
disjoint (or configurably overlapping) slice of conversation histories, from which
they extract Concepts, Claims, and Evidence and PUBLISH graph deltas; the global
brain is the funnel sink. The centralized baseline is a single brain store ingesting
all conversations with no merge protocol.

\textbf{Testbed: a three-tier computing continuum.}
The evaluation runs across the same computing continuum \smp{} is designed to keep
coherent, on operational University of Oulu and Finnish national infrastructure. The
distributed protocol experiments (sync latency, Section~\ref{sec:eval-s}; routing,
Section~\ref{sec:eval-r}; partition-heal convergence, Section~\ref{sec:eval-c}) run on
the edge tier: the University of Oulu 5G Test Network (5GTN)~\cite{piri2016gtn}, an operator-grade 5G test network whose edge-computing capacity we use directly, as four
nodes (one broker and the peer brains, $N\in\{2,3,4\}$), 4 CPU cores and 8\,GB RAM each,
Docker CE on Ubuntu~24.04, on the same 5GTN substrate used by prior federated
publish/subscribe work~\cite{loven2026neuralpubsub}. The headline merge-quality
experiment (Section~\ref{sec:eval-h}) runs five open-weight models (three
families spanning a $\sim\!20\times$ size range) on the HPC tier, CSC
(the Finnish national supercomputing centre). The
deterministic logic experiments (merge classification, the order-independence
demonstration, and the gate/conflict ablations) run locally on a single host.
Within the 5GTN edge, intra-site latency is under $1$\,ms (shared LAN); the wide-area
latency regimes are emulated on the edge nodes via \texttt{tc qdisc netem} (below).
Beyond these per-tier experiments, \smp{} is additionally deployed live end-to-end across
all three tiers at once, over the standard Apache Kafka publish/subscribe fabric, as the
cross-tier feasibility deployment of Section~\ref{sec:eval-scope}.

\textbf{Datasets.}
Three datasets serve distinct roles, all verified at source. HotpotQA distractor
(open) is the primary memory substrate for the recall headline and the systems experiments.
LongMemEval~\cite{wu2024longmemeval} is a second, long-horizon substrate that
independently corroborates the recall headline (Appendix~E). A balanced, constructed
$104$-pair gold is the diagnostic instrument for merge classification. (GAIA
level-2 is named only as a deferred harder mem-delta, Section~\ref{sec:disc-future}.)

\subsection{Recall under decentralized merge}
\label{sec:eval-h}
This is the utility check: whether a global brain populated by distributed agents
that sync and semantically merge their knowledge matches a centralized single store on
recall, at acceptable cost, i.e.\ whether memory can be decentralized without losing recall. Recall parity is only a precondition for the contribution, since coherence is worth having only if it costs no recall.
We evaluate it on HotpotQA distractor multi-hop QA as the memory substrate, splitting
the supporting paragraphs of each item across $N{=}3$ sovereign brains with overlap
fraction $0.5$ (paraphrased duplicates, so the same-enough gate fires), and the global
brain then answers under each memory condition. Each run delivers claims in one fixed,
seeded order per condition; the sensitivity of the resulting graph structure, though
not of the converged per-claim status, to delivery order is measured directly in the
electronic supplement (Appendix~M). The primary observable is recall@$k$
($k{=}5$) of the gold supporting facts in the status-live read (the layer the protocol
acts on), with answer token-F1 secondary. Thresholds are calibrated, the encoder is
all-MiniLM-L6-v2, and the run is repeated across five open-weight models spanning a
$\sim\!20\times$ size range and three architecturally distinct families (Qwen2.5-Instruct
at $1.5$B, $7$B, and $32$B; Mistral-7B-Instruct-v0.3; Gemma-3-12B-it), $N{=}200$
questions per condition per model, each served as a Q4-quantized local checkpoint under
an identical decode configuration. Two baselines isolate the merge's contribution: the centralized single store, and an ablated \smp{} using naive last-writer-wins/set-union with no merge procedure. Simulation (multi-process agents, single-host
broker) suffices for the recall claim; real multi-host infrastructure carries the
latency and scale numbers (Section~\ref{sec:eval-s}). This experiment tests merge
correctness relative to centralized operation only; it does not by itself validate the
transport integration, which the routing ablation treats (Section~\ref{sec:eval-r}).

\begin{figure*}[t]
  \centering
  \includegraphics[width=0.72\textwidth]{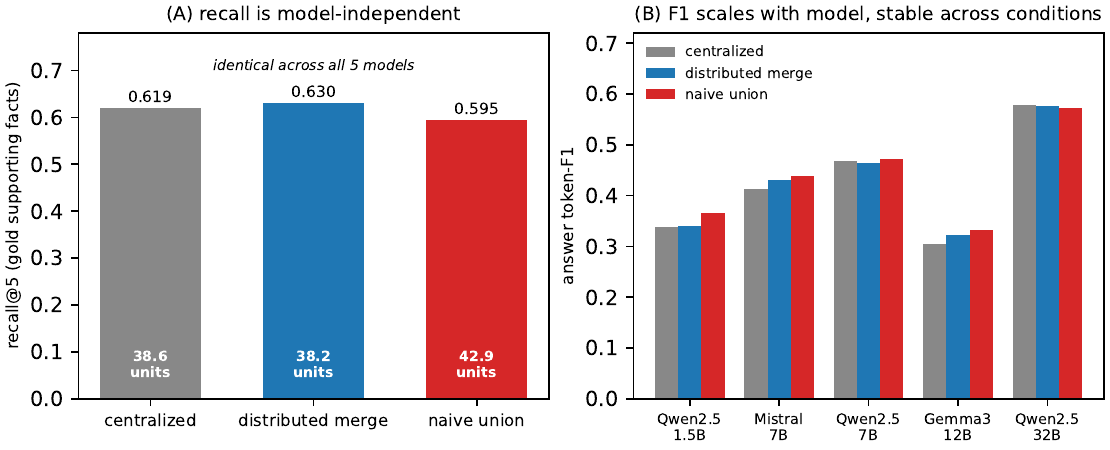}
  \caption{Recall under decentralized merge (HotpotQA distractor; $N{=}200$/condition/model; $k{=}5$, overlap
  $0.5$, $N{=}3$ brains, all-MiniLM-L6-v2). Distributed merge matches the centralized store on
  recall@$5$ (two one-sided-test equivalence, $\Delta{=}0.05$) and exceeds naive union
  ($+0.035$, CI excluding zero) at $\approx\!11\%$ fewer live units. recall@$k$ is a
  store/encoder property, hence identical across all five models (text).}
  \label{fig:m1-hotpot}
\end{figure*}

\textbf{Result.} The distributed merge is recall-non-inferior to the centralized
store: recall@5 is $0.630$ (merge) versus $0.619$ (centralized), with a two
one-sided-test of equivalence at a pre-specified margin $\Delta{=}0.05$ (one
recall@5 band, set before the run as the smallest effect the study treats as
material) satisfied
(difference $+0.011$, $90\%$ CI $[-0.001,+0.024]\subset[-\Delta,+\Delta]$). Decentralizing
memory therefore does not cost recall (Figure~\ref{fig:m1-hotpot}). Against the naive-union ablation
the merge is recall-superior: $+0.035$ ($95\%$ CI $[+0.017,+0.053]$, excluding
zero) while holding a smaller live store ($38.2$ vs $42.9$ recallable units,
$\approx 11\%$ fewer, against $40.5$ stored). By consolidating paraphrased
duplicates the merge stops them crowding the gold facts out of the fixed top-$k$ read, so it
recalls more at lower live storage. Because recall@$k$ is computed on the status-live
store upstream of the answering LLM, it is model-independent by construction.
The five-model run checks this model-independence across models. It does not constitute five independent
replications of the merge result. The result is identical across all
five models ($0.619/0.630/0.595$ centralized/merge/union, across Qwen2.5 $1.5$/$7$/$32$B,
Mistral-7B, Gemma-3-12B): retrieval quality is a property of the store and encoder, not the
LLM. Residual model-dependence is confined to the secondary answer-F1, which scales with
size (token-F1 $0.34\!\to\!0.58$) and shows a large competence floor (full-context over
no-memory, every $95\%$ CI excluding zero, $+0.289$ to $+0.445$); confining matching to a
fixed encoder rather than the LLM trades the latter's discrimination-capacity
crossover~\cite{loven2026neuralrouter} for a model-independent, edge-deployable result.
The procedure raised $117$ contradiction flags across the $200$ items
(claims the inference signal classified as conflicting, without manual validation), none silently merged. Restricting merge candidacy to active
local claims lets these reach adjudication instead of being absorbed into
retired duplicates.

Against the centralized store the live footprints are near-identical ($38.2$ vs $38.6$):
the merge recalls more at lower live storage without a decentralization premium, and since the
per-query walk ranks over the status-live set, the smaller live store is proportionally
cheaper retrieval. The remaining data-management cost term, bytes synced, is bounded by
routing, with semantic delivering $195$ deltas/cell against broadcast's $585$
(Section~\ref{sec:eval-r}), while absolute wire bytes and wall-clock re-fetch under
sustained load remain future work.

\textbf{Overlap and $k$ sensitivity.} A matched overlap$\,\times\,k$ sweep with an
exact-dedup baseline confirms the win is semantic consolidation, not store-size
reduction: at the tight $k{=}5$ read distributed merge holds recall flat across overlaps
while naive union degrades, the gap widening with overlap and vanishing by $k{=}20$,
and exact-dedup tracks union rather than merge (full sweep in the electronic supplement,
Appendix~K).

The recall result reproduces on a second, long-horizon substrate, the LongMemEval
chat-memory benchmark~\cite{wu2024longmemeval} ($n{=}40$, $\approx\!550$ turns each):
distributed merge's recall@$5$ is at least centralized's ($0.675$ vs $0.575$) and
matches or exceeds naive union at $\approx\!26\%$ fewer live units, with exact-dedup again
tracking union, so the win is semantic and not store-size. This is a descriptive corroboration
at modest $n$ (no powered equivalence test claimed) on the LLM-independent recall metric
(full sweep in the electronic supplement, Appendix~E). Sync latency and
throughput are reported on real multi-host infrastructure (Section~\ref{sec:eval-s}).

\subsection{Merge classification quality}
\label{sec:eval-m2}
This experiment measures whether the decision procedure (Section~\ref{sec:merge}) classifies claim pairs correctly into the four decision classes: same-enough (merge), overlaps (relate), conflicts (conflict), and unrelated (no-relation). We report per-class precision, recall, and F1, the macro-F1, and the false-merge rate (the dangerous error), together with a receiver-operating sweep over $\theta_{\mathrm{merge}}$ and two ablation baselines, run in simulation. M-2 evaluates the pair-adjudication classifier, the part of the procedure that runs once a candidate pair exists. Its fourth verdict, \textsc{no-relation}, maps to \textsc{insert} at the protocol level, since a claim with no candidate at or above the relatedness floor is novel rather than rejected (Section~\ref{sec:merge-procedure}). \textsc{insert} has no pair to score and is therefore out of this experiment's scope by construction; the admission-gate \textsc{reject} is exercised separately in the gate ablation.

\textbf{Gold set.} We use a reproducible, balanced \textbf{$104$-pair} gold set built from a curated $26$-seed fact bank: each seed contributes one pair to each class (a paraphrase for \textsc{merge}, an overlapping fact for \textsc{relate}, a contradiction for \textsc{conflict}, an unrelated fact for \textsc{no-relation}), giving $26$ pairs per class with distinct claim keys so the $\sigma$/$\chi$ path is exercised (all pairs are same-frame, so distinct keys follow from distinct content). Nine seeds are tagged hard: their paraphrases and contradictions are lexically divergent (e.g.\ ``the Sun is mostly hydrogen and helium'' vs ``most of the Sun's mass is carbon''), the regime a surface-similarity gate is most likely to miss. The set is a controlled diagnostic instrument, with ground truth established by construction. It is not a naturalistic claim distribution. The operating point ($\theta_{\mathrm{merge}}$, $\sigma_{\mathrm{lo}}$) is calibrated on a small separate $9$-pair development set (Section~\ref{sec:merge}), which draws on the same fact vocabulary as the gold and shares one of its pairs; because the calibrated thresholds are mildly set-dependent, the scores at this operating point are best read as in-distribution, and the threshold-independent evidence is the receiver-operating sweep below (AUC $0.968$, separation across all thresholds) rather than any single point.

\textbf{Result.} With $\sigma$ from all-MiniLM-L6-v2 and $\chi$ from the \texttt{cross-encoder/nli-deberta-v3-small} cross-encoder NLI, at the calibrated operating point ($\theta_{\mathrm{merge}}{=}0.90$, $\sigma_{\mathrm{lo}}{=}0.12$) the procedure attains macro-F1 $0.845$ ($95\%$ CI $[0.766,0.912]$) and a \textbf{false-merge rate of $0.013$} ($1$ of $78$ non-mergeable pairs; exact Clopper--Pearson $95\%$ upper bound $0.069$, percentile-bootstrap $95\%$ CI $[0.000,0.042]$), with per-class \textsc{merge} P/R/F1 $0.95/0.77/0.85$, \textsc{relate} $0.78/0.96/0.86$, \textsc{conflict} $0.76/0.96/0.85$, and \textsc{no-relation} $1.00/0.69/0.82$. The receiver-operating sweep over $\theta_{\mathrm{merge}}$ (the open interval above $\sigma_{\mathrm{lo}}$) gives \textbf{AUC $0.968$} for \textsc{merge}-versus-rest: the same-enough decision is well separated across thresholds, so the false-merge rate is a choice of operating point, not a separability limit.

\textbf{Surfacing lexically divergent contradictions.} Gating \textsc{conflict} on $\theta_{\mathrm{merge}}$ (the high merge threshold) routed contradictions carried by divergent surface forms ($\sigma$ in $[\sigma_{\mathrm{lo}},\theta_{\mathrm{merge}})$) past both the \textsc{conflict} and the \textsc{relate} branch, silently to \textsc{no-relation}; on this set that placement scores \textsc{conflict} recall $0.23$ (easy $0.29$, hard $0.11$). We gate \textsc{conflict} on $\sigma_{\mathrm{lo}}$ instead (a contradiction need not be near-identical to be a contradiction, Section~\ref{sec:merge-procedure}): \textsc{conflict} recall rises to $0.96$ (easy $1.00$, hard $0.89$) and macro-F1 from $0.69$ to $0.845$ with the false-merge rate unchanged ($0.013$). The change re-routes contradictions from \textsc{no-relation} to \textsc{conflict}, never to \textsc{merge}. A false \textsc{conflict} is the error that is fail-safe with respect to the false-merge invariant: it is surfaced for authority adjudication (Section~\ref{sec:cons-trust}) instead of merged silently, at the cost of spurious graph structure and adjudication work. \textsc{conflict} precision falls to $0.76$.

\textbf{Signal ablation (additive ladder).} Building the signal set up confirms the value is in the combination, not any one signal. Key only ($\kappa$) reaches macro-F1 $0.100$ at \textsc{merge} recall $0$: $\kappa$ alone cannot match distinct-key paraphrases (it is inactive on this distinct-key gold by construction). Key$+$embedding ($\sigma\ge\theta_{\mathrm{merge}}$, no contradiction guard) lifts macro-F1 to $0.315$ but at a false-merge rate of $0.090$ ($7$ merges): similarity finds paraphrases yet silently merges high-similarity contradictions. Key$+$embedding$+$inference, the full $\kappa{+}\sigma{+}\chi$ procedure, reaches macro-F1 $0.845$ at false-merge $0.013$ ($1$): the inference gate keeps contradictions from merging. The freshness and Context gates are ablated separately and each shown individually necessary (Section~\ref{sec:eval-extra}).

\textbf{Robustness and signal generality.} The macro-F1 and false-merge results are
neither a clean-input nor a one-backbone artefact. Under heavy embedding noise or $50\%$
NLI corruption macro-F1 degrades gracefully (to $0.684$ / $0.653$) with false-merge staying
$\le 0.049$; and across three encoders $\times$ two NLI cross-encoders macro-F1 stays
$0.765$--$0.845$, with the two learned signals doing distinct jobs: the encoder sets
discrimination (macro-F1 $0.845$ vs $0.784$ for MiniLM vs mpnet at a fixed NLI) while the
inference model sets the safety floor (false-merge $\le 0.013$ under
\texttt{deberta-v3} for every encoder, $0.026$--$0.038$ under RoBERTa). The
inference gate, not the embedding, is the contradiction guard. The operating point transfers across the MiniLM
encoders but not to mpnet, the per-encoder calibration burden of
Section~\ref{sec:disc-limits} (full sweep in the electronic supplement, Appendix~L).
Because recall@$k$ is read upstream of the LLM (Section~\ref{sec:eval-h}), it is immune by
construction to any model's empty or unparseable outputs; those touch only the secondary
answer-F1.

\textbf{Field contradiction-miss rate.} The complementary deployment risk is the NLI
missing a real contradiction and merging silently. On a naturalistic HotpotQA
sample ($48$ pairs, $8$ value-swapped contradictions), the deployed NLI misses \textbf{$0$
of $8$} (recall $1.000$); its errors are safe-direction over-flags of neutrals (precision
$0.381$). Small-$n$ bounds rather than pins the field rate, but the dangerous error does
not appear on naturalistic pairs at this scale.

\textbf{Candidate-selection cost.} The classifier above scores an already-selected
candidate pair; candidate selection itself is an exhaustive scan of the receiving
brain's live claims, and it costs a median $0.23$\,ms (p95 $0.30$\,ms) with a candidate set of at
most $7$ claims at the $95$th percentile on the $16$-claim gold-derived workload, and
it scales linearly in the store: $3.8$\,ms per $1{,}000$ active claims, crossing the
NLI signal's cost only near $3{,}500$--$6{,}200$ claims in one cell (electronic
supplement, Appendix~M).

\subsection{Sync latency, throughput, and fan-out scaling}
\label{sec:eval-s}
This experiment measures whether \smp{} reaches global status consistency in bounded
time and how that latency scales with cluster fan-out. We report
time-to-convergence (last status update until all brains agree on status) and p50 and p95
publish-to-merge latency against centralized write latency and broadcast-all baselines,
varying the number of agents, brains, and claims (the three scale axes). The numbers are
taken on real multi-host infrastructure: a federated-broker testbed with broker-death
and partition injection. We measure saturation throughput (claims merged per second at the
single global merge sink, the serialization point) directly below, since the
sink's embedding+NLI pipeline is the bottleneck resource: this converts the analytic sink
bound (Section~\ref{sec:disc-future}) into an evidenced knee.

\begin{figure}[t]
  \centering
  \includegraphics[width=0.55\columnwidth]{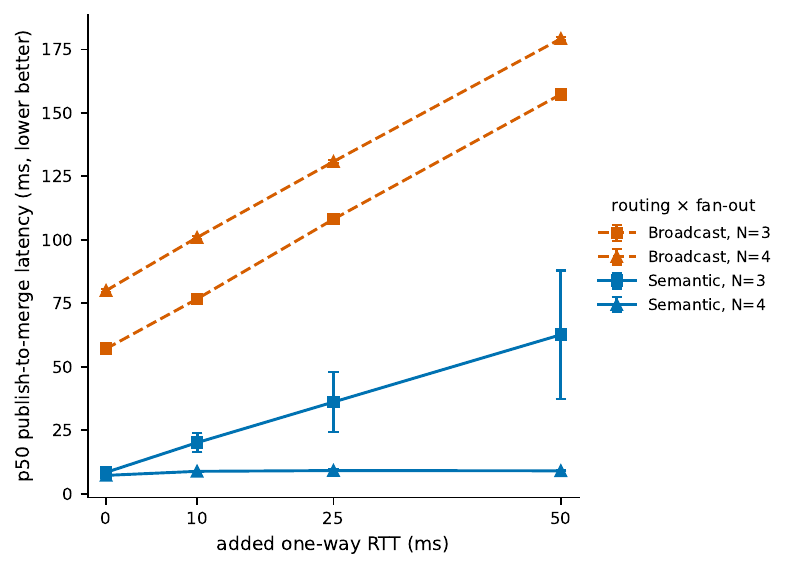}
  \caption{Sync-latency surface: p50 publish-to-merge latency vs added one-way
  RTT, one line per (routing $\times$ fan-out $N$), $N \in \{3, 4\}$; error bars are
  $95\%$ bootstrap CIs (HotpotQA, $15$ seeds/cell). Semantic routing (solid) stays low
  and falls as $N$ grows, where broadcast (dashed) rises with $N$. The
  $N{=}2$ cells are omitted from this figure: they spike under WAN in both modes, an
  emulation artifact (netem queueing on the low-fan-out replication-$2$ workload,
  without the replication-factor control that would isolate it) rather than an \smp{}
  property, and no claim rests on them. They are reported, with this confound named,
  in the electronic supplement's exact-values table (Appendix~H) for completeness
  only.}
  \label{fig:eval-s}
\end{figure}

\textbf{Result (latency).} On the multi-host testbed we measure publish-to-merge
p50 latency across a $4\!\times\!3$ grid of added one-way RTT $\in\{0,10,25,50\}$\,ms and
fan-out $N\in\{2,3,4\}$ brains ($15$ seeds per cell, HotpotQA workload, $360$ cells,
zero failures); Figure~\ref{fig:eval-s} gives the surface (exact per-cell p50 with
$95\%$ CIs in the electronic supplement, Appendix~H). Semantic routing confirms the near-flat-in-$N$ direction: p50 is flat-to-falling
in $N$ at every RTT tier (it delivers to only the $\sim\!1$ needed peer regardless
of cluster size), whereas broadcast grows with fan-out as every brain receives every
delta. The four added-one-way-RTT tiers map onto continuum hops: $0$\,ms is co-located
edge (intra-site LAN), $10$\,ms edge-to-near-edge, $25$\,ms edge-to-regional, and
$50$\,ms edge-to-core (midhaul/backhaul), anchored to the Neural Pub/Sub testbed's
within-site ($<1$\,ms) and cross-site ($\sim\!50$\,ms) boundary~\cite{loven2026neuralpubsub}.
Read this way, broadcast's growth with fan-out is the deployment-relevant point: selective
routing matters most at the scarce wide-area/backhaul edge.

\textbf{Result (sink saturation).} The single global sink saturates at a knee of
$\approx\!54$ merges/s: the per-delta embed$+$NLI merge cost (tens of ms/delta,
Section~\ref{sec:eval-scope}) is the bottleneck, so the sink is compute-bound and the knee
scales with its hardware; p95 climbs $137\!\to\!407$\,ms past the knee. The full throughput
curve and the node-scaling axes are in Section~\ref{sec:eval-scale} and the electronic
supplement (Appendices~F and~G).

\subsection{Convergence under concurrent edits and partition heal}
\label{sec:eval-c}
This experiment demonstrates that under concurrent conflicting status updates with reordered
and delayed delivery, all brains converge to the same per-claim statuses, and a
partition heals back to a single status set, as \smp{}'s status CRDT
guarantee requires (Section~\ref{sec:cons-crdt}). We report the fraction of brains agreeing on
per-claim status over time, and divergence followed by reconvergence after a partition
heals, against last-writer-wins (diverges and loses overrules) and no-CRDT union
baselines, varying concurrency degree, message-reorder rate, and partition duration.
The strong version uses real fault injection on the multi-host testbed. This demonstrates
convergence of claim status (Section~\ref{sec:cons-crdt}).

\textbf{Result (in-process CRDT demonstration).} A fixed contested-lifecycle event
set (four claims, each overruled or superseded by several peers issuing a
different status effect, with idempotent re-delivery of one link id) applied in
$200$ random delivery orderings yields a \textbf{status-agreement fraction of $1.000$}
under the status CRDT: all $200$ orderings compute one and the same per-claim status
map (e.g.\ a claim deprecated by one peer and overruled by another converges to
overruled under every order), which demonstrates empirically the order-independence
the lattice-join provides by construction (Section~\ref{sec:cons-crdt}). The agreement
fraction here is the largest set of orderings sharing one full per-claim status map,
divided by all orderings. The last-writer-wins baseline on the same event set
scores only $0.085$: delivery order picks the surviving status, fragmenting the outcome
into $24$ distinct status maps across the orderings (one per last-writer assignment)
and losing overrules, the divergence the CRDT removes. This is an in-process demonstration
(single host, reordered application); we additionally ran the multi-host version live
across real tiers (Section~\ref{sec:eval-scope}): on a real edge-link partition the
tiers diverge and then reconverge to a single status after heal, order-independent,
over wall-clock time.

\textbf{Result (robustness sweeps, summarized).} Two further in-process sweeps confirm the
order-independence is not an artefact of the single setting. Sweeping partition duration
$\{2,4,8,16\}$ against reorder rate $\{0,0.25,0.5,0.75,1.0\}$ ($200$ orderings/cell), the
status CRDT holds modal agreement $1.000$ in all $20$ cells while last-writer-wins matches
it only in-order and otherwise collapses to as low as $0.020$. On a reject-race scenario
(competing overrule and revoke links plus a stale link), \smp{}'s materialization-read gate
admits every link unconditionally and applies the staleness/overrule gate at read
time, as a pure function of the converged set. This reconverges on \textbf{all
$200/200$ orderings}, where a naive admission-time gate splits $\approx\!50/50$ between
overruled and revoked: the gate placement, not the gate itself,
preserves strong eventual consistency. Full per-cell results are in the electronic
supplement (Appendices~A and~B).

\textbf{Result (completeness under lossy routing).} The anti-entropy digest
(Section~\ref{sec:cons-wire}) makes per-claim convergence hold without assuming
complete delivery. Delivering the contested-lifecycle link set to $12$ brains under
independent per-delivery loss: without the digest, agreement degrades with loss
($0.967$ at the protocol's $\approx\!0.3\%$ rate, $0.117$ at $10\%$, fragmenting into up
to $12$ distinct status maps); with periodic reconciliation it returns to $1.000$
at every rate, in a single round at $\approx\!34$\,KB/round, so the federation reconverges
under lossy routing and not only after a partition heals (sweep and the link-survival caveat
in the electronic supplement, Appendix~J).

\textbf{Result (multi-host testbed).} On the real multi-host testbed (three brain
nodes on separate hosts, with a control plane that applies status updates and reads per-claim
status over HTTP) we run $30$ seeded partition--heal trials
(Table~\ref{tab:eval-proto}, bottom). While partitioned each
side applies only its own overrule link, so the per-claim status diverges (the
$1\!\mid\!2$ split, on every seed); after the partition heals and
both links are delivered to every node in a per-node seed-randomized order, \smp{}'s
status CRDT reconverges to a single status on \textbf{all $30/30$ trials} at full
healed agreement, order-independently (exact Clopper--Pearson one-sided $95\%$ lower
bound on the reconvergence rate $0.905$). The
last-writer-wins baseline, read off the same delivery order, reconverges on only
$\mathbf{11/30}$ trials: when the two
sides' last-applied effects differ it keeps different survivors and stays partly split,
the divergence the lattice-join removes. We model the partition as the absence
of cross-side link delivery, which is faithful because \smp{} status reconciliation is
delivery-driven (brains do not gossip status, so a partition is precisely no
cross-side propagation); the network-level isolation primitive (per-peer \texttt{tc}
loss between host groups) is implemented and unit-tested, and confirmed to install a
real $100\%$-loss partition on the testbed.

\subsection{Routing ablation: semantic routing versus broadcast}
\label{sec:eval-r}
This ablation measures whether learned semantic routing makes \smp{} sync efficient,
addressing the transport-integration question raised in Section~\ref{sec:impl-transport}.
We report deltas delivered per useful merge (routing precision), wasted-delivery rate,
latency, and throughput against naive broadcast (every PUBLISH to every brain) and
random/flooding gossip, varying the number of brains, interest selectivity, and the
embedding-match threshold. The strong version runs on real infrastructure. A positive
result supports the transport binding; it does not by itself prove the transport
optimal, and the integration remains promising and to be hardened.

\textbf{Result.} On the multi-host testbed (HotpotQA workload, $n=180$ cells per arm
across the RTT\,$\times$\,fan-out grid, $15$ seeds each), semantic routing delivers a
status update to the needed brains and almost no others (Table~\ref{tab:eval-proto},
top). The experiment's premise is the workload's interest structure, a property of the setup that the experiment does not measure: each published delta is needed by only about a third of the brains. A routing-free broadcast, which sends every delta to every brain, therefore wastes about two-thirds of its deliveries. That two-thirds follows from the workload's construction and is not itself a measured result. The measured question is how close content-aware routing comes to delivering only where needed. Semantic routing delivers $\sim\!3\times$ fewer messages than broadcast ($195$ vs $585$ deltas per cell) at matched merge recall ($0.997$ vs $1.000$), for a measured wasted-delivery rate of $0.003$: it recovers nearly all of the structurally-wasted deliveries at no recall cost. Two standard non-semantic placement baselines~\cite{loven2026neuralpubsub}, random and round-robin, confirm the mechanism: content-awareness, not merely a smaller delivery budget, recovers this waste. On a separable three-domain probe (in-process, ground-truth targets by construction) both baselines spend a budget matched to the needed-peer count without regard to content, reaching only a third of the needed brains (recall $0.33$) at the same two-thirds waste, whereas routing on the merge signal reaches every needed brain at zero waste. The advantage widens with the
number of brains: broadcast's delivery cost (and its p50 sync latency,
Figure~\ref{fig:eval-s}) grows with fan-out, while semantic routing's stays flat-to-falling, tracking only the needed peers, whose count does not grow with the cluster. A dense sweep to $N{=}128$ quantifies this
(Section~\ref{sec:eval-scale}; electronic supplement, Appendix~F). Formally, delivery
cost per published claim is $O(r)$ in the number $r$ of interested recipients. The
observed flat-in-$N$ behavior holds because this workload's interest structure keeps
$r$ roughly constant as $N$ grows. This does not show that routing cost is $O(1)$ in $N$ generally.
Semantic routing is
therefore a sound and efficient sync substrate for \smp{}, the property the transport
binding of Section~\ref{sec:impl-transport} claims.

\begin{table}[t]
  \centering
  \footnotesize
  \caption{Distributed-protocol results: routing and convergence. Top: routing ablation ($n{=}180$ cells per arm): semantic routing delivers $\sim\!3\times$ fewer deltas per cell at matched merge recall, for a measured wasted-delivery rate of $0.003$; a routing-free broadcast wastes $\approx\!2/3$ by construction on this workload (Section~\ref{sec:eval-r}). Bottom: partition-heal convergence ($n{=}30$ seeds): the status CRDT reconverges on all trials to full healed agreement, while last-writer-wins reconverges on a minority and stays split. 
  Bracketed values are $95\%$ bootstrap CIs.}
  \label{tab:eval-proto}
  \setlength{\tabcolsep}{4pt}
  \renewcommand{\arraystretch}{1.15}
  \begin{tabular}{l c c c}
    \toprule
    \multicolumn{4}{l}{\textbf{Routing} ($n{=}180$/arm)} \\
    \midrule
    \textbf{Mode} & \textbf{Wasted-delivery} & \textbf{Recall} & \textbf{Deltas/cell} \\
    \midrule
    Broadcast & $\approx\!2/3$ (by constr.) & $1.000$ & $585$ \\
    Semantic  & $0.003$ \,[0.003, 0.004] & $0.997$ & $195$ \\
    \midrule
    \multicolumn{4}{l}{\textbf{Convergence} (partition heal, $n{=}30$ seeds)} \\
    \midrule
    \textbf{Baseline} & \textbf{Reconverges} & \multicolumn{2}{c}{\textbf{Healed agreement}} \\
    \hline
    Status-CRDT       & $30/30 = 1.000$ & \multicolumn{2}{c}{$1.000$ \,[1.000, 1.000]} \\
    Last-writer-wins  & $11/30 = 0.367$ & \multicolumn{2}{c}{$0.789$ \,[0.733, 0.844]} \\
    \bottomrule
  \end{tabular}
\end{table}

\subsection{Node-scaling}
\label{sec:eval-scale}
\smp{}'s overhead scales with the rate of novel content. Per-claim sync overhead is
flat in $N$ under semantic routing where broadcast is $O(N)$ ($128\times$ at $N{=}128$, routing
recall $1.000$); and when $N$ brains re-publish a fixed corpus, merge-on-receive consolidates the
copies so the sink's live store stays corpus-sized ($26$ units, flat to $N{=}128$) where naive
union grows to $26N$. Both the per-claim sync cost and the per-sink state are thus $O(1)$ in
cluster size, the empirical form of ``load grows with the rate of novel claims, not the
number of brains'', leaving the single sink's per-delta compute (the saturation knee of
Section~\ref{sec:eval-s}) as the one serialization limit. This $O(1)$-in-$N$ figure is the
$O(r)$ routing cost of Section~\ref{sec:eval-r} evaluated under this sweep's
separable-domain interest structure, where each claim interests exactly one brain
($r{=}1$ regardless of $N$); it is not a claim that routing cost is independent of $N$
under an arbitrary interest distribution. These node-count sweeps run
in-process to reach $N{=}128$ on separable domains, so $O(1)$-in-$N$ is a scaling
property of the protocol logic, not a multi-host wall-clock claim (the live latency surface of
Section~\ref{sec:eval-s} is measured at $N\le 4$); routing quality on realistic domains
is the Section~\ref{sec:eval-r} ablation. The full sweeps, with the side-by-side small-multiples
figure, are in the electronic supplement (Appendices~F and~G).

\subsection{Additional ablations}
\label{sec:eval-extra}
Four further ablations isolate individual design choices as deterministic, model-free checks on small constructed probes: necessity and invariant demonstrations. They do not report rates over a distribution; the measured false-merge rate is the merge-classification result (Section~\ref{sec:eval-m2}). Two bear on safe adaptation: the authority tie-break picks the canonical brain's claim by construction (the basis for ``authority gating prevents unsafe state change''), and the gate-necessity probe shows each of the Context, inference, and staleness
gates individually necessary. Dropping any one flips exactly its own distinct-key
probe to a false MERGE. The contradiction-representation and signature-verification probes,
and the consolidated probe-by-probe table, are in the electronic supplement (Appendix~C).

\subsection{Cross-tier deployment and per-delta cost}
\label{sec:eval-scope}
The deterministic claims (accuracy, classification, overhead, gate ablations) reproduce on a
multi-process single host. Latency, throughput, and convergence-under-partition need real
infrastructure and run on the multi-host testbed, each controlled experiment on its
representative tier: merge delta on national HPC, latency, routing and partition on the 5G
edge, classification and ablations locally. The 5GTN deployment is genuine operator-grade 5G
edge infrastructure; the sync-latency wide-area RTT tiers ($10/25/50$\,ms) are emulated on
those nodes via \texttt{tc qdisc netem} rather than over-the-air, and the model-family runs
use CSC GPU nodes. The partition experiment (Section~\ref{sec:eval-c}) models a partition as
the absence of cross-side delivery, faithful because \smp{} status reconciliation is
delivery-driven. Beyond these, we deployed
\smp{} live end-to-end across all
three real tiers (local host $+$ 5GTN edge $+$ CSC HPC, brains over real WAN): the merge
procedure and status CRDT ran at every tier, the running-example decisions reproduced
identically, and per-claim status converged to one value, at real publish-to-merge
round-trips of $\approx\!11$--$13$\,ms p50 (the well-provisioned Oulu--CSC research network,
below the $50$\,ms tier the sync-latency experiment emulates). This single-session feasibility deployment
(dense signals on the hub, characterized in the merge-classification experiment) ran over two transports: a minimal relay
and, to confirm operation over the standard published fabric, Apache Kafka
(Section~\ref{sec:impl-transport}, KRaft, $N{=}3$, all converging to the OVERRULED join);
the partition-heal is quantified over $10$ trials below. It is an existence proof of
cross-tier operation and convergence; saturation throughput and larger-$N$ scale are open
(Section~\ref{sec:disc-future}).

\textbf{Per-delta admission cost, and what the sync-latency measurement includes.} The deployment computed
dense signals at the hub, so we characterize the per-delta admission cost separately. On
edge-class CPU the common one-pair admission path costs $23.5$\,ms per delta ($\pm 13.3$, $n{=}200$: $4.0$ embed $+$
$19.4$ NLI; HMAC verify sub-ms); on a CSC V100 GPU \textbf{$9.8$\,ms} ($1.8$ embed $+$ $8.0$
NLI), a $\approx\!2.4\times$ speedup. The gate-aware fallback of Section~\ref{sec:merge-procedure} adds one inference check when it fires. NLI is $83\%$ of cost and the dominant term at the evaluated store sizes; it bounds
the $\approx\!54$ merges/s sink knee (Section~\ref{sec:eval-s}). At stores beyond a few
thousand active claims the linear candidate scan overtakes it (Appendix~M), where
indexed or approximate-nearest-neighbour retrieval is the natural extension.
The sync latencies are thus transport plus merge with hub-precomputed
signals; an edge node running the full admission path per receive adds the $23.5$\,ms above,
so ``edge-deployable'' covers embed$+$NLI$+$verify at tens of ms/delta on commodity CPU, not
only transport.
\textbf{Quantitative cross-tier partition-heal (real Kafka WAN).} Beyond the single
existence-proof session, a seeded multi-trial partition-heal over the real Kafka WAN (laptop
$+$ CSC HPC $+$ 5GTN edge brains) \textbf{reconverged on all $10/10$ trials} (success
fraction $1.000$, every trial diverging first; Clopper--Pearson one-sided $95\%$ lower
bound $0.741$) at heal latency p50~$\approx\!3.0$\,s /
p95~$\approx\!3.4$\,s (WAN reconnection and missed-offset re-consumption, not the merge step;
method and per-trial data in the electronic supplement, Appendix~D).

The evaluation therefore runs on a real computing
continuum~\cite{beckman2020continuum,parashar2025everywhere}: per-tier experiments and
the seeded cross-tier partition-heal are the quantitative evidence, the live deployment
is a feasibility existence proof, and the latency regimes are representative mobile-edge
emulation rather than live radio. Section~\ref{sec:disc-limits} bounds these claims.

\section{Related work}
\label{sec:related}

\smp{} composes mature ingredients; the novelty is the composition, a wire-level
semantic-merge discipline for sovereign agent brains. Eight threads position it.

\textbf{Agent communication and context protocols.}
Knowledge-sharing agent languages (KQML~\cite{finin1994kqml}, FIPA-ACL~\cite{fipa2002acl})
carried an \texttt{ontology} field but assumed it already shared, with no alignment or merge
step; the contemporary protocols are orthogonal in payload: the Model Context Protocol
exposes tools and resources~\cite{anthropic2024mcp} and Agent2Agent exposes tasks
and capabilities~\cite{google2025a2a}, neither merged knowledge state. The Open Knowledge
Format (OKF)~\cite{google2026okf} is the closest recent entrant and the sharpest contrast.
It standardizes the representation of curated knowledge for agents, a bundle of
markdown concepts with typed-by-prose links and frontmatter carrying provenance, trust, and
a concept-level lifecycle (\texttt{status}, \texttt{stale\_after}), and is explicitly
non-prescriptive about storage, serving, and query infrastructure. It therefore defines no
reconciliation across sovereign writers: two bundles asserting the same fact differently,
or contradicting one another, are reconciled by git review workflows, with a human as the
merge procedure. Representation is being standardized, but reconciliation is not.
\smp{} occupies that empty slot: MCP exposes tools, A2A exposes tasks, OKF standardizes how
knowledge is written down, and \smp{} exposes meanings, claims, and merges. These are
complementary rather than competing: an MCP tool call yields an observation \smp{} can
publish as Evidence, an \smp{}-merged Claim can parameterize a later MCP call, and an OKF
bundle is a representation \smp{} can carry. A harness speaks MCP to tools and \smp{} to
brains; MCP governs an agent's available actions, and \smp{} keeps every brain's
knowledge coherent.

\textbf{Conflict-free replication and eventual consistency.}
CRDTs give deterministic convergence for lattice-mergeable
state~\cite{shapiro2011crdt}, extended to delta efficiency~\cite{almeida2018deltacrdt}, nested
JSON~\cite{kleppmann2017jsoncrdt}, local-first operation~\cite{kleppmann2019localfirst}, and
Byzantine peers~\cite{kleppmann2020byzantine}, but a CRDT converges to a value by a
fixed algebraic rule, with no notion of whether two nodes mean the same thing. \smp{} runs a semantic verdict above the lattice, reusing a status CRDT only for per-claim convergence (Section~\ref{sec:consistency}).

\textbf{Ontology alignment, schema matching, entity resolution.}
Deciding whether two symbols are equivalent, related, or disjoint is mature (canonical
theory~\cite{euzenat2013ontologymatching}, surveys and a standing
benchmark~\cite{shvaiko2013ontologymatching,pour2023oaei}, record
linkage~\cite{elmagarmid2007duplicate,christen2012datamatching}, an LLM-driven
turn~\cite{chen2024llmalign}), but produces correspondences offline, batch, and
pairwise. \smp{} runs the same verdict as a runtime primitive on live, multi-owner
state, and can plug such matchers in as its \texttt{same-enough} engine.

\textbf{Knowledge fusion, truth discovery, and shared memory.}
Web-scale fusion assigns correctness probabilities and resolves conflicts by estimating source dependence~\cite{dong2014knowledgevault,dong2009integrating}, but for a centralized integrator that owns all sources and emits one truth; shared-memory coordination recurs from blackboards~\cite{hayesroth1985blackboard,nii1986blackboard} through federated linked data~\cite{sambra2016solid} to multi-agent LLM memory, closest in Collaborative Memory~\cite{rezazadeh2025collaborativememory}, which names the divergent-belief problem but gives no merge verdicts, trust hierarchy, or wire protocol. \smp{} is the decentralized, sovereign-peer inversion of both: pairwise semantic merges over a protocol, a declared trust hierarchy in place of inferred source reliability, and \texttt{overlaps}/\texttt{conflicts} kept first-class rather than collapsed to one fused value.

\textbf{Sync transports: gossip and publish/subscribe.}
Anti-entropy replication converges replicas by content-agnostic pairwise
exchange~\cite{demers1987epidemic}; publish/subscribe decouples producers and
consumers~\cite{eugster2003pubsub} with content-based routing~\cite{carzaniga2001siena}.
Both route bytes and events; a subscription is a filter, not a merge. \smp{} makes
publish/subscribe/merge the protocol, on a learned-semantic
fabric~\cite{loven2026neuralpubsub,loven2026neuralrouter}.

\textbf{Versioned structured knowledge and provenance.}
``Git for data/knowledge'' (Dolt~\cite{dolt}, TerminusDB~\cite{terminusdb}) provides branch/merge/diff but only syntactic three-way merge within one repository, no semantic verdict; W3C PROV~\cite{moreau2013provdm} standardizes the provenance vocabulary but not a merge protocol or trust model consuming it. \smp{} combines PROV-style Evidence on every Claim, semantic (not textual) merge, and a canonical-brain trust hierarchy, promoting git-for-knowledge from a repository feature to an inter-brain protocol.

\textbf{Agent-memory stores.}
\smp{} synchronizes the versioned, typed-link agent-memory stores of Zep/Graphiti~\cite{rasmussen2025zep}, A-MEM~\cite{xu2025amem}, GraphRAG~\cite{edge2024graphrag}, HippoRAG~\cite{gutierrez2024hipporag}, and Mem0~\cite{chhikara2025mem0}, which make typed links (and, in Zep/Graphiti, validity intervals) first class within a single store but define no merge protocol across sovereign stores, \smp{}'s slot.

\textbf{Autonomic and self-adaptive systems.} \smp{}'s per-brain merge loop (observe a delta, decide a verdict, reconfigure local knowledge, self-healing and self-protecting; Section~\ref{sec:contributions}) is a MAPE-style loop in the autonomic-computing~\cite{kephart2003vision} and self-adaptive-software~\cite{salehie2009self,delemos2013roadmap} traditions, differing on two axes that keep it within that tradition. Where most self-adaptive systems centralize the loop under a single or hierarchical manager (e.g.\ Rainbow~\cite{garlan2004rainbow}), \smp{} is decentralized control~\cite{weyns2013patterns}: every brain runs the full loop, coordinating on the wire with no central manager. The artefact it adapts is also a model (the shared knowledge graph as a \emph{model@run.time}~\cite{blair2009models}), so \smp{} adapts knowledge where self-adaptive LLM multi-agent systems~\cite{nascimento2023selfadaptive} adapt behaviour or configuration. The same two axes separate \smp{} from the federated-broker substrate it rides~\cite{loven2026neuralpubsub}, which closes a MAPE-K loop per broker over health, load, and clearing prices, and whose partition result is transport availability; \smp{} closes its loop per brain over the knowledge graph, and its partition result is per-claim status convergence. The two are complementary layers of one federation, not competing designs.

\section{Discussion and limitations}
\label{sec:discussion}

\subsection{Limitations}
\label{sec:disc-limits}
\begin{itemize}
  \item \textbf{Merge is evaluated relative to a centralized baseline.} Distributed
  merge matches centralized memory and beats naive union (Section~\ref{sec:eval-h});
  the routing ablation evidences the transport binding without establishing it optimal
  (Section~\ref{sec:eval-r}).

  \item \textbf{Status convergence does not adjudicate truth.} The CRDT guarantees order-independent agreement
  on claim status ($30/30$ after a real partition heal, against $11/30$ for
  last-writer-wins; Section~\ref{sec:eval-c}), not the correctness of any overrule. The conflict-of-law
  problem, which of two contradictory active claims governs, remains open. \smp{}
  records the contradiction as a first-class link and converges its status; the
  normative resolution is deferred to adjudication rather than settled by a protocol
  default.

  \item \textbf{The merge graph is order-sensitive.} Only the status
  layer is order-independent. The graph the per-pair procedure builds against already-held
  state is not, and the electronic supplement (Appendix~M) quantifies it: under randomized
  delivery orders, $91/200$ orderings reproduce the running example's depicted graph, a
  $16$-claim workload yields $177$ distinct joint graph fingerprints across $200$
  orderings, and top-$k$ retrieval differs across orderings on every probe query. Two
  brains fed the same deltas in different orders thus agree on every status while holding
  different link structure and representatives; converging the graph itself, e.g.\ by
  deterministic re-adjudication over the converged claim set, is an open extension.

  \item \textbf{First-class relational structure rests on an open empirical premise:} that
  link structure carries information beyond node contents ($H(X_L \mid X_N)>0$) on the
  graphs that matter; a negative finding would weaken typed links as a primitive, not
  the rest of the protocol.

  \item \textbf{Merge thresholds are empirically calibrated.} The thresholds
  (Section~\ref{sec:merge}) come from a small separate set, not from the embedding
  geometry, and are mildly set-dependent; we report the ROC over $\theta_{\mathrm{merge}}$
  (AUC $0.968$) and a degradation sweep rather than claiming derived optima. A tuned
  threshold drifts with encoder and domain (the operating point fails to transfer to
  mpnet, Section~\ref{sec:eval-m2}), a calibration burden the deployer inherits.
\end{itemize}

\subsection{Future work}
\label{sec:disc-future}
Several evaluation points are realized only in part. A LongMemEval answer-accuracy run is confirmatory (answer-F1 already sits at merge-versus-centralized parity, Section~\ref{sec:eval-h}); a harder GAIA level-2 delta needs a tool-using agent loop, out of scope here. Open on the systems side: absolute bytes synced under sustained load and the latency surface beyond $N\le 4$ (Sections~\ref{sec:eval-h} and~\ref{sec:eval-scale}). A contradiction-tuned inference model would lift recall on lexically divergent contradictions (Section~\ref{sec:eval-m2}). Finally, learning the signal combination within the hard safety constraints (a detected contradiction is never merged), plus a grounding-evidence signal where provenance exists, is the natural next step, both sensitive to the inference signal's reliability.

\subsection{Deployment outlook}
\label{sec:disc-deploy}

Because the merge procedure and status CRDT run identically at every brain, \smp{} is topology-agnostic: the same guarantees hold under a global-brain star (Section~\ref{sec:eval-h}), a peer-to-peer mesh (Section~\ref{sec:eval-c}), or a hybrid per-domain deployment peering across boundaries under the Context and authority gates (Figure~\ref{fig:running}); the global brain is optional.

\section{Conclusion}
\label{sec:conclusion}

Autonomous agents need to merge what they know as well as exchange tools, and a tool-invocation protocol cannot provide this, because it moves capability while leaving two brains that hold the same fact differently unreconciled. \smp{} fills that gap with an operational merge decision procedure, an auditable Patch, and a publish/subscribe binding kept coherent by a per-claim status CRDT. In the evaluation (\cref{sec:eval}), distributed merge matches a centralized store on recall and exceeds naive union at less live storage; the status CRDT reconverges $30/30$ after a real partition heal against $11/30$ for last-writer-wins; and semantic routing delivers $\approx\!3\times$ fewer messages at matched recall. \smp{} therefore keeps sovereign brains coherent in per-claim status across the computing continuum, while each brain adjudicates its semantic graph structure locally; the limitations that future work should harden are stated in \cref{sec:disc-limits}.

\begin{acks}
\small
This work was supported by the Research Council of Finland through the 6G Flagship program (grant 318927), as well as the CO2CREATION Strategic Research Council project (grant 372355), by the EC through HEU NEUROCLIMA project (GA 101137711) and the HEU ARGENTIC project (GA 101298599), the Interreg Aurora ResilientEdge project (Grant Number: 20373282), the ERDF (project numbers A81568, A91867), by the Business Finland through the Neural pub/sub research project (diary number 8754/31/2022).
\end{acks}

\bibliographystyle{ACM-Reference-Format}
\bibliography{refs}


\begin{thebibliography}{48}


\ifx \showCODEN    \undefined \def \showCODEN     #1{\unskip}     \fi
\ifx \showISBNx    \undefined \def \showISBNx     #1{\unskip}     \fi
\ifx \showISBNxiii \undefined \def \showISBNxiii  #1{\unskip}     \fi
\ifx \showISSN     \undefined \def \showISSN      #1{\unskip}     \fi
\ifx \showLCCN     \undefined \def \showLCCN      #1{\unskip}     \fi
\ifx \shownote     \undefined \def \shownote      #1{#1}          \fi
\ifx \showarticletitle \undefined \def \showarticletitle #1{#1}   \fi
\ifx \showURL      \undefined \def \showURL       {\relax}        \fi
\providecommand\bibfield[2]{#2}
\providecommand\bibinfo[2]{#2}
\providecommand\natexlab[1]{#1}
\providecommand\showeprint[2][]{arXiv:#2}

\bibitem[Almeida et~al\mbox{.}(2018)]%
        {almeida2018deltacrdt}
\bibfield{author}{\bibinfo{person}{Paulo~S{\'e}rgio Almeida},
  \bibinfo{person}{Ali Shoker}, {and} \bibinfo{person}{Carlos Baquero}.}
  \bibinfo{year}{2018}\natexlab{}.
\newblock \showarticletitle{Delta State Replicated Data Types}.
\newblock \bibinfo{journal}{\emph{J. Parallel and Distrib. Comput.}}
  \bibinfo{volume}{111} (\bibinfo{year}{2018}), \bibinfo{pages}{162--173}.
\newblock
\href{https://doi.org/10.1016/j.jpdc.2017.08.003}{doi:\nolinkurl{10.1016/j.jpdc.2017.08.003}}
\newblock
\shownote{arXiv:1603.01529}.


\bibitem[{Anthropic}(2024)]%
        {anthropic2024mcp}
\bibfield{author}{\bibinfo{person}{{Anthropic}}.}
  \bibinfo{year}{2024}\natexlab{}.
\newblock \bibinfo{title}{Introducing the Model Context Protocol}.
\newblock \bibinfo{howpublished}{Anthropic (announced 2024-11-25; spec version
  2024-11-05)}.
\newblock
\urldef\tempurl%
\url{https://www.anthropic.com/news/model-context-protocol}
\showURL{%
\tempurl}


\bibitem[Beckman et~al\mbox{.}(2020)]%
        {beckman2020continuum}
\bibfield{author}{\bibinfo{person}{Pete Beckman} {et~al\mbox{.}}}
  \bibinfo{year}{2020}\natexlab{}.
\newblock \bibinfo{booktitle}{\emph{Harnessing the Computing Continuum for
  Programming Our World}}.
\newblock \bibinfo{publisher}{John Wiley \& Sons, Ltd}, Chapter~7,
  \bibinfo{pages}{215--230}.
\newblock
\href{https://doi.org/10.1002/9781119551713.ch7}{doi:\nolinkurl{10.1002/9781119551713.ch7}}


\bibitem[Blair et~al\mbox{.}(2009)]%
        {blair2009models}
\bibfield{author}{\bibinfo{person}{Gordon Blair} {et~al\mbox{.}}}
  \bibinfo{year}{2009}\natexlab{}.
\newblock \showarticletitle{Models@run.time}.
\newblock \bibinfo{journal}{\emph{Computer}} \bibinfo{volume}{42},
  \bibinfo{number}{10} (\bibinfo{year}{2009}), \bibinfo{pages}{22--27}.
\newblock
\href{https://doi.org/10.1109/MC.2009.326}{doi:\nolinkurl{10.1109/MC.2009.326}}


\bibitem[Bowman et~al\mbox{.}(2015)]%
        {bowman2015snli}
\bibfield{author}{\bibinfo{person}{Samuel~R. Bowman} {et~al\mbox{.}}}
  \bibinfo{year}{2015}\natexlab{}.
\newblock \showarticletitle{A large annotated corpus for learning natural
  language inference}. In \bibinfo{booktitle}{\emph{Proc. Conf. Empirical
  Methods in Natural Language Processing (EMNLP)}}. \bibinfo{pages}{632--642}.
\newblock


\bibitem[Carzaniga et~al\mbox{.}(2001)]%
        {carzaniga2001siena}
\bibfield{author}{\bibinfo{person}{Antonio Carzaniga},
  \bibinfo{person}{David~S. Rosenblum}, {and} \bibinfo{person}{Alexander~L.
  Wolf}.} \bibinfo{year}{2001}\natexlab{}.
\newblock \showarticletitle{Design and Evaluation of a Wide-Area Event
  Notification Service}.
\newblock \bibinfo{journal}{\emph{ACM Transactions on Computer Systems}}
  \bibinfo{volume}{19}, \bibinfo{number}{3} (\bibinfo{year}{2001}),
  \bibinfo{pages}{332--383}.
\newblock
\href{https://doi.org/10.1145/380749.380767}{doi:\nolinkurl{10.1145/380749.380767}}


\bibitem[Chen et~al\mbox{.}(2024)]%
        {chen2024llmalign}
\bibfield{author}{\bibinfo{person}{Xuan Chen}, \bibinfo{person}{Tong Lu}, {and}
  \bibinfo{person}{Zhichun Wang}.} \bibinfo{year}{2024}\natexlab{}.
\newblock \bibinfo{title}{LLM-Align: Utilizing Large Language Models for Entity
  Alignment in Knowledge Graphs}.
\newblock
\showeprint[arxiv]{2412.04690}~[cs.CL]
\newblock
\shownote{arXiv:2412.04690}.


\bibitem[Chhikara et~al\mbox{.}(2025)]%
        {chhikara2025mem0}
\bibfield{author}{\bibinfo{person}{Prateek Chhikara} {et~al\mbox{.}}}
  \bibinfo{year}{2025}\natexlab{}.
\newblock \bibinfo{title}{Mem0: Building Production-Ready AI Agents with
  Scalable Long-Term Memory}.
\newblock
\showeprint[arxiv]{2504.19413}~[cs.AI]
\newblock
\shownote{arXiv:2504.19413}.


\bibitem[Christen(2012)]%
        {christen2012datamatching}
\bibfield{author}{\bibinfo{person}{Peter Christen}.}
  \bibinfo{year}{2012}\natexlab{}.
\newblock \bibinfo{booktitle}{\emph{Data Matching: Concepts and Techniques for
  Record Linkage, Entity Resolution, and Duplicate Detection}}.
\newblock \bibinfo{publisher}{Springer}.
\newblock
\href{https://doi.org/10.1007/978-3-642-31164-2}{doi:\nolinkurl{10.1007/978-3-642-31164-2}}


\bibitem[de~Lemos et~al\mbox{.}(2013)]%
        {delemos2013roadmap}
\bibfield{author}{\bibinfo{person}{Rog\'{e}rio de Lemos} {et~al\mbox{.}}}
  \bibinfo{year}{2013}\natexlab{}.
\newblock \showarticletitle{Software Engineering for Self-Adaptive Systems: A
  Second Research Roadmap}.
\newblock In \bibinfo{booktitle}{\emph{Software Engineering for Self-Adaptive
  Systems II}}. Vol.~\bibinfo{volume}{7475}. \bibinfo{publisher}{Springer},
  \bibinfo{pages}{1--32}.
\newblock
\href{https://doi.org/10.1007/978-3-642-35813-5_1}{doi:\nolinkurl{10.1007/978-3-642-35813-5_1}}


\bibitem[Demers et~al\mbox{.}(1987)]%
        {demers1987epidemic}
\bibfield{author}{\bibinfo{person}{Alan Demers} {et~al\mbox{.}}}
  \bibinfo{year}{1987}\natexlab{}.
\newblock \showarticletitle{Epidemic Algorithms for Replicated Database
  Maintenance}. In \bibinfo{booktitle}{\emph{Proc. ACM Symp. Principles of
  Distributed Computing (PODC)}}. \bibinfo{pages}{1--12}.
\newblock
\href{https://doi.org/10.1145/41840.41841}{doi:\nolinkurl{10.1145/41840.41841}}


\bibitem[{DoltHub}({[n.\,d.]})]%
        {dolt}
\bibfield{author}{\bibinfo{person}{{DoltHub}}.}
  \bibinfo{year}{[n.\,d.]}\natexlab{}.
\newblock \bibinfo{title}{Dolt: Git for Data}.
\newblock \bibinfo{howpublished}{Software,
  \url{https://github.com/dolthub/dolt}}.
\newblock


\bibitem[Dong et~al\mbox{.}(2014)]%
        {dong2014knowledgevault}
\bibfield{author}{\bibinfo{person}{Xin~Luna Dong} {et~al\mbox{.}}}
  \bibinfo{year}{2014}\natexlab{}.
\newblock \showarticletitle{Knowledge Vault: A Web-Scale Approach to
  Probabilistic Knowledge Fusion}. In \bibinfo{booktitle}{\emph{Proc. ACM
  SIGKDD Conf. Knowledge Discovery and Data Mining (KDD)}}.
  \bibinfo{pages}{601--610}.
\newblock
\href{https://doi.org/10.1145/2623330.2623623}{doi:\nolinkurl{10.1145/2623330.2623623}}


\bibitem[Dong et~al\mbox{.}(2009)]%
        {dong2009integrating}
\bibfield{author}{\bibinfo{person}{Xin~Luna Dong}, \bibinfo{person}{Laure
  Berti-{\'E}quille}, {and} \bibinfo{person}{Divesh Srivastava}.}
  \bibinfo{year}{2009}\natexlab{}.
\newblock \showarticletitle{Integrating Conflicting Data: The Role of Source
  Dependence}.
\newblock \bibinfo{journal}{\emph{Proceedings of the VLDB Endowment}}
  \bibinfo{volume}{2}, \bibinfo{number}{1} (\bibinfo{year}{2009}),
  \bibinfo{pages}{550--561}.
\newblock
\href{https://doi.org/10.14778/1687627.1687690}{doi:\nolinkurl{10.14778/1687627.1687690}}


\bibitem[Edge et~al\mbox{.}(2024)]%
        {edge2024graphrag}
\bibfield{author}{\bibinfo{person}{Darren Edge} {et~al\mbox{.}}}
  \bibinfo{year}{2024}\natexlab{}.
\newblock \bibinfo{title}{From Local to Global: A Graph RAG Approach to
  Query-Focused Summarization}.
\newblock
\showeprint[arxiv]{2404.16130}~[cs.CL]
\newblock
\shownote{arXiv:2404.16130 (Microsoft GraphRAG)}.


\bibitem[Elmagarmid et~al\mbox{.}(2007)]%
        {elmagarmid2007duplicate}
\bibfield{author}{\bibinfo{person}{Ahmed~K. Elmagarmid},
  \bibinfo{person}{Panagiotis~G. Ipeirotis}, {and}
  \bibinfo{person}{Vassilios~S. Verykios}.} \bibinfo{year}{2007}\natexlab{}.
\newblock \showarticletitle{Duplicate Record Detection: A Survey}.
\newblock \bibinfo{journal}{\emph{IEEE Transactions on Knowledge and Data
  Engineering}} \bibinfo{volume}{19}, \bibinfo{number}{1}
  (\bibinfo{year}{2007}), \bibinfo{pages}{1--16}.
\newblock
\href{https://doi.org/10.1109/TKDE.2007.250581}{doi:\nolinkurl{10.1109/TKDE.2007.250581}}


\bibitem[Eugster et~al\mbox{.}(2003)]%
        {eugster2003pubsub}
\bibfield{author}{\bibinfo{person}{Patrick~Th. Eugster} {et~al\mbox{.}}}
  \bibinfo{year}{2003}\natexlab{}.
\newblock \showarticletitle{The Many Faces of Publish/Subscribe}.
\newblock \bibinfo{journal}{\emph{Comput. Surveys}} \bibinfo{volume}{35},
  \bibinfo{number}{2} (\bibinfo{year}{2003}), \bibinfo{pages}{114--131}.
\newblock
\href{https://doi.org/10.1145/857076.857078}{doi:\nolinkurl{10.1145/857076.857078}}


\bibitem[Euzenat and Shvaiko(2013)]%
        {euzenat2013ontologymatching}
\bibfield{author}{\bibinfo{person}{J{\'e}r{\^o}me Euzenat} {and}
  \bibinfo{person}{Pavel Shvaiko}.} \bibinfo{year}{2013}\natexlab{}.
\newblock \bibinfo{booktitle}{\emph{Ontology Matching} (\bibinfo{edition}{2nd}
  ed.)}.
\newblock \bibinfo{publisher}{Springer}.
\newblock
\href{https://doi.org/10.1007/978-3-642-38721-0}{doi:\nolinkurl{10.1007/978-3-642-38721-0}}


\bibitem[Finin et~al\mbox{.}(1994)]%
        {finin1994kqml}
\bibfield{author}{\bibinfo{person}{Tim Finin} {et~al\mbox{.}}}
  \bibinfo{year}{1994}\natexlab{}.
\newblock \showarticletitle{KQML as an Agent Communication Language}. In
  \bibinfo{booktitle}{\emph{Proc. Int. Conf. Information and Knowledge
  Management (CIKM)}}. \bibinfo{pages}{456--463}.
\newblock
\href{https://doi.org/10.1145/191246.191322}{doi:\nolinkurl{10.1145/191246.191322}}


\bibitem[{Foundation for Intelligent Physical Agents (FIPA)}(2002)]%
        {fipa2002acl}
\bibfield{author}{\bibinfo{person}{{Foundation for Intelligent Physical Agents
  (FIPA)}}.} \bibinfo{year}{2002}\natexlab{}.
\newblock \bibinfo{title}{FIPA ACL Message Structure Specification (SC00061)
  and FIPA Communicative Act Library}.
\newblock \bibinfo{howpublished}{FIPA Specification SC00061}.
\newblock


\bibitem[Garlan et~al\mbox{.}(2004)]%
        {garlan2004rainbow}
\bibfield{author}{\bibinfo{person}{David Garlan} {et~al\mbox{.}}}
  \bibinfo{year}{2004}\natexlab{}.
\newblock \showarticletitle{Rainbow: Architecture-Based Self-Adaptation with
  Reusable Infrastructure}.
\newblock \bibinfo{journal}{\emph{Computer}} \bibinfo{volume}{37},
  \bibinfo{number}{10} (\bibinfo{year}{2004}), \bibinfo{pages}{46--54}.
\newblock


\bibitem[{Google} and {Linux Foundation}(2025)]%
        {google2025a2a}
\bibfield{author}{\bibinfo{person}{{Google}} {and} \bibinfo{person}{{Linux
  Foundation}}.} \bibinfo{year}{2025}\natexlab{}.
\newblock \bibinfo{title}{Agent2Agent (A2A) Protocol}.
\newblock \bibinfo{howpublished}{Open protocol announced 2025-04-09;
  Apache-2.0; governed by the Linux Foundation}.
\newblock
\urldef\tempurl%
\url{https://a2a-protocol.org/}
\showURL{%
\tempurl}


\bibitem[{Google Cloud}(2026)]%
        {google2026okf}
\bibfield{author}{\bibinfo{person}{{Google Cloud}}.}
  \bibinfo{year}{2026}\natexlab{}.
\newblock \bibinfo{title}{Open Knowledge Format (OKF) Specification, Version
  0.2}.
\newblock \bibinfo{howpublished}{Vendor-neutral open specification; announced
  2026-06-12; accessed 2026-07-26}.
\newblock
\urldef\tempurl%
\url{https://github.com/GoogleCloudPlatform/knowledge-catalog/tree/main/okf}
\showURL{%
\tempurl}


\bibitem[Guti{\'e}rrez et~al\mbox{.}(2024)]%
        {gutierrez2024hipporag}
\bibfield{author}{\bibinfo{person}{Bernal~Jim{\'e}nez Guti{\'e}rrez}
  {et~al\mbox{.}}} \bibinfo{year}{2024}\natexlab{}.
\newblock \showarticletitle{HippoRAG: Neurobiologically Inspired Long-Term
  Memory for Large Language Models}. In \bibinfo{booktitle}{\emph{Advances in
  Neural Information Processing Systems (NeurIPS)}}.
\newblock
\newblock
\shownote{arXiv:2405.14831}.


\bibitem[Hayes-Roth(1985)]%
        {hayesroth1985blackboard}
\bibfield{author}{\bibinfo{person}{Barbara Hayes-Roth}.}
  \bibinfo{year}{1985}\natexlab{}.
\newblock \showarticletitle{A Blackboard Architecture for Control}.
\newblock \bibinfo{journal}{\emph{Artificial Intelligence}}
  \bibinfo{volume}{26}, \bibinfo{number}{3} (\bibinfo{year}{1985}),
  \bibinfo{pages}{251--321}.
\newblock
\href{https://doi.org/10.1016/0004-3702(85)90063-3}{doi:\nolinkurl{10.1016/0004-3702(85)90063-3}}


\bibitem[Kephart and Chess(2003)]%
        {kephart2003vision}
\bibfield{author}{\bibinfo{person}{Jeffrey~O. Kephart} {and}
  \bibinfo{person}{David~M. Chess}.} \bibinfo{year}{2003}\natexlab{}.
\newblock \showarticletitle{The Vision of Autonomic Computing}.
\newblock \bibinfo{journal}{\emph{Computer}} \bibinfo{volume}{36},
  \bibinfo{number}{1} (\bibinfo{year}{2003}), \bibinfo{pages}{41--50}.
\newblock
\href{https://doi.org/10.1109/MC.2003.1160055}{doi:\nolinkurl{10.1109/MC.2003.1160055}}


\bibitem[Kleppmann et~al\mbox{.}(2019)]%
        {kleppmann2019localfirst}
\bibfield{author}{\bibinfo{person}{Martin Kleppmann} {et~al\mbox{.}}}
  \bibinfo{year}{2019}\natexlab{}.
\newblock \showarticletitle{Local-First Software: You Own Your Data, in spite
  of the Cloud}. In \bibinfo{booktitle}{\emph{Proc. ACM SIGPLAN Int. Symp. New
  Ideas, New Paradigms, and Reflections on Programming and Software
  (Onward!)}}. \bibinfo{pages}{154--178}.
\newblock
\href{https://doi.org/10.1145/3359591.3359737}{doi:\nolinkurl{10.1145/3359591.3359737}}


\bibitem[Kleppmann and Beresford(2017)]%
        {kleppmann2017jsoncrdt}
\bibfield{author}{\bibinfo{person}{Martin Kleppmann} {and}
  \bibinfo{person}{Alastair~R. Beresford}.} \bibinfo{year}{2017}\natexlab{}.
\newblock \showarticletitle{A Conflict-Free Replicated JSON Datatype}.
\newblock \bibinfo{journal}{\emph{IEEE Transactions on Parallel and Distributed
  Systems}} \bibinfo{volume}{28}, \bibinfo{number}{10} (\bibinfo{year}{2017}),
  \bibinfo{pages}{2733--2746}.
\newblock
\href{https://doi.org/10.1109/TPDS.2017.2697382}{doi:\nolinkurl{10.1109/TPDS.2017.2697382}}
\newblock
\shownote{arXiv:1608.03960}.


\bibitem[Kleppmann and Howard(2020)]%
        {kleppmann2020byzantine}
\bibfield{author}{\bibinfo{person}{Martin Kleppmann} {and}
  \bibinfo{person}{Heidi Howard}.} \bibinfo{year}{2020}\natexlab{}.
\newblock \bibinfo{title}{Byzantine Eventual Consistency and the Fundamental
  Limits of Peer-to-Peer Databases}.
\newblock
\showeprint[arxiv]{2012.00472}~[cs.DC]
\newblock
\shownote{arXiv:2012.00472}.


\bibitem[Lov\'en et~al\mbox{.}(2026a)]%
        {loven2026neuralrouter}
\bibfield{author}{\bibinfo{person}{Lauri Lov\'en}, \bibinfo{person}{Alexander
  Engelhardt}, \bibinfo{person}{Abhishek Kumar}, \bibinfo{person}{Alaa Saleh},
  \bibinfo{person}{Roberto Morabito}, \bibinfo{person}{Xiaoli Liu},
  \bibinfo{person}{Naser~Hossein Motlagh}, {and} \bibinfo{person}{Sasu
  Tarkoma}.} \bibinfo{year}{2026}\natexlab{a}.
\newblock \bibinfo{title}{Neural Router: Semantic Content Matching for Agentic
  AI}.
\newblock
\showeprint[arxiv]{2605.25701}~[cs.DC]


\bibitem[Lov\'en et~al\mbox{.}(2026b)]%
        {loven2026neuralpubsub}
\bibfield{author}{\bibinfo{person}{Lauri Lov\'en}, \bibinfo{person}{Roberto
  Morabito}, \bibinfo{person}{Abhishek Kumar}, \bibinfo{person}{Susanna
  Pirttikangas}, \bibinfo{person}{Jukka Riekki}, {and} \bibinfo{person}{Sasu
  Tarkoma}.} \bibinfo{year}{2026}\natexlab{b}.
\newblock \bibinfo{title}{Autonomic Federated-Market Orchestration for the
  Edge--Cloud Continuum}.
\newblock
\showeprint[arxiv]{2605.27106}~[cs.DC]


\bibitem[Moreau and Missier(2013)]%
        {moreau2013provdm}
\bibfield{author}{\bibinfo{person}{Luc Moreau} {and}
  \bibinfo{person}{Paolo~(eds.) Missier}.} \bibinfo{year}{2013}\natexlab{}.
\newblock \bibinfo{booktitle}{\emph{{PROV-DM}: The {PROV} Data Model}}.
\newblock \bibinfo{type}{W3C Recommendation}. \bibinfo{institution}{W3C}.
\newblock
\urldef\tempurl%
\url{https://www.w3.org/TR/2013/REC-prov-dm-20130430/}
\showURL{%
\tempurl}
\newblock
\shownote{30 April 2013}.


\bibitem[Nascimento et~al\mbox{.}(2023)]%
        {nascimento2023selfadaptive}
\bibfield{author}{\bibinfo{person}{Nathalia Nascimento} {et~al\mbox{.}}}
  \bibinfo{year}{2023}\natexlab{}.
\newblock \showarticletitle{Self-Adaptive Large Language Model ({LLM})-Based
  Multiagent Systems}. In \bibinfo{booktitle}{\emph{IEEE Int. Conf. Autonomic
  Comp. and Self-Organizing Systems Companion (ACSOS-C)}}.
  \bibinfo{pages}{104--109}.
\newblock
\href{https://doi.org/10.1109/ACSOS-C58168.2023.00048}{doi:\nolinkurl{10.1109/ACSOS-C58168.2023.00048}}


\bibitem[Nii(1986)]%
        {nii1986blackboard}
\bibfield{author}{\bibinfo{person}{H.~Penny Nii}.}
  \bibinfo{year}{1986}\natexlab{}.
\newblock \showarticletitle{Blackboard Systems}.
\newblock \bibinfo{journal}{\emph{AI Magazine}} \bibinfo{volume}{7},
  \bibinfo{number}{2--3} (\bibinfo{year}{1986}).
\newblock
\newblock
\shownote{Part 1: vol.~7(2); Part 2: vol.~7(3)}.


\bibitem[Parashar(2025)]%
        {parashar2025everywhere}
\bibfield{author}{\bibinfo{person}{Manish Parashar}.}
  \bibinfo{year}{2025}\natexlab{}.
\newblock \showarticletitle{Everywhere and Nowhere: Envisioning a Computing
  Continuum for Science}.
\newblock \bibinfo{journal}{\emph{Computing in Science \& Engineering}}
  \bibinfo{volume}{27}, \bibinfo{number}{1} (\bibinfo{year}{2025}),
  \bibinfo{pages}{51--56}.
\newblock
\href{https://doi.org/10.1109/MCSE.2025.3543924}{doi:\nolinkurl{10.1109/MCSE.2025.3543924}}


\bibitem[Piri et~al\mbox{.}(2016)]%
        {piri2016gtn}
\bibfield{author}{\bibinfo{person}{Esa Piri} {et~al\mbox{.}}}
  \bibinfo{year}{2016}\natexlab{}.
\newblock \showarticletitle{{5GTN}: A test network for {5G} application
  development and testing}. In \bibinfo{booktitle}{\emph{2016 European
  Conference on Networks and Communications (EuCNC)}}. IEEE,
  \bibinfo{pages}{313--318}.
\newblock
\href{https://doi.org/10.1109/EuCNC.2016.7561054}{doi:\nolinkurl{10.1109/EuCNC.2016.7561054}}


\bibitem[Pour et~al\mbox{.}(2023)]%
        {pour2023oaei}
\bibfield{author}{\bibinfo{person}{Mina Abd~Nikooie Pour},
  \bibinfo{person}{Alsayed Algergawy}, {et~al\mbox{.}}}
  \bibinfo{year}{2023}\natexlab{}.
\newblock \showarticletitle{Results of the Ontology Alignment Evaluation
  Initiative 2023}. In \bibinfo{booktitle}{\emph{Proc. Int. Workshop on
  Ontology Matching (OM @ ISWC 2023)}}, Vol.~\bibinfo{volume}{3591}.
  \bibinfo{pages}{97--139}.
\newblock


\bibitem[Rasmussen et~al\mbox{.}(2025)]%
        {rasmussen2025zep}
\bibfield{author}{\bibinfo{person}{Preston Rasmussen} {et~al\mbox{.}}}
  \bibinfo{year}{2025}\natexlab{}.
\newblock \bibinfo{title}{Zep: A Temporal Knowledge Graph Architecture for
  Agent Memory}.
\newblock
\showeprint[arxiv]{2501.13956}~[cs.AI]
\newblock
\shownote{arXiv:2501.13956}.


\bibitem[Reimers and Gurevych(2019)]%
        {reimers2019sbert}
\bibfield{author}{\bibinfo{person}{Nils Reimers} {and} \bibinfo{person}{Iryna
  Gurevych}.} \bibinfo{year}{2019}\natexlab{}.
\newblock \showarticletitle{Sentence-{BERT}: Sentence Embeddings using
  {Siamese} {BERT}-Networks}. In \bibinfo{booktitle}{\emph{Proc. Conf.
  Empirical Methods in Natural Language Processing (EMNLP-IJCNLP)}}.
  \bibinfo{pages}{3982--3992}.
\newblock


\bibitem[Rezazadeh et~al\mbox{.}(2025)]%
        {rezazadeh2025collaborativememory}
\bibfield{author}{\bibinfo{person}{Alireza Rezazadeh} {et~al\mbox{.}}}
  \bibinfo{year}{2025}\natexlab{}.
\newblock \bibinfo{title}{Collaborative Memory: Multi-User Memory Sharing in
  LLM Agents with Dynamic Access Control}.
\newblock
\showeprint[arxiv]{2505.18279}~[cs.AI]
\newblock
\shownote{arXiv:2505.18279}.


\bibitem[Salehie and Tahvildari(2009)]%
        {salehie2009self}
\bibfield{author}{\bibinfo{person}{Mazeiar Salehie} {and}
  \bibinfo{person}{Ladan Tahvildari}.} \bibinfo{year}{2009}\natexlab{}.
\newblock \showarticletitle{Self-Adaptive Software: Landscape and Research
  Challenges}.
\newblock \bibinfo{journal}{\emph{ACM Transactions on Autonomous and Adaptive
  Systems (TAAS)}} \bibinfo{volume}{4}, \bibinfo{number}{2}
  (\bibinfo{year}{2009}), \bibinfo{pages}{Article 14}.
\newblock
\href{https://doi.org/10.1145/1516533.1516538}{doi:\nolinkurl{10.1145/1516533.1516538}}


\bibitem[Sambra et~al\mbox{.}(2016)]%
        {sambra2016solid}
\bibfield{author}{\bibinfo{person}{Andrei~V. Sambra} {et~al\mbox{.}}}
  \bibinfo{year}{2016}\natexlab{}.
\newblock \bibinfo{booktitle}{\emph{Solid: A Platform for Decentralized Social
  Applications Based on Linked Data}}.
\newblock \bibinfo{type}{{T}echnical {R}eport}. \bibinfo{institution}{MIT CSAIL
  and Qatar Computing Research Institute}.
\newblock


\bibitem[Shapiro et~al\mbox{.}(2011)]%
        {shapiro2011crdt}
\bibfield{author}{\bibinfo{person}{Marc Shapiro} {et~al\mbox{.}}}
  \bibinfo{year}{2011}\natexlab{}.
\newblock \showarticletitle{Conflict-Free Replicated Data Types}. In
  \bibinfo{booktitle}{\emph{Stabilization, Safety, and Security of Distributed
  Systems (SSS 2011)}}, Vol.~\bibinfo{volume}{6976}.
  \bibinfo{publisher}{Springer}, \bibinfo{pages}{386--400}.
\newblock
\href{https://doi.org/10.1007/978-3-642-24550-3_29}{doi:\nolinkurl{10.1007/978-3-642-24550-3_29}}


\bibitem[Shvaiko and Euzenat(2013)]%
        {shvaiko2013ontologymatching}
\bibfield{author}{\bibinfo{person}{Pavel Shvaiko} {and}
  \bibinfo{person}{J{\'e}r{\^o}me Euzenat}.} \bibinfo{year}{2013}\natexlab{}.
\newblock \showarticletitle{Ontology Matching: State of the Art and Future
  Challenges}.
\newblock \bibinfo{journal}{\emph{IEEE Transactions on Knowledge and Data
  Engineering}} \bibinfo{volume}{25}, \bibinfo{number}{1}
  (\bibinfo{year}{2013}), \bibinfo{pages}{158--176}.
\newblock
\href{https://doi.org/10.1109/TKDE.2011.253}{doi:\nolinkurl{10.1109/TKDE.2011.253}}


\bibitem[{TerminusDB}({[n.\,d.]})]%
        {terminusdb}
\bibfield{author}{\bibinfo{person}{{TerminusDB}}.}
  \bibinfo{year}{[n.\,d.]}\natexlab{}.
\newblock \bibinfo{title}{TerminusDB: A Git-like Knowledge Graph Database}.
\newblock \bibinfo{howpublished}{Software, \url{https://terminusdb.com/}}.
\newblock


\bibitem[Weyns et~al\mbox{.}(2013)]%
        {weyns2013patterns}
\bibfield{author}{\bibinfo{person}{Danny Weyns} {et~al\mbox{.}}}
  \bibinfo{year}{2013}\natexlab{}.
\newblock \showarticletitle{On Patterns for Decentralized Control in
  Self-Adaptive Systems}.
\newblock In \bibinfo{booktitle}{\emph{Software Engineering for Self-Adaptive
  Systems II}}. Vol.~\bibinfo{volume}{7475}. \bibinfo{publisher}{Springer},
  \bibinfo{pages}{76--107}.
\newblock
\href{https://doi.org/10.1007/978-3-642-35813-5_4}{doi:\nolinkurl{10.1007/978-3-642-35813-5_4}}


\bibitem[Wu et~al\mbox{.}(2025)]%
        {wu2024longmemeval}
\bibfield{author}{\bibinfo{person}{Di Wu} {et~al\mbox{.}}}
  \bibinfo{year}{2025}\natexlab{}.
\newblock \showarticletitle{{LongMemEval}: Benchmarking Chat Assistants on
  Long-Term Interactive Memory}. In \bibinfo{booktitle}{\emph{International
  Conference on Learning Representations (ICLR)}}.
\newblock
\newblock
\shownote{arXiv:2410.10813}.


\bibitem[Xu et~al\mbox{.}(2025)]%
        {xu2025amem}
\bibfield{author}{\bibinfo{person}{Wujiang Xu} {et~al\mbox{.}}}
  \bibinfo{year}{2025}\natexlab{}.
\newblock \showarticletitle{A-MEM: Agentic Memory for LLM Agents}. In
  \bibinfo{booktitle}{\emph{Advances in Neural Information Processing Systems
  (NeurIPS)}}.
\newblock


\end{thebibliography}

\appendix

\noindent These appendices collect extended results that the main text states in
summary form: the two convergence
robustness sweeps that exercise the partition axes (Appendices~\ref{supp:sweep}
and~\ref{supp:gate}), the consolidated constructed-probe ablations
(Appendix~\ref{supp:ablations}), the cross-tier and second-substrate results
(Appendices~\ref{supp:m3a},~\ref{supp:lme}), the node-scaling and latency surfaces
(Appendices~\ref{supp:scale}--\ref{supp:latency}), an experiment inventory
(Appendix~\ref{supp:inventory}), the anti-entropy digest that makes convergence
independent of the semantic-routing path's delivery completeness (Appendix~\ref{supp:antientropy}), the
recall and merge-classifier sensitivity sweeps
(Appendices~\ref{supp:overlapk},~\ref{supp:robustness}), the order-sensitivity of the
merge graph under randomized delivery orderings (Appendix~\ref{supp:order}), and an
architecture and control-loop reference (Appendix~\ref{supp:arch}).

\medskip
\noindent Notation (as in the main paper). The merge decision procedure
(main paper, Section~4, ``Merge semantics'') returns one of five outcomes for an
incoming claim: \textsc{insert}, \textsc{merge}, \textsc{relate}, \textsc{conflict}, or
\textsc{reject} (the admission-gate drop). Its pair-adjudication core classifies a
candidate claim pair as \textsc{merge}, \textsc{relate}, \textsc{conflict}, or
\textsc{no-relation} from three signals (claim-key identity $\kappa$, over canonical
content and discrete scope; a similarity signal $\sigma$; and an inference (NLI) signal
$\chi$) at the calibrated operating point
$\theta_{\mathrm{merge}}{=}0.90$, $\sigma_{\mathrm{lo}}{=}0.12$. The status-CRDT
(main paper, Section~5, ``Consistency and trust'') is a grow-only set of status links
whose per-claim status is a lattice-join read over the active\,$\sqsubseteq$\,deprecated\,$\sqsubseteq$\,overruled\,$\sqsubseteq$\,revoked
order. These supplement the experiments of the main paper's Section~7 (``Evaluation'').

\section{Convergence: partition-duration $\times$ reorder-rate sweep}
\label{supp:sweep}
The main paper's in-process CRDT demonstration (Section~7.4, ``Convergence under
concurrent edits and partition heal'') reports a single contested-lifecycle setting; here
we exercise the two axes that experiment names. We sweep partition duration (queued status
updates per claim before heal, $\{2,4,8,16\}$) against reorder rate (the fraction of the
heal delivery that is permuted, $\{0,0.25,0.5,0.75,1.0\}$), $200$ orderings per cell, six
claims. The status-CRDT holds \textbf{modal agreement $1.000$ in every one of the $20$
cells} (a single distinct outcome) regardless of how long the partition ran or how
reordered the heal is; last-writer-wins matches it only in the in-order column ($1.000$ at
reorder $0$) and otherwise collapses to as low as $0.020$, fragmenting into up to $155$
distinct status maps. Order-independence is invariant to both axes for the CRDT and to
neither for the baseline. This demonstrates, across both axes, the order-independence
the lattice-join provides by construction (main paper, Section~5).

\section{Convergence: reject-gate placement}
\label{supp:gate}
\smp{}'s wire discipline gates stale and overruled deltas (main paper, Section~5,
``Consistency and trust''). The gate must not become an order-dependent admission filter:
were a brain to drop an incoming link based on the claim's local status at receipt
time, two brains receiving the same links in different orders would admit different subsets
and diverge. \smp{} therefore admits every link into the grow-only set unconditionally and
applies the gate at materialization-read, a pure function of the converged set. On a
reject-race scenario (a claim with competing overrule and revoke links plus one stale link,
where the stale/overrule gate races the status transition), the materialization-read gate
reconverges to a single status on \textbf{all $200/200$ orderings} (one outcome,
revoked); the naive admission-time gate diverges, splitting $\approx\!50/50$ between
overruled and revoked ($101$ vs $99$ of $200$, two distinct outcomes). Read-time
placement preserves strong eventual consistency.

\section{Additional ablations (constructed probes)}
\label{supp:ablations}
Four further ablations isolate individual design choices, all as deterministic, model-free
checks on small constructed probes (hashing encoder, lexical inference; thresholds
as calibrated). These are necessity and invariant demonstrations, not error rates over a
distribution: each probe is built to need exactly one mechanism, so the figures in
Table~\ref{supp:tab-ablations} are properties of the construction, reproducible by
inspection, not measurements of a population (the measured false-merge rate over a
populated benchmark is the main paper's merge-classification result, Section~7.2, ``Merge classification
quality''). The conflict-and-authority ablation checks that contradictions become
first-class CONTRADICTS edges and that the authority tie-break selects the canonical brain.
The gate-necessity ablation drops the Context, inference, and staleness gates in turn over three
distinct-key probes, each constructed so that exactly one gate's input blocks an
otherwise high-similarity merge. The echo-safety probe checks that the wire discipline
never re-publishes a received claim. The MAC-admission probe grounds the one
cryptographic claim (main paper, Section~5.4, threat model). The token-cost instance
is a single motivating data point on memory economics, not a contribution.

\textbf{Result} (Table~\ref{supp:tab-ablations}). Contradiction representation and authority. \smp{} represents every
injected contradiction as a CONTRADICTS edge rather than silently merging it, where an
auto-merge baseline represents none; the authority tie-break then selects the canonical
brain's claim by construction, which is correct exactly when the canonical brain holds
ground truth and wrong when a peer does; authority settles precedence, not correctness.
Gate necessity. With all gates active no probe false-merges or stale-promotes; dropping each
gate flips exactly its own probe to a (semantic, distinct-key) MERGE and no other, so each
gate is individually necessary and the three guard disjoint failures, none
redundant. The Context gate guards the cross-frame RELATE merge, the inference gate the
contradiction merge, and the staleness gate both a false merge and the promotion of
a deprecated node. This is a logic-necessity check on one probe per gate, not an empirical
rate. Wire discipline. A received claim is re-published zero times, so the echo-safety invariant
holds. MAC verification. A delta carrying a forged MAC and a Patch
tampered after MAC computation are each dropped before any signal is read, while a
correctly MAC-authenticated delta is admitted, a deterministic HMAC probe, the
verification step preceding the decision procedure. Memory economics. Against a cleaned, linked brain the recalled-context token
cost is $32$ versus $49$ tokens per query at $k{=}20$ ($-34.7\%$); the ratio is constant
because it is the same query repeated, an economics data point, not a scaling result.

\begin{table}[t]
  \centering
  \footnotesize
  \caption{Consolidated ablations: deterministic, model-free checks on
  small constructed probes. Each row reports what a mechanism does on a probe built
  to need it, against the ablated or baseline alternative. These are necessity and
  invariant demonstrations, not error rates over a distribution (the measured false-merge
  rate is the main paper's merge-classification result, Section~7.2).}
  \label{supp:tab-ablations}
  \setlength{\tabcolsep}{5pt}
  \renewcommand{\arraystretch}{1.2}
  \begin{tabular}{@{}lll@{}}
    \hline
    Probe (size) & \smp{} & Ablated / baseline \\
    \hline
    \multicolumn{3}{@{}l}{Conflict representation and authority}\\
    contradiction flagging         & $100\%$ CONTRADICTS & auto-merge: $0\%$ \\
    authority tie-break            & picks canonical$^{\dagger}$ & recency-only \\
    \multicolumn{3}{@{}l}{Gate necessity: each gate necessary (distinct-key, same-scope probes)}\\
    Context, cross-frame ($1$)     & RELATE              & drop $\to$ false MERGE \\
    inference, contradiction ($1$) & CONFLICT            & drop $\to$ false MERGE \\
    staleness, stale delta ($1$)   & REJECT (admission)  & drop $\to$ MERGE\,$+$\,stale-promote \\
    \multicolumn{3}{@{}l}{Wire discipline / MAC verification / memory economics}\\
    echo-safety ($1$ claim, $3$ peers) & $0$ re-publishes & invariant holds \\
    MAC admission (forged, tampered) & both dropped & valid: admitted \\
    token cost ($k{=}20$ query)    & $32$ tok            & union $49$ ($-34.7\%$) \\
    \hline
  \end{tabular}\\[2pt]
  {\footnotesize $^{\dagger}$resolves status, not truth: correct iff the canonical
  brain holds ground truth.}
\end{table}

\section{Quantitative cross-tier partition-heal over the real Kafka WAN}
\label{supp:m3a}
The main paper (Section~7.8, ``Cross-tier deployment and per-delta cost'') reports a
seeded multi-trial partition-heal run directly over the real Apache Kafka WAN, upgrading the
single-session cross-tier existence proof to a reconvergence distribution. This section gives
the method and the per-trial data.

\textbf{Topology and partition model.} A laptop hub runs the Kafka broker (KRaft mode) and a
local brain; a CSC HPC brain and a 5GTN edge brain each reach the broker over a real
wide-area network through an \texttt{ssh -R} reverse tunnel. Following the main paper's
delivery-driven partition model (Section~5, ``Consistency and trust''), each trial
(i)~deprecates a fresh claim and waits until all three tiers converge to \textsc{deprecated};
(ii)~partitions the edge brain by dropping its broker link; (iii)~overrules the
claim, which the two connected tiers apply (diverging from the cut-off edge, which still
reads \textsc{deprecated}); (iv)~heals by restoring the link, whereupon the edge brain
re-consumes the missed overrule from its grow-only set and recomputes the lattice-join. We
record, per trial, whether the partition produced a genuine divergence first, whether the
edge reconverged, and the heal latency (from link restoration to the edge reading
\textsc{overruled}).

\textbf{Result.} Over $10$ seeded trials, every trial diverged first and then reconverged to
a single status: reconvergence-success fraction $\mathbf{1.000}$ ($10/10$), with heal latency
$\mathrm{p50}=2990$\,ms, $\mathrm{p95}=3364$\,ms, mean $2712$\,ms
(Table~\ref{supp:tab-m3a}). The heal latency is dominated by WAN reconnection and
missed-offset re-consumption, not the merge step (the per-delta merge cost is tens of
milliseconds; main paper, Section~7.2). The success fraction carries the
status-CRDT's strong-eventual-consistency guarantee onto the real continuum, now as a
distribution over independent trials rather than a single session.

\begin{table}[t]
  \centering
  \footnotesize
  \caption{Per-trial cross-tier partition-heal over the real Kafka WAN (laptop $+$ CSC
  HPC $+$ 5GTN edge). Every trial diverged under the partition and reconverged after the
  heal; heal latency is the link-restoration-to-reconvergence time.}
  \label{supp:tab-m3a}
  \setlength{\tabcolsep}{6pt}
  \begin{tabular}{@{}cccc@{}}
    \hline
    Trial & Diverged first & Reconverged & Heal (ms) \\
    \hline
    1  & yes & yes & 2147.1 \\
    2  & yes & yes & 3007.5 \\
    3  & yes & yes & 2135.4 \\
    4  & yes & yes & 2989.9 \\
    5  & yes & yes & 2145.7 \\
    6  & yes & yes & 2993.1 \\
    7  & yes & yes & 3363.7 \\
    8  & yes & yes & 2161.6 \\
    9  & yes & yes & 3182.4 \\
    10 & yes & yes & 2991.2 \\
    \hline
    \multicolumn{2}{@{}l}{success $1.000$ ($10/10$)} & p50 $2990$ & p95 $3364$ \\
    \hline
  \end{tabular}
\end{table}

\section{Second-substrate recall: LongMemEval}
\label{supp:lme}
The main paper (Section~7.1) reports that the headline distributed-versus-centralized recall
result reproduces on a second, long-horizon substrate. This section gives the method and the
full sweep.

\textbf{Method.} LongMemEval~(Wu et al., ICLR 2025; arXiv 2410.10813) pairs each question with
a haystack of chat sessions (distractors plus the gold session) in which the evidence turns are
flagged \texttt{has\_answer}. We map each question to a retrieval task (one context paragraph
per haystack session, with its turns as units, and each \texttt{has\_answer} turn as a gold item) and
run the same recall machinery as the HotpotQA headline (main paper, Section~7.1): recall@$k$
of the gold evidence turns in the status-live store, upstream of any LLM, under the
centralized, distributed-merge, naive-union, and exact-dedup conditions. We use $40$ questions
($\approx\!550$ turns each), the all-MiniLM-L6-v2 encoder, overlap fractions $\{0,0.5\}$, and
$k\in\{5,10,20\}$. As on HotpotQA, recall@$k$ is computed on the store and encoder and is
LLM-independent. We report recall@$k$ rather than end-task accuracy by design: the headline run
established that the merge-vs-centralized answer-F1 is at parity and model-independent
(the merge delta is in recall and storage, not answer quality), so an LLM answer-accuracy
run on this substrate would re-confirm parity rather than test a distinct claim; we leave that
(billable) confirmation for future work.

\textbf{Result} (Table~\ref{supp:tab-lme}, overlap $0$; overlap $0.5$ is within $\le 0.013$ at
every cell). Distributed merge is recall-non-inferior to centralized at every $k$, and in fact
superior (recall@$5$ $0.675$ vs $0.575$; @$10$ $0.806$ vs $0.731$; @$20$ $0.881$ vs $0.794$),
because the centralized store keeps near-duplicate turns that crowd the gold out of a fixed
top-$k$, the consolidation effect the main paper reports on HotpotQA. Against naive union,
merge leads at the tight $k{=}5$ read ($0.675$ vs $0.650$), ties at $k{=}10$, and trails by $0.013$
at the generous $k{=}20$, but at \textbf{$\approx\!26\%$ fewer live units} ($363$ vs $488$)
throughout. Exact-dedup tracks union, not merge, at every cell: dropping byte-identical
turns recovers none of the storage gap, so the compaction is semantic, as on HotpotQA. The
sample is modest ($n{=}40$) and the metric is retrieval recall, not end-task accuracy; within those
bounds the headline holds on a second, structurally different substrate.

\begin{table}[t]
  \centering
  \footnotesize
  \caption{Second-substrate recall on LongMemEval ($n{=}40$, overlap $0$, all-MiniLM-L6-v2;
  mean recall@$k$ / mean live units). Distributed merge is recall-non-inferior to centralized at
  every $k$ and holds $\approx\!26\%$ fewer live units than union; exact-dedup tracks union, so the
  compaction is semantic.}
  \label{supp:tab-lme}
  \setlength{\tabcolsep}{6pt}
  \begin{tabular}{@{}lccc@{}}
    \hline
    Condition & recall@5 & recall@10 & recall@20 \\
    \hline
    Centralized        & $0.575$ & $0.731$ & $0.794$ \\
    Distributed merge  & $\mathbf{0.675}$ & $0.806$ & $0.881$ \\
    Naive union        & $0.650$ & $0.806$ & $0.894$ \\
    Exact-dedup        & $0.650$ & $0.806$ & $0.894$ \\
    \hline
    \multicolumn{4}{@{}l}{live units: merge $363$ \quad central $375$ \quad union/dedup $488$}\\
    \hline
  \end{tabular}
\end{table}

\section{Node-scaling of sync overhead}
\label{supp:scale}
The main paper (Section~7.5) reports that the semantic-routing advantage widens with the number
of brains, and Section~7.3 sweeps fan-out only to $N{=}4$. This section sweeps cluster size over
a wide, dense range and measures how per-claim delivery overhead scales.

\textbf{Method.} For each $N$ we instantiate $N$ brains, each owning one distinct interest
domain, and a workload of claims each needed by exactly the brain owning its domain (ground
truth by construction). Per published claim we count the brains each policy delivers to:
broadcast delivers to all $N$; semantic routing (top-1 over the merge signal, the
Section~7.5 router) delivers only to the brain that needs it. We report routing recall (needed
brain reached), deliveries per claim, and the broadcast-to-semantic overhead ratio. The domains
are constructed to be cleanly separable so the sweep isolates the delivery-count scaling
of the routing policy; the encoder's discrimination on realistic, overlapping domains is the
separate concern addressed by the Section~7.5 ablation (real domains, with random and
round-robin baselines) and the merge-classification experiment.

\textbf{Result} (Table~\ref{supp:tab-scale}). Across $N\in\{2,3,4,6,8,12,16,24,32,48,64,96,128\}$,
routing recall is $1.000$ throughout: semantic routing reaches the needed brain even among $127$
distractors. Semantic deliveries per claim stay \textbf{flat at $1.0$} (the needed-count,
independent of $N$), while broadcast grows as $N$; the broadcast-to-semantic overhead ratio is
therefore \textbf{linear in $N$}, reaching $128\times$ at $N{=}128$ ($99.2\%$ of broadcast
deliveries wasted). \smp{}'s transport overhead per claim is thus constant in cluster size where
a routing-free fabric's is linear. This extends the Section~7.5 result across cluster size and
gives the empirical form of ``sink load grows with the rate of novel claims, not the number of brains.''

\begin{table}[t]
  \centering
  \footnotesize
  \caption{Node-scaling of sync overhead (representative $N$; full set
  $\{2,3,4,6,8,12,16,24,32,48,64,96,128\}$ in \texttt{scale-sweep.json}). Semantic deliveries
  per claim are flat in $N$ at routing recall $1.000$; broadcast grows as $N$, so the overhead
  ratio is linear in cluster size.}
  \label{supp:tab-scale}
  \setlength{\tabcolsep}{6pt}
  \begin{tabular}{@{}rcccc@{}}
    \hline
    $N$ & recall & semantic deliv/claim & broadcast deliv/claim & broadcast/semantic \\
    \hline
    2   & $1.000$ & $1.0$ & $2$   & $2\times$ \\
    4   & $1.000$ & $1.0$ & $4$   & $4\times$ \\
    8   & $1.000$ & $1.0$ & $8$   & $8\times$ \\
    16  & $1.000$ & $1.0$ & $16$  & $16\times$ \\
    32  & $1.000$ & $1.0$ & $32$  & $32\times$ \\
    64  & $1.000$ & $1.0$ & $64$  & $64\times$ \\
    128 & $1.000$ & $1.0$ & $128$ & $128\times$ \\
    \hline
  \end{tabular}
\end{table}

\begin{figure*}[t]
  \centering
  \includegraphics[width=\textwidth]{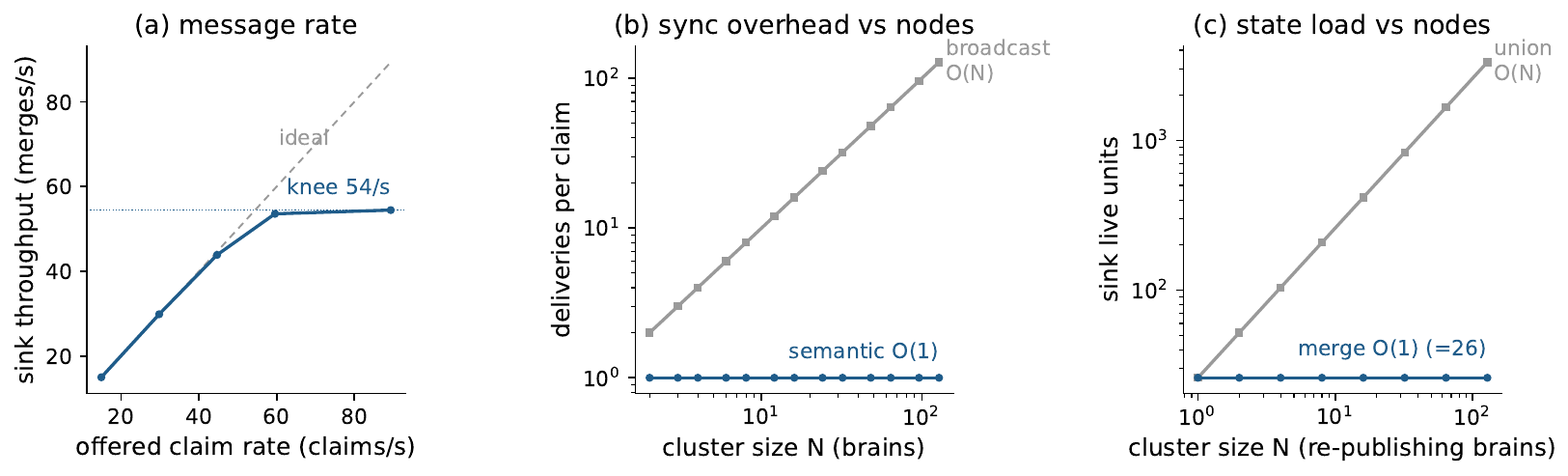}
  \caption{\smp{} scaling, side by side. (a)~sink
  throughput vs offered claim rate: achieved throughput tracks the ideal (grey dashed) to a
  knee at $\approx\!54$ merges/s, then saturates. (b)~per-claim sync overhead vs cluster
  size $N$ (log--log): semantic routing is $O(1)$ (flat) where broadcast is $O(N)$ (unit slope;
  the sweep of this appendix). (c)~sink live-store vs $N$ (log--log): merge-on-receive is
  $O(1)$ (flat at the $26$-fact corpus) where naive union is $O(N)$ (Appendix~G).}
  \label{supp:fig-scaling}
\end{figure*}

\section{Node-scaling of sink state-load}
\label{supp:sinkload}
Companion to the node-scaling of sync overhead (Section~\ref{supp:scale}): where that experiment measures the per-claim sync
overhead as the cluster grows, this measures the sink's state load. The single global
merge sink is a serialization point; the main paper (Sections~7.6,~9) argues its load tracks
the rate of novel claims and is not driven by cluster size $N$ alone. We test the state side directly:
$N$ brains re-publish a fixed novel corpus (replication factor $N$, the redundant-rediscovery
case), and the merge-on-receive sink consolidates the re-publications.

\textbf{Method.} A fixed corpus of $26$ distinct facts (the curated merge-classification fact bank) is written
into one merge sink $N$ times; we report the sink's status-live unit count against the
naive-union baseline ($N\times$ the corpus). The merge engine and encoder are the merge-classification setup.

\textbf{Result} (Table~\ref{supp:tab-sinkload}). The merge sink's live store stays at the
$26$-fact corpus size for every $N$ from $1$ to $128$ ($O(1)$ in cluster size), because
merge-on-receive consolidates each re-publication; naive union grows as $N$ (to $26\times128=3328$
units at $N{=}128$). The sink's state footprint is therefore set by novel content and does
not change with the number of brains feeding it. This is the storage-side complement of the constant per-claim sync
overhead measured above, and of the per-delta compute knee of the saturation-throughput experiment (Section~7.3), which is the sole
serialization limit.

\begin{table}[t]
  \centering
  \footnotesize
  \caption{Sink state-load vs cluster size (corpus $=26$ novel facts; full set in
  \texttt{sink-load.json}). The merge sink stays corpus-sized for all $N$; naive union grows
  as $N$.}
  \label{supp:tab-sinkload}
  \setlength{\tabcolsep}{8pt}
  \begin{tabular}{@{}rcc@{}}
    \hline
    $N$ & merge live units & union live units \\
    \hline
    1   & $26$ & $26$ \\
    2   & $26$ & $52$ \\
    8   & $26$ & $208$ \\
    32  & $26$ & $832$ \\
    128 & $26$ & $3328$ \\
    \hline
  \end{tabular}
\end{table}

\section{Sync-latency surface (exact values)}
\label{supp:latency}
The main paper (Section~7.3, ``Sync latency, throughput, and fan-out scaling'') plots the
publish-to-merge latency surface (Figure~5 there); Table~\ref{supp:tab-latency} gives the exact
per-cell values with confidence intervals. p50 publish-to-merge latency (ms) by added one-way
RTT (continuum hop: $0$ co-located edge, $10$ near-edge, $25$ regional, $50$ edge-to-core) and
fan-out $N$, per routing mode (HotpotQA workload, $15$ seeds/cell, $360$ cells, $0$ failures;
each cell is the across-seed mean of the per-seed p50, $95\%$ bootstrap CIs in brackets).
Semantic p50 is flat-to-falling in $N$ at every RTT tier (each publish reaches the $\sim\!1$
needed peer); broadcast grows with $N$ and is non-monotone under WAN ($N{=}2$ worst at
$50$\,ms), an emulation-confounded artifact of the replication-$2$ workload (scoped out in the
main paper), not an \smp{} property.

\begin{table}[t]
  \centering
  \footnotesize
  \caption{Sync-latency surface: p50 publish-to-merge latency (ms) by added one-way RTT and
  fan-out $N$, per routing mode ($95\%$ bootstrap CIs in brackets).}
  \label{supp:tab-latency}
  \setlength{\tabcolsep}{4pt}
  \begin{tabular}{l l c c c}
    \hline
    Routing & RTT (ms) & $N{=}2$ & $N{=}3$ & $N{=}4$ \\
    \hline
    Broadcast & $0$  & $36.0$ \,[32.8, 42.2]   & $57.1$ \,[56.7, 57.7]   & $80.0$ \,[79.5, 80.5] \\
              & $10$ & $74.6$ \,[74.3, 75.0]   & $76.7$ \,[76.4, 77.1]   & $100.9$ \,[100.3, 101.5] \\
              & $25$ & $134.8$ \,[134.5, 135.1] & $108.1$ \,[107.4, 108.8] & $130.8$ \,[130.2, 131.4] \\
              & $50$ & $234.7$ \,[234.5, 235.0] & $157.3$ \,[156.7, 158.0] & $179.2$ \,[178.6, 179.9] \\
    \hline
    Semantic & $0$  & $10.4$ \,[10.3, 10.6]   & $8.4$ \,[8.2, 8.7]     & $7.2$ \,[7.1, 7.4] \\
             & $10$ & $52.4$ \,[52.2, 52.6]   & $20.1$ \,[16.3, 23.8]  & $8.8$ \,[8.6, 9.1] \\
             & $25$ & $111.8$ \,[111.7, 111.9] & $36.1$ \,[24.2, 47.9]  & $9.1$ \,[8.8, 9.6] \\
             & $50$ & $212.4$ \,[212.2, 212.7] & $62.6$ \,[37.3, 87.9]  & $9.0$ \,[8.8, 9.2] \\
    \hline
  \end{tabular}
\end{table}

\section{Experiment inventory}
\label{supp:inventory}
Table~\ref{supp:tab-inventory} inventories every experiment in the main paper at a glance:
its question, dataset(s), the infrastructure tier it ran on, and its headline result.
Per-experiment detail is in the main paper's Evaluation and in Appendices~\ref{supp:sweep}--\ref{supp:order}.

\begin{table}[h]
  \centering
  \footnotesize
  \caption{Experiment inventory for \smp{}. Datasets and tiers as
  used here; ``sim'' = multi-process single host; ``5GTN'' = University of Oulu 5G test-network
  edge; ``CSC'' = Finnish national HPC.}
  \label{supp:tab-inventory}
  \setlength{\tabcolsep}{4pt}
  \renewcommand{\arraystretch}{1.2}
  \begin{tabular}{p{2.7cm} p{2.0cm} p{1.9cm} p{4.7cm}}
    \toprule
    Experiment & Dataset(s) & Tier & Headline result \\
    \midrule
    Distributed-vs-centralized merge (headline) & HotpotQA distractor; LongMemEval & CSC / sim &
      Recall non-inferior to centralized ($0.630$ vs $0.619$); $+0.035$ vs naive union at
      $\approx\!11\%$ less live storage \\
    Merge classification & $104$-pair gold & local &
      Macro-F1 $0.845$; false-merge $0.013$; ROC AUC $0.968$ \\
    Overlap $\times$ $k$ sensitivity & HotpotQA distractor & sim &
      Merge holds recall flat as overlap rises where union degrades; exact-dedup tracks
      union, so the win is semantic \\
    Classifier robustness \& signal generality & $104$-pair gold & local &
      Macro-F1 $0.845\!\to\!0.684$ (embedding noise) / $0.653$ ($50\%$ NLI corruption),
      false-merge $\le\!0.049$ throughout; across six encoder\,$\times$\,NLI pairings
      macro-F1 $0.765$--$0.845$, with the NLI (not the encoder) setting the false-merge rate \\
    Sync latency \& fan-out & HotpotQA workload & 5GTN (multi-host) &
      p50 flat-to-falling in $N$ (semantic) vs $O(N)$ broadcast \\
    Saturation throughput & HotpotQA workload & edge / local &
      Single-sink knee $\approx\!54$ merges/s (compute-bound) \\
    Convergence + partition heal & constructed events; HotpotQA & 5GTN (multi-host) &
      $30/30$ reconverge after heal vs $11/30$ for last-writer-wins \\
    Overhead vs centralized & HotpotQA & sim / edge &
      $\approx\!11\%$ fewer live units; $\approx\!3\times$ fewer deltas synced \\
    Routing ablation & HotpotQA workload & 5GTN (multi-host) &
      Semantic $0.003$ wasted delivery at matched recall (broadcast $\approx\!2/3$ by constr.) \\
    Node-scaling (sync + sink state) & separable domains & in-process ($N\le 128$) &
      $O(1)$ in cluster size vs $O(N)$ broadcast / naive union \\
    Cross-tier feasibility + Kafka-WAN heal & running example & laptop + 5GTN + CSC &
      Single-session converges end-to-end; $10/10$ WAN reconverge \\
    Additional ablations (four probes) & constructed probes & local &
      Each design choice shown individually necessary \\
    Order-sensitivity of the merge graph & running example; $16$-claim gold-derived & local &
      Running example: $45.5\%$ of $200$ reorderings reproduce the depicted graph;
      $16$-claim workload: $2.0\%$ joint-fingerprint agreement; retrieval-visible in both \\
    Completeness under lossy routing & constructed events & in-process &
      Anti-entropy digest restores agreement $0.117\!\to\!1.000$ at $10\%$ loss \\
    \bottomrule
  \end{tabular}
\end{table}

\section{Completeness under lossy routing: the anti-entropy digest}
\label{supp:antientropy}
Per-claim status convergence (main paper, Section~5) is a strong-eventual-consistency
guarantee given eventually-complete delivery. Under semantic routing at a recall below
one, a status link delivered to no brain is missing state. The anti-entropy digest
closes this gap. A brain summarizes its grow-only status-link set as a per-claim,
order-independent digest: for each claim key, the sorted tuple of its link ids and a $16$-hex
set-hash over them, serialized as canonical sorted-key JSON. Because the digest is a pure
function of the converged link set, two brains with the same links emit byte-identical digests,
so a mismatch is exactly a missing link. An anti-entropy round between two brains
exchanges digests, computes the per-claim symmetric difference of link ids, and each side pulls
and applies the links it lacks; re-applying is safe (the set is a join-semilattice, idempotent
under union), so a round drives both brains to the same link set and the same effective
per-claim status, independent of routing recall.

\textbf{Loss-injection experiment.} We deliver the contested-lifecycle link set
(Appendix~\ref{supp:sweep}) to $12$ brains, dropping each delivery independently with
probability loss (the lossy semantic routing). Table~\ref{supp:tab-antientropy} reports
per-claim status agreement without the digest (lost links are permanent) and with
periodic anti-entropy reconciliation, over a loss sweep with seeds $0$--$4$. Without the digest,
agreement degrades monotonically with loss and the federation fragments into up to $12$ distinct
status maps; with the digest it returns to $1.000$ at every rate (including the protocol's own
$\approx\!0.3\%$ rate) in a single round, at $\approx\!34$\,KB of digest exchanged per round.

\begin{table}[h]
  \centering \footnotesize
  \caption{Anti-entropy under independent per-delivery loss ($12$ brains, $24$ claims, $168$
  links; seeds $0$--$4$). Without the digest, per-claim agreement falls and the federation
  fragments; with periodic reconciliation it is $1.000$ throughout, in one round.}
  \label{supp:tab-antientropy}
  \setlength{\tabcolsep}{6pt}\renewcommand{\arraystretch}{1.15}
  \begin{tabular}{r r r r r}
    \toprule
    Loss & No-digest agree. & (distinct maps) & With-digest agree. & Links repaired \\
    \midrule
    $0.000$ & $1.000$ & $1$ & $1.000$ & $0.0$ \\
    $0.003$ & $0.967$ & $3$ & $1.000$ & $1.4$ \\
    $0.010$ & $0.850$ & $4$ & $1.000$ & $4.2$ \\
    $0.050$ & $0.433$ & $9$ & $1.000$ & $21.6$ \\
    $0.100$ & $0.117$ & $12$ & $1.000$ & $45.6$ \\
    \bottomrule
  \end{tabular}
\end{table}

\textbf{Scope.} The loss model drops each delivery independently per brain, so at these rates
($\le 10\%$ over $12$ brains) each link survives on at least one brain with overwhelming
probability, and a single ring-gossip pass reconciles the federation; the table reports
rounds-to-converge (one, throughout) so this is transparent rather than assumed. A link lost by
every brain (a full disconnection rather than independent routing loss) is recovered
instead by log re-consumption on reconnect (the cross-tier partition-heal of the main paper,
Section~7), not by anti-entropy. The experiment is in-process and deterministic in its seed.

\section{Overlap and $k$ sensitivity (recall)}
\label{supp:overlapk}
This sweep isolates why distributed merge exceeds naive union on recall: semantic
consolidation of paraphrases, not store-size reduction. We sweep overlap fraction
$\{0,0.25,0.5,0.75\}$ against read depth $k\in\{5,10,20\}$ with a matched exact-dedup
baseline (byte-identical duplicates removed, paraphrases kept). This is a separate
matched-setup sweep, so its absolute level differs from the recall experiment of the main
paper (Section~7, at overlap $0.5$, $k{=}5$).

At the tight $k{=}5$ read, distributed merge holds recall flat at $0.714$ across all overlap
fractions, while naive union falls from $0.714$ to $0.644$ as overlap rises (its live store
growing $45\to47$ recallable units against merge's $43$). The gap thus widens with
overlap: a fixed-top-$k$ budget effect, in which un-consolidated paraphrases crowd gold
facts out of the read, and which vanishes by $k{=}20$ (recall $\approx 0.97$ for all
conditions, the budget no longer binding). Exact-dedup tracks union, not
merge (removing byte-identical copies does not recover the gap), so the advantage is the
semantic consolidation of paraphrased duplicates, which only the merge procedure performs.

\section{Merge-classifier robustness and signal generality}
\label{supp:robustness}
Two sweeps establish that the merge-classification result (Section~7) is neither a
clean-input nor a single-backbone artefact.

\textbf{Robustness to degraded signals.} Stressing the upstream signals, macro-F1 degrades
gracefully from $0.845$ to $0.684$ under heavy embedding noise (additive $\sigma$
perturbation, standard deviation $0.30$) and to $0.653$ under $50\%$ NLI corruption, while
the false-merge rate stays below $0.05$ throughout: the safety property survives degraded
inputs rather than being a clean-input artefact.

\textbf{Signal generality (encoder $\times$ NLI sweep).} Sweeping both signals over three
encoders (all-MiniLM-L6-v2, all-MiniLM-L12-v2, all-mpnet-base-v2) and two NLI
cross-encoders (\texttt{nli-deberta-v3-small}, RoBERTa-MNLI) at the fixed operating point,
macro-F1 stays $0.765$--$0.845$ and false-merge $0.000$--$0.038$ across all six pairings:
not a one-backbone artefact. The two signals separate cleanly by role. Discrimination
tracks the encoder: swapping it at a fixed NLI moves macro-F1 by up to $0.061$
($0.845$ MiniLM vs $0.784$ mpnet), where swapping the NLI at a fixed encoder moves it by
$0.018$--$0.019$. Safety tracks the NLI: \texttt{deberta-v3} holds false-merge
$\le 0.013$ for every encoder, RoBERTa raises it to $0.026$--$0.038$, and the encoder
barely moves it, confirming the inference gate, not the embedding, is the contradiction
guard. The operating
point transfers across the two MiniLM encoders (both reproduce $0.845$/$0.013$) but
not to mpnet, whose different cosine scale shifts the optimum (macro-F1 $0.784$): the
per-encoder calibration burden noted in the main paper.

\section{Order-sensitivity of the merge graph}
\label{supp:order}
Section~5 (``Consistency and trust'') states the scope of \smp{}'s
order-independence claim: the per-claim \emph{status} converges
regardless of delivery order because it is a state-based CRDT join over a
grow-only link set, but the merge graph itself, built by the per-pair
decision procedure (Section~4) against already-held state, is not claimed
order-independent. Appendices~\ref{supp:sweep} and~\ref{supp:gate} demonstrate
the status half of that sentence for a fixed link set. This appendix
measures the second half: when the same incoming claims are delivered in
different orders, how much of the resulting graph structure differs, and
whether a caller can see it.

We deliver a workload of claim deltas to a fresh sink brain in $K{=}200$
random orders (one seeded, reproducible shuffle per ordering) through the real
merge-on-receive path (dense rung: minilm encoder, a cross-encoder NLI, the
calibrated operating point $\theta_{\mathrm{merge}}{=}0.90$,
$\sigma_{\mathrm{lo}}{=}0.12$), and compare a canonical fingerprint of the
resulting graph across orderings: the live claim set, the effective per-claim
status map, the typed link multiset, and which claim is the surfaced canonical
representative within each \textsc{supersedes}-consolidated group. Two
workloads: the paper's own four-message running example (Figure~2, the running
example, main paper Section~3), delivered against a receiver that already holds
$c_1$ active as the figure states; and a $16$-claim workload built from four
gold seeds (main paper Section~7.2), each contributing a canonical,
paraphrase, related, and contradicting claim, ingested from three synthetic
source brains and funnelled to one sink.

On the running example, only one outcome ever varies across the $200$
orderings: whether message~1 (the EU paraphrase) merges into $c_1$, as
depicted, or instead relates against message~2's claim. This happens because
\texttt{\_best\_candidate} returns the single highest-similarity local
candidate, not every candidate above the floor: message~1's text is a
paraphrase of $c_1$ ($\sigma\!\approx\!0.9$) but byte-identical to message~2's
text ($\sigma{=}1.0$). If message~2 has already been delivered (a
\textsc{relate} keeps the incoming claim status-live, so it remains a
candidate), it outranks $c_1$ on similarity alone; message~2 is scoped
$\langle\text{US}\rangle$, message~1 is scoped $\langle\text{EU}\rangle$, the
Context gate fails, and the pair falls to \textsc{relate} instead of merging.
The figure's own linear delivery order reproduces the depicted graph exactly,
but only \textbf{$91/200$ ($45.5\%$)} of random reorderings do; the remaining
\textbf{$109/200$ ($54.5\%$)} relate message~1 rather than merging it, a
near coin flip in the paper's own three-claim illustration. Both probe
queries used for retrieval agreement (recall by graph-walk read, the method
the main paper's evaluation, Section~7, uses) show this two-way split: message~1's
claim appears as a third retrieved item only in the \textsc{relate} branch.

On the $16$-claim gold-derived workload the divergence is far larger.
\textbf{$177/200$ ($88.5\%$)} of orderings produce a joint fingerprint seen
nowhere else in the run; the largest identical-fingerprint cluster is
$4/200$ ($2.0\%$). The live claim set itself differs across $112/200$ orderings, not only the
link structure: delivery order changes which claims survive as the store's
content. The status map
tracks this ($131/200$ distinct) because it is a function of which claim got
deprecated under which \textsc{supersedes} choice, not an independent signal;
this is not a counterexample to the status-CRDT theorem of Section~5, which
concerns a fixed link set's join, because here the orderings produce
different link sets to begin with. The representative choice (which
claim a consolidated group surfaces as canonical) is the least scattered
component but still far from order-independent ($65/200$ distinct). Every
probe query (the four seeds' canonical texts) shows retrieval disagreement
across orderings under the same status-aware top-$k$ read the main paper's
recall evaluation uses: divergence is visible to a caller, not confined to
internal bookkeeping.

Candidate-cost instrumentation on the same $200$-ordering gold-derived run
(pooled over $3{,}200$ \texttt{\_best\_candidate} calls, warm embedding
cache) shows the per-call mechanism is cheap: candidate-set size (local
claims scanned) has median $4$, $95$th percentile $7$; scan latency has
median $0.23$\,ms, $95$th percentile $0.30$\,ms. Novel rate (the
\textsc{insert} fraction of the Patch stream) has median $0.19$ per ordering.
The order-sensitivity above is therefore a structural property of comparing
against the single best local candidate seen so far, not a performance
pathology: every individual decision remains a deterministic, correct, cheap
application of the classifier to whichever candidate the delivery order happened
to present.

The measurement above holds for the store size it was drawn from: a median
of only $4$ (p95 $7$) simultaneously ACTIVE local claims. A separate sweep
makes candidate-set size the controlled variable: a store fixed at
$N_{\mathrm{local}}\in\{16,64,256,1024,4096\}$ ACTIVE claims, scored against
$K{=}200$ incoming claims per size (a seeded paraphrase/novel mix) through
the same instrumented \texttt{\_best\_candidate}; paraphrases reword rather
than repeat a held claim verbatim, so the exact-key shortcut never fires and
candidate-set size scanned equals $N_{\mathrm{local}}$ exactly. Scan latency
is linear (Table~\ref{supp:tab-candscale}; median-latency fit
$r^2{=}0.999998$): $\mathbf{3.80}$\,ms per $\mathbf{1{,}000}$ claims scanned
on a near-zero intercept. Through $1{,}024$ claims the scan ($3.9$\,ms)
stays a minority of the NLI step's per-delta cost ($19.4$ of $23.5$\,ms
edge-CPU total, main paper Section~7.8); only past roughly
$\mathbf{3{,}500}$--$\mathbf{6{,}200}$ simultaneously ACTIVE claims, bracketed
directly by the sizes swept, does the scan reach parity with the NLI step.

\begin{table}[h]
  \centering
  \footnotesize
  \caption{Claim-count scaling of candidate selection ($K{=}200$ incoming
  claims per store size, seed $0$, dense rung). Candidate-set size scanned
  equals $N_{\mathrm{local}}$ by construction; scan latency grows linearly
  ($3.80$\,ms/$1$k claims); the NLI-dominated residual (matched deltas, the
  rest of the per-delta call) stays flat, so the two curves cross once the
  scan reaches the residual's scale.}
  \label{supp:tab-candscale}
  \setlength{\tabcolsep}{6pt}\renewcommand{\arraystretch}{1.15}
  \begin{tabular}{r r r r}
    \toprule
    $N_{\mathrm{local}}$ & Scan median (ms) & Scan p95 (ms) & NLI residual, matched (ms) \\
    \midrule
    $16$    & $0.089$  & $0.101$  & $11.99$ \\
    $64$    & $0.282$  & $0.323$  & $11.72$ \\
    $256$   & $1.024$  & $1.176$  & $12.15$ \\
    $1{,}024$ & $3.925$  & $4.269$  & $13.24$ \\
    $4{,}096$ & $15.591$ & $17.550$ & $16.63$ \\
    \bottomrule
  \end{tabular}
\end{table}

\section{Architecture and control-loop reference}
\label{supp:arch}
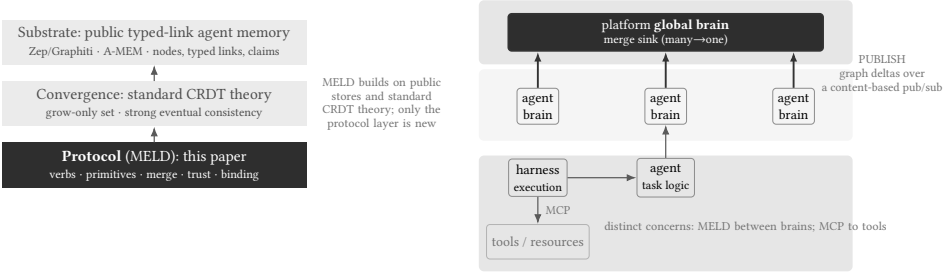
\begin{figure*}[t]
  \centering
  \resizebox{0.90\textwidth}{!}{%
  \begin{tikzpicture}[
    font=\small,
    band/.style={align=center, inner sep=4pt, minimum height=0.9cm,
                 minimum width=6.2cm},
    newband/.style={band, fill=black!82, text=white},
    oldband/.style={band, fill=black!8, text=black!70},
    note/.style={font=\scriptsize, text=black!55, align=center},
    box/.style={draw=black!45, line width=0.5pt, rounded corners=2pt,
                align=center, inner sep=3pt, font=\footnotesize, minimum height=0.7cm},
    newbox/.style={box, draw=black!80, line width=1.0pt},
    flow/.style={-{Latex[length=2.0mm]}, black!65, line width=0.7pt},
    sync/.style={-{Latex[length=2.0mm]}, black!80, line width=1.1pt}
  ]
    \node[newband] (proto) at (0,0)
      {\textbf{Protocol} (\smp{}): this paper\\[-1pt]
       {\scriptsize verbs $\cdot$ primitives $\cdot$ merge $\cdot$ trust $\cdot$ binding}};
    \node[oldband] (theory) at (0,1.25)
      {Convergence: standard CRDT theory\\[-1pt]
       {\scriptsize grow-only set $\cdot$ strong eventual consistency}};
    \node[oldband] (subst) at (0,2.5)
      {Substrate: public typed-link agent memory\\[-1pt]
       {\scriptsize Zep/Graphiti $\cdot$ A-MEM $\cdot$ nodes, typed links, claims}};

    \draw[flow] (proto.north) -- (theory.south);
    \draw[flow] (theory.north) -- (subst.north |- subst.south);
    \node[note, anchor=west, text width=2.5cm] at (3.25,1.25)
      {\smp{} builds on public stores and standard CRDT theory; only the protocol layer is new};

    \begin{scope}[xshift=10.4cm]
      \begin{scope}[on background layer]
        \fill[black!10, rounded corners=3pt] (-3.8, 2.05) rectangle (3.8, 3.35);
        \fill[black!4,  rounded corners=3pt] (-3.8, 0.50) rectangle (3.8, 1.95);
        \fill[black!10, rounded corners=3pt] (-3.8,-2.15) rectangle (3.8, 0.20);
      \end{scope}
      \node[newbox, fill=black!82, text=white, minimum width=6.4cm] (gb) at (0,2.7)
        {platform \textbf{global brain}\\[-2pt]{\scriptsize merge sink (many$\to$one)}};

      \node[box] (b1) at (-2.6,1.2) {agent\\brain};
      \node[box] (b2) at ( 0.0,1.2) {agent\\brain};
      \node[box] (b3) at ( 2.6,1.2) {agent\\brain};
      \draw[sync] (b1.north) -- (gb.south -| b1.north);
      \draw[sync] (b2.north) -- (gb.south);
      \draw[sync] (b3.north) -- (gb.south -| b3.north);
      \node[note, anchor=west] at (3.05,1.85)
        {PUBLISH\\graph deltas over\\a content-based pub/sub};

      \node[box] (har) at (-2.6,-0.25) {harness\\{\scriptsize execution}};
      \node[box] (ag)  at ( 0.0,-0.25) {agent\\{\scriptsize task logic}};
      \draw[flow] (har.east) -- (ag.west);
      \draw[flow] (ag.north) -- (b1.south -| ag.north) ;
      \node[box, draw=black!30, text=black!55] (tools) at (-2.6,-1.55)
        {tools / resources};
      \draw[flow] (har.south) -- (tools.north)
        node[midway, right=1pt, note] {MCP};
      \node[note, anchor=west, text width=6.0cm] at (-1.5,-1.25)
        {distinct concerns: \smp{} between brains; MCP to tools};
    \end{scope}
  \end{tikzpicture}}
  \caption{The layer stack and the deployment topology. Left: \smp{} is the new protocol layer over a public typed-link agent-memory substrate (Zep/Graphiti, A-MEM), with its status convergence resting on standard CRDT theory; both foundations are public/standard, and only the protocol layer, drawn dark, is new.
  Right: harnesses (execution), agents (task logic), and agent brains (knowledge) are kept distinct; many agent brains PUBLISH graph deltas, content-routed over a pub/sub fabric, into a platform global brain, whose sink is the merge procedure, not an opaque aggregator. It runs the same procedure as any brain and holds no special authority; the federation converges without it. On a harness, the agent speaks MCP to tools and \smp{} to brains.}
  \label{fig:architecture}
\end{figure*}

Figure~\ref{fig:architecture} shows the three-layer stack and the deployment topology
described in the main paper's Section~6.1: harnesses, agents, and brains as distinct
layers, with the protocol layer new and everything beneath it reused, and the
global-brain funnel as an aggregation role over sovereign peers.

\begin{table}[h]
  \centering
  \small
  \caption{\smp{} as a self-managing control loop: one authenticated Patch is the only
  state mutation. Self-managing, not self-optimizing (thresholds calibrated
  offline).}
  \label{tab:control}
  \setlength{\tabcolsep}{4pt}
  \begin{tabular}{@{}p{0.26\columnwidth}p{0.66\columnwidth}@{}}
    \toprule
    \textbf{Control element} & \textbf{In \smp{}} \\
    \midrule
    Monitored state & an incoming graph delta versus the held claim it candidate-matches \\
    Uncertainty & embedding similarity $\sigma$ and an NLI contradiction verdict $\chi$ (reliability-bounded) \\
    Decision policy & five-outcome procedure \textsc{insert}/\textsc{merge}/\textsc{relate}/\textsc{conflict}/\textsc{reject} under Context, authority, and freshness gates (main paper, Section~4) \\
    Adaptation action & one authenticated \emph{Patch}, the sole object that mutates local state \\
    Invariant & a detected contradiction is never silently merged; per-claim status is strong-eventually-consistent; a canonical brain is not silently rewritten \\
    Failure model & benign-fault (crash, partition, reorder, loss, stale/overruled deltas); not Byzantine \\
    \bottomrule
  \end{tabular}
\end{table}

\end{document}